\documentclass[onecolumn,11pt]{article}
\usepackage{mymacros}

\title{Holographic complexity of the extended Schwarzschild-de Sitter space}
\author[a]{Sergio E. Aguilar-Gutierrez,}
\author[b]{Stefano Baiguera}
\author[c]{and Nicolò Zenoni}
\affiliation[a]{Institute for Theoretical Physics, KU Leuven,\\ Celestijnenlaan 200D, B-3001 Leuven, Belgium}
\affiliation[b]{Department of Physics, Ben-Gurion University of the Negev,\\ David Ben Gurion Boulevard 1, Beer Sheva 84105, Israel}
\affiliation[c]{Department of Physics, Osaka University,\\ Toyonaka, Osaka 560-0043, JAPAN}
\emailAdd{sergio.ernesto.aguilar@gmail.com, stefano.baiguera92@gmail.com, nicolo@hetmail.phys.sci.osaka-u.ac.jp}
\abstract{According to static patch holography, de Sitter space admits a unitary quantum description in terms of a dual theory living on the stretched horizon, that is a timelike surface close to the cosmological horizon.
In this manuscript, we compute several holographic complexity conjectures in a periodic extension of the Schwarzschild-de Sitter black hole.
We consider multiple configurations of the stretched horizons to which geometric objects are anchored.
The holographic complexity proposals admit a hyperfast growth when the gravitational observables only lie in the cosmological patch, except for a class of complexity=anything observables that admit a linear growth. 
All the complexity conjectures present a linear increase when restricted to the black hole patch, similar to the AdS case.
When both the black hole and the cosmological regions are probed, 
codimension-zero proposals are time-independent, while codimension-one proposals can have non-trivial evolution with linear increase at late times.
As a byproduct of our analysis, we find that codimension-one spacelike surfaces are highly constrained in Schwarzschild-de Sitter space. 
Therefore, different locations of the stretched horizon give rise to different behaviours of the complexity conjectures.}

\begin{document}

\begin{flushright}
 	\hfill{\small OU-HET-1220}
\end{flushright}

\maketitle
\newpage
\section{Introduction}
\label{sec:Intro}

The holographic principle, asserting that the physical properties of a gravitational system are encoded in its boundary \cite{tHooft:1993dmi,Susskind:1994vu}, has garnered substantial support from tools in quantum information, including the interpretation of the entropy of black holes as an area \cite{Ryu:2006bv}.
Most of the explicit tests of this conjecture have been realized in the framework of AdS/CFT correspondence \cite{Maldacena:1997re}.
However, there has recently been a pressing need to extend holography to de Sitter (dS) space, which describes the early and late stages of our universe (\eg see \cite{Galante:2023uyf} for a review).
This idea is supported by the fact that the cosmological horizon of dS space also carries information about the temperature and the horizon entropy, as observed by Gibbons and Hawking \cite{PhysRevD.15.2738}.

\paragraph{Static patch holography.}
In holography, the \textit{central dogma} states that black holes can be described in terms of unitary quantum systems carrying $\exp \le A/4 G_N \ri$ degrees of freedom, as observed from the outside of an event horizon with area $A$ \cite{Almheiri:2020cfm}.
In a similar way, it was conjectured that regions delimited by a cosmological horizon can be described from their inside as unitary quantum systems with a finite number of degrees of freedom \cite{Bousso:1999dw,Bousso:2000nf,Banks:2006rx,Anninos:2011af,Banks:2018ypk,Shaghoulian:2021cef}.
In particular, this applies to the \textit{static patch} of dS space, which is the region in causal contact with an observer in this geometry.
The so-called \textit{static patch holography} assumes that the putative dual theory lives on the \textit{stretched horizon}, a timelike codimension-one surface located just inside the cosmological horizon \cite{Susskind:2021esx,Susskind:2022dfz,Lin:2022nss,Susskind:2022bia,Susskind:2023hnj,Susskind:2023rxm,Rahman:2022jsf} (see also \cite{Nomura:2017fyh,Nomura:2019qps,Murdia:2022giv})\footnote{We consider that both the interior and exterior geometry with respect to a static patch observer can be described by stretched horizon holography, as recently studied in the bilayer proposal \cite{Shaghoulian:2022fop,Franken:2023pni,Franken:2023jas}. We thank Victor Franken for discussions on this point.}.
While there exist alternative proposals for a holographic dual of dS space, in this work we will not focus on them.\footnote{We refer the reader to the review \cite{Galante:2023uyf} and references therein for other approaches to dS holography, including the dS/CFT correspondence \cite{Strominger:2001pn} and centaur geometries \cite{Anninos:2017hhn}. }

There are multiple reasons to believe that static patch holography may provide a well-defined framework.
First, it has been recently understood that timelike boundaries play an important role in dS space.
For instance, in three bulk dimensions, it is possible to provide a dual interpretation in terms of irrelevant $T\bar{T}$ deformations of two-dimensional CFTs, followed by the inclusion of an additional term that introduces a positive cosmological constant in the geometry \cite{Shyam:2021ciy}.
In this way, one can perform a microstate counting inside the static patch \cite{Lewkowycz:2019xse,Coleman:2021nor}, which captures the leading and the logarithmic contributions to the entropy \cite{Anninos:2020hfj}.
A further hint of the role played by timelike boundaries is that they define sensible thermodynamics \cite{Banihashemi:2022htw}.
Second, the role of the observer in the presence of gravity has recently been stressed in \cite{Chandrasekaran:2022cip,Witten:2023qsv,Witten:2023xze,Mirbabayi:2023vgl,Aguilar-Gutierrez:2023odp,Kudler-Flam:2023qfl}. 
In particular, the static patch of dS space provides a setting where an observer needs to be included to define a sensible algebra of observables.

\paragraph{Holographic complexity conjectures.}
An exciting advantage within the framework of static patch holography is that many lessons from the AdS/CFT correspondence carry over. This has allowed several developments in quantum information applied to dS space, including holographic complexity. 
Complexity heuristically quantifies the difficulty of reaching a target state by acting with a sequence of unitary operators on a reference state (\eg see \cite{Chapman:2021jbh} for a review).
In quantum information, complexity is used to establish the advantages of quantum over classical computation, and it is applied as a measure of quantum chaos in many-body systems. 
In the holographic context, several geometric observables have emerged, with the idea to capture the interior growth of black holes and to match with the computational complexity of a dual state in the quantum theory \cite{Maldacena:2013xja,Susskind:2014moa}.

We briefly summarize the holographic conjectures. Consider a $(d+1)$--dimensional bulk background with a codimension-one boundary containing a slice $\Sigma$ where a dual quantum state is defined.
Complexity=volume (CV) relates complexity to the maximal volume $V$ of a codimension-one hypersurface $\mathcal{B}$ anchored at the $(d-1)$--dimensional boundary slice $\Sigma$ \cite{Susskind:2014rva,Stanford:2014jda}
\begin{equation}\label{eq:CV 1st}
    \mathcal{C}_V (\Sigma) =\max_{\Sigma=\partial \mathcal{B}}\frac{{V}(\mathcal{B})}{G_N\ell} \, ,
\end{equation}
where $G_N$ is Newton's constant and $\ell$ is an arbitrary length scale, usually identified as the (A)dS radius.
Complexity=volume 2.0 (CV2.0) associates complexity to the spacetime volume $V_{\rm WDW}$ of the Wheeler-De Witt (WDW) patch, \ie the bulk domain of dependence of a spacelike surface anchored at the boundary slice $\Sigma$ \cite{Couch:2016exn}
\begin{equation}
    \mathcal{C}_{\rm 2.0V}(\Sigma)=\frac{V_{\rm WDW}}{G_N\ell^2}~.
\end{equation}
Complexity=action (CA) is defined in terms of the on-shell gravitational action $I_{\rm WDW}$ of the WDW patch \cite{Brown:2015bva, Brown:2015lvg}
\begin{equation}
    \mathcal{C}_A (\Sigma)=\frac{I_{\rm WDW}}{\pi} \, .
\end{equation}
Finally, there is a plethora of additional holographic conjectures, the complexity=anything (CAny), which define a class of observables 
displaying linear growth at late times, as well as the switchback effect in the presence of shock waves in planar AdS black hole geometries \cite{Belin:2021bga, Belin:2022xmt,Omidi:2022whq,Jorstad:2023kmq}.\footnote{The switchback effect is a delay in the growth of complexity arising as a consequence of inserting a perturbation in the system \cite{Stanford:2014jda}. }
These observables are evaluated in a spacetime region identified by the extremization of a certain functional. 
The characteristic features of linear growth and switchback effect unify all the previous proposals, which can be obtained as special limits of CAny.
In this manuscript, we will be interested in a particular class
of codimension-one CAny observables defined on constant-mean curvature (CMC) slices, introduced in \cite{Belin:2022xmt} (see section \ref{sec:CAny CMC} for the precise details).
Depending on the extrinsic curvature on the CMC slices, they will generally admit a different behaviour compared to other complexity proposals.

\paragraph{Holographic complexity in de Sitter space.}
Several intuitions and a plethora of examples involving all the holographic proposals have been considered in asymptotically AdS space.
In parallel, several investigations have been performed on the quantum side of AdS/CFT duality, \eg in the form of tensor networks with renormalization group-inspired techniques, or as continuous circuits in quantum mechanics (QM) and quantum field theory (QFT).\footnote{We refer the reader to the references in the review \cite{Chapman:2021jbh} for more details.  }

The status of complexity in asymptotically dS space is less clear since holography is less understood, and it is more difficult to find a dual quantum circuit reproducing the expected features, despite the existence of different toy models \cite{Bao:2017iye,Bao:2017qmt,Niermann:2021wco,Cao:2023gkw}. 
However, as long as one can define diffeomorphism-invariant observables in a covariant manner, it is likely that they will have a holographic dual realization. 
For this reason, one might speculate that the holographic complexity observables can have a holographic realization in dS space too.
Furthermore, by construction, all the holographic complexity proposals have very similar properties in AdS holography. Applying them outside of such a setting will allow us to identify the differences between various proposals, as well as their universal aspects.

This point of view has motivated different studies about holographic complexity in dS space. 
In this case, the prescriptions introduced above require all the geometric objects to be anchored to the timelike boundary of the static patch, \ie the stretched horizon \cite{Susskind:2021esx}. 
The time coordinate of a putative boundary theory is taken to run along the stretched horizon itself.
Arguably, the most striking feature of the complexity proposals in asymptotically dS space is the existence of a divergent (also called \textit{hyperfast}) growth of complexity \cite{Jorstad:2022mls,Auzzi:2023qbm,Anegawa:2023wrk} at some finite time scale.\footnote{This feature is not shared by the volume conjecture in the context of two-dimensional centaur geometries, where the growth of complexity is finite and stops at some fixed time \cite{Chapman:2021eyy}.}
This behaviour is in sharp contrast with the eternal evolution of gravitational quantities in asymptotically AdS space, approaching a linear increase at late times.
However, a certain subset of codimension-one observables inside the class of CAny proposal is still able to grow forever \cite{Aguilar-Gutierrez:2023zqm,Aguilar-Gutierrez:2023tic}. 
Once we incorporate black holes in dS space, \ie we consider Schwarzschild-de Sitter (SdS) space, holographic complexity in the inflating region of the geometry does not qualitatively change.
Despite the different outcomes described so far, all the proposals display a switchback similar to black holes once shock waves are introduced in the background \cite{Anegawa:2023dad,Baiguera:2023tpt,Aguilar-Gutierrez:2023pnn}.
The similar realization of the switchback effect for all the complexity conjectures may be understood as a consequence of the universal feature that the insertion of a perturbation consistent with the null energy condition creates causal contact between regions that were originally separated in the geometry \cite{Gao:2000ga}.
This ultimately allows for the transfer of information \cite{Aalsma:2021kle,Aguilar-Gutierrez:2023ymx}.

\begin{figure}[t!]
    \centering
\includegraphics[width=\textwidth]{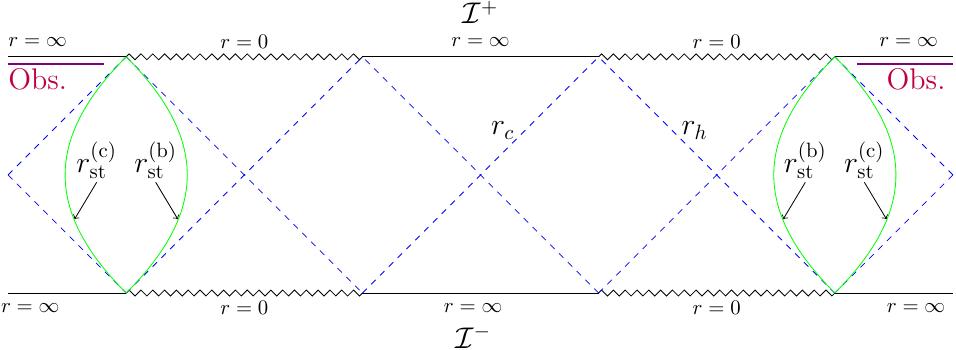}
\caption{SdS$_{d+1}^n$ space (illustrated for $n=2$) where $r_{\rm st}^{\rm (h)}$, $r_{\rm st}^{\rm (c)}$ (both in green) denote the stretched horizons close to the black hole and the cosmological horizon, respectively. $r_h$ and $r_c$ denote the black hole and cosmological horizon radii. The purple line region near $\mathcal{I}^+$ indicates a spacelike slice where an observer could collect information encoded in the inflating region. }
    \label{fig:extSdS}
\end{figure}

\paragraph{Motivations and new observables.} 
The distinction between hyperfast and linear growth at late times is a consequence of the extremal surfaces (where complexity is evaluated) reaching or not timelike infinity $\mathcal{I}^{\pm}$. 
So far, the explorations of holographic complexity in the SdS background only focused on the inflating patch (the region outside the cosmological horizon $r_c$ in Fig.~\ref{fig:extSdS}) \cite{Baiguera:2023tpt,Aguilar-Gutierrez:2023zqm,Aguilar-Gutierrez:2023pnn}.
Instead, no investigations have focused on the black hole patch (the region inside the black hole horizon $r_h$ in Fig.~\ref{fig:extSdS}).
As anticipated above, the central dogma in (A)dS space gives great importance to the region located outside the black hole and inside the cosmological horizons, since it is believed to have a unitary description in terms of a quantum system.
An observer in such a region may have the possibility to describe both the interior of a black hole and the exterior of a cosmological horizon.
In this paper, we study the holographic complexity conjectures for several different choices of the stretched horizons to which the extremal surfaces of interest can be anchored.
From the lenses of dS holography, this procedure might allow us to predict the behaviour of new observables for the putative dual quantum theory.
The background of interest is the periodic extension of the SdS black hole with $n$ copies (that we will denote as SdS$^n$), see Fig.~\ref{fig:extSdS}. 
The possible choices for the timelike surfaces to which geometric observables can be anchored is partially motivated by the proposal of holographic screens by Bousso \cite{Bousso:1999cb,Bousso:1999dw}, which has been previously used to compute holographic complexity in FRLW cosmologies \cite{Caginalp:2019fyt}.

Our motivations for studying complexity in the extended SdS space are two-fold. First, this type of spacetime has been used to construct multiverse models. One instance is in the context of dS space proliferation \cite{Bousso:1998bn,Bousso:1999ms}, where a near-extremal extended Nariai black hole can lead to different disconnected universes due to quantum effects. Other toy models of eternal inflation use the extended SdS background to study the von Neumann entropy for a subregion of $\mathcal{I}^+$ in lower-dimensional quantum gravity \cite{Aguilar-Gutierrez:2021bns,Shaghoulian:2022fop} ({see \cite{Langhoff:2021uct} for a higher dimensional implementation).}, which has also motivated certain recent braneworld multiverse models \cite{Yadav:2023qfg,Aguilar-Gutierrez:2023zoi}. 
In the quantum cosmology context, the typical setting involves a meta-observer living on a spacelike surface (see the horizontal purple lines in Fig.~\ref{fig:extSdS}), from which they are able to collect coarse-grained information on the history of the spacetime. In these latter approaches, the extended SdS black hole has proven to be a useful tool (\eg see \cite{Hartle:2016tpo}).
Second, we want to accumulate evidence for universal and non-universal features of the complexity conjectures in more general asymptotically dS geometries, an effort initiated in \cite{Aguilar-Gutierrez:2024xfi}.

\begin{figure}[t!]
    \centering
    \subfloat[Case 1]{ \label{subfig:case1} \includegraphics[width=0.3\textwidth]{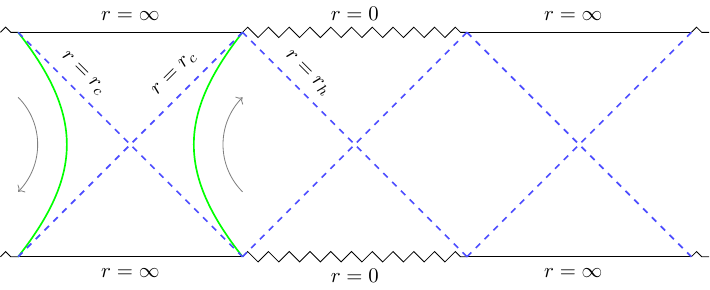}}
\qquad \subfloat[Case 2]{ \label{subfig:case2} \includegraphics[width=0.3\textwidth]{Figures/Fig4}} \\
 \subfloat[Case 3]{ \label{subfig:case3} \includegraphics[width=0.3\textwidth]{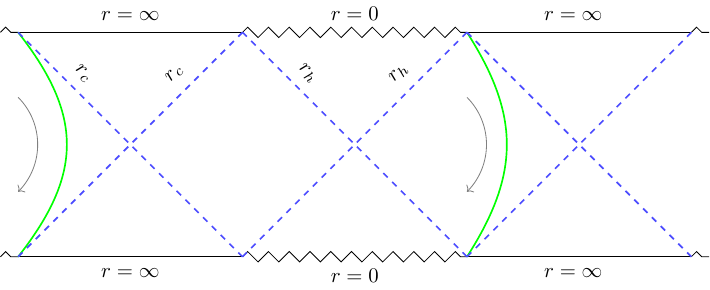}}
\hfill \subfloat[Case 4]{\label{subfig:case4} \includegraphics[width=0.3\textwidth]{Figures/Fig6}}
\hfill \subfloat[Case 5]{\label{subfig:case5} \includegraphics[width=0.3\textwidth]{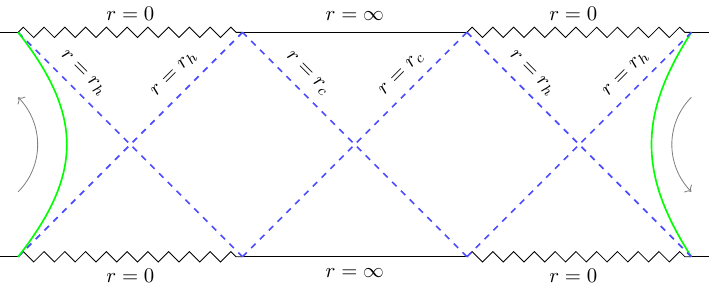}}
    \caption{Summary of the possible configurations for the stretched horizons. Holographic observables are located between the stretched horizons.  }
    \label{fig:summary_cases}
\end{figure}

\paragraph{Main results.}
The extended SdS background allows for several possible configurations of the stretched horizons, which are summarized in Fig.~\ref{fig:summary_cases} (the details will be given in section \ref{ssec:stretched_horizons}).
The setting denoted as case 1 contains a single copy of the SdS geometry where the geometric quantities only probe the cosmological patch. This is the kind of holographic complexity already studied in the literature in empty dS space, which here we extend to the case of a black hole.
The configuration in case 2 focuses on the black hole region only, since it probes the interior of the corresponding event horizon.
This is similar to the investigations of holographic complexity performed in asymptotically AdS geometries.
The main novelty of this work is provided by cases 3-5, where gravitational observables probe both the interior of the black hole and the exterior of the cosmological horizon. This is possible when the stretched horizons belong to static patches in non-consecutive regions of the Penrose diagram.

\begin{table}[t!]   
\begin{center}    
\begin{tabular}  {|c|c|c|} \hline  &  \textbf{Case 1 (Fig.~\ref{subfig:case1})} & \textbf{Case 2 (Fig.~\ref{subfig:case2})}  \\ \hline
\rule{0pt}{4.9ex} $\mathcal{C}_{V2.0}$  & Hyperfast \eqref{eq:rateCV20_case1}  & Linear \eqref{eq:late_times_CV20_case2}  \\
\rule{0pt}{4.9ex} $\mathcal{C}_A$ &  Hyperfast \eqref{eq:approx_rate_CA_case1} & Linear \eqref{eq:rate_CA_case2_latelimit}  \\ 
 \rule{0pt}{4.9ex}  $\mathcal{C}_V$ &  Hyperfast \eqref{eq:ratelate_CV_case1}  & Linear \eqref{eq:ratelate_CV_case2}  \\ 
 \rule{0pt}{4.9ex} $\mathcal{C}^{\epsilon} $ &   
\eqref{eq:Late time growth} 
$ \begin{cases}
    \mathrm{Hyperfast} & \mathrm{if} \,\, |K_{\epsilon}| < K_{\rm crit} \\ 
     \mathrm{Linear} & \mathrm{if} \,\, |K_{\epsilon}| \geq K_{\rm crit}
\end{cases} $
 &  Linear \eqref{eq:Late time growth}  \\[0.2cm]
\hline
\end{tabular}   
\vskip 3mm
\begin{tabular}  {|c|c|c|} \hline  & \textbf{Case 3} (Fig.~\ref{subfig:case3}) & \textbf{Cases 4-5 (Fig.~\ref{subfig:case4}--\ref{subfig:case5})} \\ \hline
\rule{0pt}{4.9ex} $\mathcal{C}_{V2.0}$   &  Time-independent \eqref{eq:rate_CV20_case3} & Time-independent  \\
\rule{0pt}{4.9ex} $\mathcal{C}_A$ &Time-independent \eqref{eq:rate_CA_case3} & Time-independent  \\ 
 \rule{0pt}{4.9ex}  $\mathcal{C}_V$ & Linear  \eqref{CV_latetimes_case3}  &  Time-independent \\ 
 \rule{0pt}{4.9ex} $\mathcal{C}^{\epsilon} $ 
 &  Linear \eqref{eq:t inft dCdt CAny3}  &   Time-independent  \\[0.2cm]
\hline
\end{tabular}   
\caption{Late-time behaviour of the holographic complexity conjectures.
$\mathcal{C}^{\epsilon}$ is a class of codimension-one observables among the CAny proposals, defined on CMC slices with extrinsic curvature $K_{\epsilon}$ ($K_{\rm crit}$ denotes a critical value).} 
\label{tab:results}
\end{center}
\end{table}

We anticipate the main results of this paper in table~\ref{tab:results}, which collects the late-time behaviour of the complexity conjectures.\footnote{For visual convenience, we have split table~\ref{tab:results} in two parts. } 
The first set of comments comes from reading the table by columns, which allows us to scan for universal and distinguishing properties of the holographic proposals:
\begin{itemize}
    \item The first column (case 1) shows that hyperfast growth is a feature of the holographic complexity proposals evaluated in a single inflating patch of the SdS background. This generalizes previous results in empty dS space \cite{Jorstad:2022mls}. The only exception is provided by the observable $\mathcal{C}^\epsilon$, belonging to a class of codimension-one CAny observables evaluated on CMC slices, which present an eternal evolution with linear growth at late times whenever their extrinsic curvature is greater than a critical value (as already noticed in \cite{Aguilar-Gutierrez:2023zoi}).
    \item The second column (case 2) tells us that the growth rate of complexity at late times is universally linear whenever the geometric observables are located inside the black hole patch. This is consistent with the behaviour of black holes in asymptotically AdS space \cite{Carmi:2017jqz}.
    \item  From the column referring to case 3, we read that all the geometric observables extending across a single black hole and one cosmological patch do \textit{not} present hyperfast growth. However, there are some differences. The codimension-zero proposals are always time-independent, as a consequence of the WDW patch reaching timelike infinities $\mathcal{I}^{\pm}$ and the singularity at all times. 
    On the other hand, codimension-one conjectures evolve forever, approaching a linear rate at late times. 
    \item The last column (cases 4-5) refers to configurations with multiple copies of the SdS geometry, including both cosmological and black hole patches. In this setting, we find a universal  behaviour: all the holographic proposals are time-independent. 
    For codimension-zero observable, this is a consequence of the WDW patch reaching timelike infinities $\mathcal{I}^{\pm}$ and the singularity at all times. For codimension-one quantities, we find certain constraints that only allow for the trivial evolution generated by the time isometry of the background.  
\end{itemize}

The second interpretation of table~\ref{tab:results} comes from reading it by rows, to determine the common features of the various configurations of the stretched horizons:
\begin{itemize}
\item The first two lines show that codimension-zero proposals present the same features, which drastically change according to the location of the stretched horizons.
In particular, complexity is time-independent in cases 3-5.
 \item The codimension-one proposals also present similar features, except for the two possibilities mentioned above for $\mathcal{C}^{\epsilon}$ in case 1. 
 When there exists at least one inflating and one black hole patch (cases 3-5), we will show that any configuration with symmetric boundary times $t$ is forbidden, except for a maximal surface only defined at a unique $t$.
 While this leads to a time-independent volume in cases 4-5, instead the isometries in case 3 allow for a non-trivial evolution that approaches linear growth at late times.
\end{itemize}

In summary, our study shows that the holographic complexity of a black hole in asymptotically dS space presents different features depending on the location of the stretched horizons: one can find hyperfast, linear, or vanishing growth.
In case 3, which probes one cosmological and one black hole patch, there is a time regime such that CA is time-independent, while CV is not.
This also happens in asymptotically AdS black holes \cite{Carmi:2017jqz}. 
Furthermore, let us mention that a completely different behaviour of CV and CA is also reminiscent of the fact that defects and boundaries show a different structure of UV divergences when evaluating CA and CV proposals \cite{Chapman:2018bqj,Sato:2019kik,Braccia:2019xxi,Auzzi:2021nrj,Baiguera:2021cba,Auzzi:2021ozb}.
In this paper, the novelty is that the different behaviour arises in a dynamical situation, in asymptotically dS space, and at any boundary time, including late times.

\paragraph{Outline.}
The paper is organized as follows.
In section~\ref{sec:Geometry} we introduce the extended SdS$^n$ black hole, which is composed of $n$ pairs of black hole and cosmological patches.
We then describe the possible configurations for the stretched horizons to which complexity observables are anchored.
In section~\ref{sec:Cod0 complexity} we study the time evolution of the WDW patch and of the codimension-zero holographic proposals (CV2.0 and CA). In section~\ref{sec:Cod1 complexity} we similarly proceed with the analysis of codimension-one proposals (CV and CAny). We then conclude in section~\ref{sec:Conclusions} with a summary of our findings, and some relevant future directions.
Appendix~\ref{app:details_CA} contains some technical details on the computation of CA.

\section{Extended Schwarzschild-de Sitter spacetime}
\label{sec:Geometry}

Black hole solutions in asymptotically dS space can be analytically extended to an infinite sequence of regions that admit both a black hole and a cosmological horizon.
We describe this geometry and its causal structure in section \ref{ssec:geom_setting}.
We then define in section \ref{ssec:stretched_horizons} the stretched horizon, a timelike surface where a putative dual theory is conjectured to live, and we present a set of geometric quantities that we are going to study, and are expected to have a holographic interpretation.
Section \ref{ssec:bdy_times} defines the time evolution of the dual quantum theory in terms of the bulk time running along the stretched horizon.

\subsection{Geometric setting}
\label{ssec:geom_setting}

We consider the Schwarzschild-de Sitter (SdS) spacetime in $d+1$ dimensions, which is a spherically symmetric and static black hole solution to the equations of motion of the Einstein-Hilbert action with a positive cosmological constant:
\beq
I = \frac{1}{16 \pi G_N} \int d^{d+1} x \, \sqrt{-g} \, \le R - 2 \Lambda \ri \, , \qquad
\Lambda = \frac{d(d-1)}{2 L^2} \, .
\label{eq:action_EOM_gend}
\eeq
The metric reads
\beq
ds^2 = - f(r) dt^2 + \frac{dr^2}{f(r)} + r^2 d\Omega_{d-1}^2 \, ,  \qquad
f (r) = 1 - \frac{2 \mu}{r^{d-2}} - \frac{r^2}{L^2} \, ,
\label{eq:asympt_dS}
\eeq
where $L$ is the dS curvature radius, $f(r)$ the blackening factor, and $\mu$ a parameter related to the mass of the black hole (\eg see \cite{PhysRevD.15.2738,Balasubramanian:2001nb,Ghezelbash:2001vs} for more details).
When $\mu=0$, the geometry reduces to empty dS space. The case $d=2$ is special because the term proportional to $r^{-(d-2)}$ inside the blackening factor in eq.~\eqref{eq:asympt_dS} becomes constant.
In this special case, the black hole can be obtained as a discrete quotient of empty dS space with a conical defect at $r=0$.
In the coordinate system \eqref{eq:asympt_dS}, $r=\infty$ represents timelike infinities $\mathcal{I}^{\pm}$, while $r=0$ is the black hole singularity.

\begin{figure}[t!]
    \centering
    \includegraphics[width=0.8\textwidth]{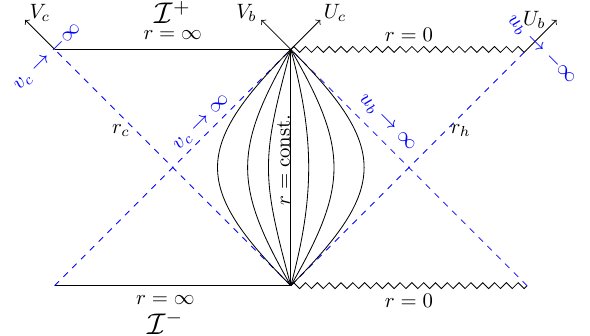}
    \caption{ Penrose diagram of SdS$_{d+1}$ space in dimensions $d\geq 3,$ in the regime where the mass parameter satisfies $ \mu \in (0, \mu_{N}).$ 
    $r_h$ denotes the black hole horizon and $r_c$ the cosmological horizon. We display the orientation of the null coordinates in eqs.~\eqref{eq:general_null_coordinates} and \eqref{eq:Kruskal_coord_sec2}, where the subscripts $\lbrace b,c \rbrace $ denote black hole and cosmological patches, respectively. } 
    \label{fig:Penrose_SdS}
\end{figure}

\paragraph{Generic dimensions.}
In this work, we will only focus on the case when $d \geq 3$ and the parameter $\mu$ belongs to the range  $\mu\in (0,\,\mu_N)$, where
\beq
\mu_{N} \equiv \frac{r_{N}^{d-2}}{d}  \, , \qquad
r_{N} \equiv L \sqrt{\frac{d-2}{d}} \, .
\label{eq:critical_mass_SdS}
\eeq
In this regime, the blackening factor admits two positive roots denoted as $r_h <  r_c$, where the larger value represents a cosmological horizon and the smaller one a black hole horizon.
The Penrose diagram corresponding to the previous range of $\mu$ is depicted in \Fig{fig:Penrose_SdS}.
In the following, we will refer to the left square of the causal diagram (containing a cosmological horizon) as the \textit{cosmological} or \textit{inflating} region of the geometry, and to the right side of the causal diagram as the \textit{black hole} region.
We will also refer to the region described by the radial coordinate $r<r_h$ as the interior of the black hole, and $r>r_h$ as its exterior.
Similarly, $r<r_c$ describes the interior of the cosmological horizon, while for $r>r_c$ the exterior.
In particular, the static patch $r_h \leq r \leq r_c$ is located in the exterior of the black hole, but inside the cosmological horizon.

The coordinate system in eq.~\eqref{eq:asympt_dS} only covers a portion of the spacetime.
In order to enlarge this region, and in view of the evaluation of geometric observables that extend behind the horizons, it is convenient to introduce the null directions 
\beq
u = t - r^* (r) \, , \qquad
v = t + r^* (r) \, ,
\label{eq:general_null_coordinates}
\eeq
where the tortoise coordinate is defined by
\beq
r^*(r) = \int_{r_{0}}^r \frac{d r'}{f(r')} \, ,
\label{eq:general_tortoise_coordinate}
\eeq
and the integration constant $r_0$ is chosen such that $r^*(r \rightarrow \infty)=0$.
In this way, it is possible to bring the metric into the Eddington-Finkelstein (EF) form 
\beq
ds^2 = - f(r) du^2 - 2 du dr + r^2 d\Omega_{d-1}^2 = 
- f(r) dv^2 + 2 dv dr + r^2 d\Omega_{d-1}^2
\, .
\label{eq:generic_null_metric_noshock}
\eeq
It is important to observe that whenever an event horizon is crossed, either the null coordinate $u$ or $v$ ends its range of validity and needs to be defined again using eq.~\eqref{eq:general_null_coordinates} in the new patch.

The Penrose diagram in \Fig{fig:Penrose_SdS}, corresponding to the maximal analytic extension of the metric, is obtained by compactifying the Kruskal coordinates $(U,V)$ defined in the central quadrant as follows\footnote{Notice that despite the existence of two sets of Kruskal coordinates in the middle quadrant in Fig.~\ref{fig:Penrose_SdS}, the null coordinates $(u,v)$ are uniquely defined in such region. Similar conventions apply to the Kruskal coordinates for Kerr black holes \cite{Heinicke:2014ipp}.  } 
\beq
 U_c =  e^{\frac{u}{\ell}}  \, , \quad
V_c = - e^{-\frac{v}{\ell}}  \, ,\qquad U_b =  -e^{-\frac{u}{\ell}}  \, , \quad
V_b = e^{\frac{v}{\ell}}  \, ,
\label{eq:Kruskal_coord_sec2} 
\eeq
where $\ell$ is an appropriate length scale that makes the exponent dimensionless (for instance, one can take $\ell=L$).
Due to the existence of two horizons, we need to introduce two sets of Kruskal coordinates: $(U_c, V_c)$ cover the region with radial coordinate $r \in (r_h, \infty) $, while $(U_b, V_b)$ cover the portion with $r \in (0,r_c)$.
Both the coordinate systems are well-defined in the central static patch with $r \in (r_h, r_c)$, where one can change the coordinate chart from one system to the other.

The cosmological and black hole horizons emit thermal radiation, with associated Hawking temperature and entropy given by\footnote{ \label{foot:extremal_temp} Notice that the Hawking temperatures in eq.~\eqref{eq:hor_cosm_temp_SdS} are defined with respect to the Killing vector $\xi = \gamma \partial_t$ with $\gamma=1$. However, a consistent extremal limit is only achieved with the Hawking-Bousso choice, corresponding to $\gamma=1/\sqrt{f(r_0)}$ with $r_0$ such that $f'(r_0)=0$. This subtlety is discussed, \eg in \cite{Bousso:1996au,Morvan:2022ybp}, but it does not play a relevant role in the present paper.  }
\beq
     T_{h(c)} = \frac{1}{4 \pi} \left|\frac{\partial f(r)}{\partial r}\right|_{r=r_{h(c)}}  
    \qquad
    S_{h(c)} =  \frac{\Omega_{d-1} r_{h(c)}^{d-1}}{4 G} \, .
\label{eq:temp_entropy_SdS_general}  
\eeq
In a general number of dimensions, it is possible to exploit the constraints $f(r_h)=f(r_c)=0$ to write all the relevant quantities defining the SdS solution in terms of the two horizons.
Therefore, we obtain the following expressions for the mass parameter and dS curvature length 
\beq
\mu = \frac{1}{2} \frac{r_c^d r_h^{d-2} - r_h^d r_c^{d-2}}{r_c^d - r_h^d} \, , \qquad
L^2 = \frac{r_c^d - r_h^d}{r_c^{d-2} - r_h^{d-2}} \, ,
\label{eq:general_relation_rhrcm}
\eeq
the blackening factor 
\beq
f(r) = \frac{1}{r_c^d - r_h^d} 
\left[ r_c^d \le 1 - \frac{r^2}{r_c^2} - \frac{r_h^{d-2}}{r^{d-2}}  \ri
- r_h^d \le 1 - \frac{r^2}{r_h^2} - \frac{r_c^{d-2}}{r^{d-2}}  \ri
\right] \, ,
\label{eq:general_blackening_SdS_rcrh}
\eeq
and the Hawking temperatures 
\beq 
T_h  = d \, \frac{r_{N}^2 - r_h^2 }{4 \pi r_h L^2}  \, , \qquad
    T_c  = d \, \frac{r_c^2 -r_{N}^2 }{4 \pi r_c L^2}  \, .
   \label{eq:hor_cosm_temp_SdS}
\eeq
For any mass parameter in the range $\mu \in (0, \mu_N)$, we get $T_h > T_c,$ which shows that the system is out of equilibrium.

In dimensions $d\geq 3,$ the Nariai geometry corresponds to the limiting case where $\mu=\mu_{N}$, such that the black hole and cosmological horizons approach each other.
However, the proper distance between the event horizons does not vanish, because the blackening factor $f(r)$ is also small in this regime.
A proper description of the spacetime in this limit is obtained by taking a near-horizon limit, \eg see section 3.1 in \cite{Anninos:2012qw} or appendix B in \cite{Svesko:2022txo}.
The Nariai geometry can be mapped to $\mathrm{dS}_2 \times S^{d-1}$ by a change of coordinates \cite{Maldacena:2019cbz,Anninos:2012qw}.

\paragraph{Four dimensions.}
When $d=3$, certain analytic expressions are available.
The blackening factor in eq.~\eqref{eq:asympt_dS} becomes 
\beq
f_{\rm SdS_4} (r) = 1 - \frac{2\mu}{r} - \frac{r^2}{L^2} \, ,
\label{eq:blackening_SdS4}
\eeq
and the critical mass in eq.~\eqref{eq:critical_mass_SdS} is $\mu_N^2 = L^2/27$.
In the regime $ \mu \in (0, \mu_N)$ when the blackening factor admits two roots, one can invert the identities  \eqref{eq:general_relation_rhrcm} specialized to $d=3$ and find a closed form for the event horizons in terms of the dS radius and the mass parameter \cite{Shankaranarayanan:2003ya,Choudhury:2004ph,Visser:2012wu}:
\beq
   r_h = r_{N} \le \cos \eta - \sqrt{3} \sin \eta \ri \, , \quad
r_C = r_{N} \le \cos \eta + \sqrt{3} \sin \eta \ri \, , \quad  
\eta \equiv \frac{1}{3} \, \mathrm{arccos}  \le \frac{\mu}{\mu_{N}} \ri \, .
\label{eq:analytic_rh_rc_SdS4}
\eeq
The inverse of the blackening factor \eqref{eq:general_blackening_SdS_rcrh} can be integrated analytically, giving the tortoise coordinate
\beq
\begin{aligned}
r^*(r) = &
- \frac{r_C^2 + r_h r_C + r_h^2}{(r_C-r_h)(2r_C+r_h) (r_C+2r_h)} 
\left[ r_C^2 \log \left| \frac{r-r_C}{r+r_C+r_h} \right| \right. \\
& \left. + 2 r_h r_C \log \left| \frac{r-r_C}{r-r_h} \right|
-  r_h^2 \log \left| \frac{r-r_h}{r+r_C+r_h} \right|
\right] \, .
\end{aligned}
\label{eq:tortoise_coord_SdS4}
\eeq
These expressions will be used for any numerical computation involving the SdS$_4$ background in the remainder of the paper.

\paragraph{Extended Schwarzschild-de Sitter background.}
It is important to observe that SdS spacetime can be analytically extended even beyond the region depicted in \Fig{fig:Penrose_SdS}, \ie it is possible to build an infinite sequence of singularities 
and timelike infinities $\mathcal{I}^{\pm}$.
The Penrose diagram is composed by a periodic extension as depicted in \Fig{fig:extSdS}, where it is allowed to have $n$-pairs of black hole and cosmological patches.
We refer to this configuration as the \textit{extended Schwarzschild-de Sitter} spacetime, denoted with SdS$^n$.
Each of the static patches can be described by the metric introduced above:
\beq
ds_{(i)}^2 = - f\qty(r_{(i)}) dt_{(i)}^2 + \frac{dr^2}{f\qty(r_{(i)})} + r_{(i)}^2 d\Omega_{d-1}^2 \, ,  \qquad
f \qty(r_{(i)}) = 1 - \frac{2 \mu}{r_{(i)}^{d-2}} - \frac{r_{(i)}^2}{L^2} \, .
\label{eq:multi_asympt_dS}
\eeq
The only caveat is that we need to introduce a copy of the coordinates for each copy of SdS, which here we denoted with the index $i \in \lbrace 1, \dots, n \rbrace$.
It is worth mentioning that the Euclidean continuation of SdS$^n$ geometry is singular, but this issue will not play a role in our classical gravitational analysis.\footnote{The Euclidean continuation is generically singular even when $n=1$ (\eg see \cite{Morvan:2022ybp}), unless one works in the extremal SdS limit, where the black hole and inflating patches have the same temperature. In the extended geometry, even at thermal equilibrium, there are conical singularities \cite{Aguilar-Gutierrez:2021bns,Levine:2022wos}.  \label{fn:1}}

\vskip 3mm
\noindent
Next, we define the relevant geometric quantities that will be studied in the extended SdS spacetime in this work.

\subsection{Stretched horizons}
\label{ssec:stretched_horizons}

In the context of static patch holography, it is conjectured that there exists a dual quantum-mechanical description of the static patch of empty dS space in terms of a theory living on a timelike surface (the \textit{stretched horizon}) located just inside the cosmological horizon \cite{Dyson:2002pf,Susskind:2011ap,Susskind:2021esx,Susskind:2021dfc,Shaghoulian:2021cef,Susskind:2021omt,Shaghoulian:2022fop}.
In the black hole geometry defined by eq.~\eqref{eq:asympt_dS}, the stretched horizon is defined
as a surface at the constant radial coordinate 
\beq
r_{\rm st} = (1-\rho) r_h + \rho r_c \, , \qquad
\rho \in [0,1] \, ,
\label{eq:stretched_horizon_SdS4}
\eeq
such that $r_{\rm st} \in [r_h, r_c]$ and the parameter $\rho$ interpolates between the two horizons.
In particular, we will be mainly interested in the limits $\rho \rightarrow 0$ and $ \rho \rightarrow 1$, when the stretched horizon approaches either the black hole or the cosmological horizon.\footnote{In this regard, a stationary observer in the static patch will generally experience an intermediate temperature between the cosmological and black hole one, since the system is out of equilibrium. However, if the stretched horizons is close to one of the horizons (either $\rho \rightarrow 0$ or $\rho \rightarrow 1$), then the corresponding flux of radiation will be dominant. We thank Shan-Ming Ruan for raising this question.   }
There are various reasons that make this choice natural.
First, according to the central dogma for black holes and for inflationary models, the degrees of freedom and the unitary evolution of the dual quantum system should be determined from the region inside the cosmological horizon, but outside the black hole one \cite{Almheiri:2020cfm,Shaghoulian:2021cef,Aguilar-Gutierrez:2023zoi}.
This region is the static patch of the SdS background. 
Secondly, the choice \eqref{eq:stretched_horizon_SdS4} defines a timelike surface where a time coordinate for the dual quantum system can be consistently defined.
Had we chosen a surface at constant $r$ located in the intervals $[0,r_h]$ or $[r_c, \infty]$, it would have been spacelike instead.
Furthermore, holographic screens can be defined in other ways, for instance, by taking a congruence of geodesics and imposing that the expansion parameter vanishes \cite{Bousso:1999cb}.
In the case of a static and spherically-symmetric background like \eqref{eq:asympt_dS}, this requirement is equivalent to locating the stretched horizon close to one of the event horizon.
Finally, a natural static sphere observer is freely falling at fixed radius $r_{\mathcal{O}}$, \ie identified by the condition $f'(r_{\mathcal{O}})=0$. This requirement fixes $r_{\mathcal{O}} \in (r_h, r_c)$, which is the range where the stretched horizon is located. 

\begin{figure}[t!]
  	\centering
   \begin{subfigure}[t]{0.7\textwidth}
   \includegraphics[width=\textwidth]{Figures/Fig3.pdf}
   \caption{\textbf{Case 1}: two cosmological stretched horizons.
   }
   \label{eq:Single CH}
   \end{subfigure}
   \begin{subfigure}[t]{0.7\textwidth}
  	\includegraphics[width=\textwidth]{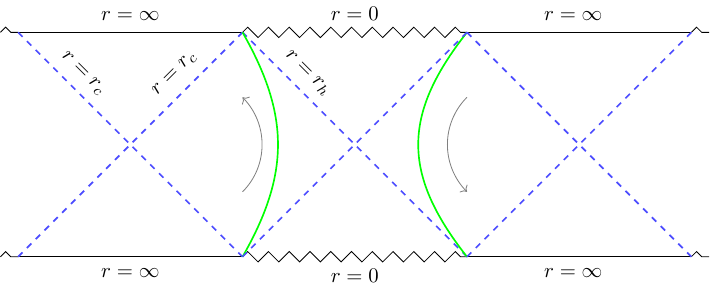}
   \caption{\textbf{Case 2}: two black hole stretched horizons.}
   \label{eq:Single BH}
   \end{subfigure}
  	\caption{ Prescriptions for the stretched horizons (in green) in a holographic setting. (a) The cosmological stretched horizons are placed in consecutive static patches. Geometric quantities anchored to the cosmological stretched horizons explore the region beyond the cosmological horizon. (b) The stretched horizons are placed in consecutive black hole patches. Geometric quantities anchored to the black hole stretched horizon explore the region behind the black hole horizon. The gray arrows represent the orientation of the Killing vector $\partial_t$. }
  	\label{fig:stretched-hor-cases12}
  \end{figure}
It turns out that the realization of the requirement \eqref{eq:stretched_horizon_SdS4} in SdS background leads to several different holographic settings:
\begin{enumerate}
\item \textbf{Two cosmological stretched horizons.} So far, all the studies of holography in asymptotically dS space in the literature focused on locating the stretched horizons just inside the cosmological horizon as depicted in Fig.~\ref{eq:Single CH}.
We refer to these surfaces at constant radii as the \textit{cosmological stretched horizons}, because they allow us to 
investigate the properties of the geometry of the cosmological region.
Indeed, it was conjectured that the Ryu-Takayanagi (RT) surface, the codimension-one extremal surfaces or the WDW patch should be anchored at the cosmological stretched horizons, such that these geometric objects lie in the inflating region of the Penrose diagram \cite{Shaghoulian:2021cef,Susskind:2021dfc,Susskind:2021esx,Shaghoulian:2022fop,Jorstad:2022mls}.
\item \textbf{Two black hole stretched horizons.} In comparison to empty dS space, the existence of a black hole event horizon in higher-dimensional SdS allows to define a so-called \textit{black hole stretched horizon} which still satisfies eq.~\eqref{eq:stretched_horizon_SdS4}, but is now located in the black hole region of the Penrose diagram.
The corresponding scenario, depicted in Fig.~\ref{eq:Single BH}, implies that the geometric quantities lie behind the black hole horizon.
This configuration is very similar to the standard holographic setting considered in asymptotically AdS space. 
\item  \textbf{One cosmological and one black hole stretched horizon.}
We depict in Fig.~\ref{fig:cosmobh-stretched-hor} a novel prescription.
We take one cosmological stretched horizon and one black hole stretched horizon, both belonging to the same copy of SdS space.

\begin{figure}[t!]
    \centering
    \begin{subfigure}[t]{0.7\textwidth}
    \includegraphics[width=\textwidth]{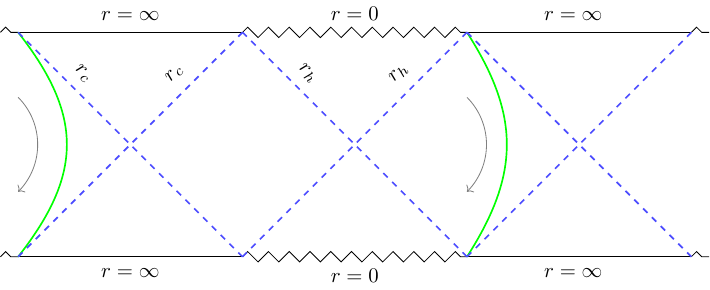}
\caption*{\textbf{Case 3}: one cosmological and one black hole stretched horizon.}
    \end{subfigure}
    \caption{New prescription for the stretched horizons in the SdS background. The left surface is a cosmological stretched horizon, while the right surface is a black hole stretched horizon. Geometric objects anchored to the stretched horizons can go behind $r_h$ and beyond $r_c$. Gray arrows represent the orientation of the Killing vector $\partial_t$. }
    \label{fig:cosmobh-stretched-hor}
\end{figure}
\item \textbf{Two cosmological stretched horizons in different copies of SdS.}
The periodicity of the SdS$^n$ geometry opens the way for more possibilities, where there are two cosmological stretched horizons belonging to different copies of the spacetime.
This setting is depicted in Fig.~\ref{fig:extended-cosmo-sec2} for the case of SdS$^2$.
\item \textbf{Two black hole stretched horizons in different copies of SdS.}
Another possibility coming from the periodicity of the Penrose diagram is to take two black hole stretched horizons in different copies, as reported in Fig.~\ref{fig:extended-bh-sec2} for the SdS$^2$ case. 
\end{enumerate}

\begin{figure}[t!]
  	\centering
   \begin{subfigure}[t]{0.7\textwidth}
   \includegraphics[width=\textwidth]{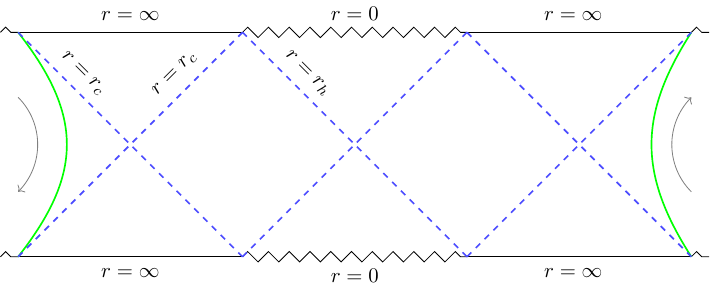}
       \caption{\textbf{Case 4}: two cosmological stretched horizons in SdS$^2$.}
       \label{fig:extended-cosmo-sec2} 
   \end{subfigure}
   \\
   \begin{subfigure}[t]{0.7\textwidth}
   \includegraphics[width=\textwidth]{Figures/Fig7.pdf}
   \caption{\textbf{Case 5}: two black hole stretched horizons in SdS$^2$.}
   \label{fig:extended-bh-sec2}
   \end{subfigure}
  	\caption{Novel prescriptions for the stretched horizons in a holographic setting with two copies of SdS space. 
   (a) The cosmological stretched horizons (in green) are located in different copies of the static patch. 
 (b) The stretched horizons (in green) are located in different copies of the black hole patch. Gray arrows represent the orientation of the Killing vector $\partial_t$.}
  	\label{fig:multiverse-hor-cases45}
  \end{figure}
In the remainder of the paper, we will denote as $r_{\rm st}^L$ ($r_{\rm st}^R$) the left (right) stretched horizons in a certain configuration.
The common idea to all the new cases 3, 4 and 5 is that geometric objects hanging from the corresponding stretched horizons probe both the black hole interior and the exterior of the cosmological horizon, which are both of physical interest.

\subsection{Boundary times}
\label{ssec:bdy_times}

We interpret the time coordinates $t_L, t_R$ running along the left or right stretched horizons of the SdS geometry as the boundary times of the corresponding dual theories.\footnote{From now on, we will refer to $t_L, t_R$ as the \textit{boundary times}, referring to the idea that they may be interpreted as the time coordinates of a putative quantum theory defined on these codimension-one surfaces.}
First of all, we define the following 
\begin{conv}
\label{conv_times}
    The boundary time $t_L (t_R)$ defined on the left (right) stretched horizon is always chosen to be oriented upward.
\end{conv}

The signs of the times $t_L, t_R$ for the left and right boundary theories, compared to the time coordinate $t_{\rm bulk}$ naturally defined in the bulk and running along the stretched horizon, are determined by the orientation of the Killing vector $\p_t$ associated with the time-translation invariance of the metric \eqref{eq:asympt_dS}. 
In particular, this orientation flips whenever an event horizon is crossed.

Let us clarify the convention \ref{conv_times} with an example.
In case 1, the SdS geometry is depicted in Fig.~\ref{fig:WDW_cosmo_stretched}.
The Killing vector $\p_t$ runs upward on the right stretched horizon and downward on the left stretched horizon.
The orange lines at $\pm 45$\textdegree \, represent surfaces at constant $u (v)$ coordinates, which compose the null boundaries of the WDW patch (that will be analyzed in section \ref{ssec:WDW_patch}).
Focusing on the top-right boundary, we find that the constant value of the $u$ coordinate is given by
\beq
u_{\rm max} = t_{\rm bulk} - r^*(r^R_{\rm st}) = t_R - r^*(r^{R}_{\rm st}) \, , \qquad (\mathrm{case} \,\,\, 1)
\label{eq:identification_times_umax}
\eeq
where we used the definition \eqref{eq:general_null_coordinates} and the identification $t_{\rm bulk} = t_R$ that comes from the upward orientation of the Killing vector $\p_t$, consistent with the upward orientation of $t_R$ according to the convention \ref{conv_times}.
On the contrary, if we evaluate the constant value of $u$ on the bottom-left boundary, we get
\beq
u_{\rm min} = t_{\rm bulk} - r^*(r^L_{\rm st}) = -t_L - r^*(r^L_{\rm st}) \, ,  \qquad (\mathrm{case} \,\,\, 1) 
\eeq
where now we identified $t_{\rm bulk} = -t_L$ because the Killing vector $\p_t$ is oriented downward, while the convention \ref{conv_times} requires $t_L$ to be oriented upward.
A similar reasoning will be applied in the remainder of the paper to define the boundary times in any other configuration.

Next, our goal is to introduce a time dependence into the setting, in order to probe the exterior of the cosmological horizon and the interior
of the black hole.\footnote{This guiding principle was originally introduced to study the evolution of the entanglement entropy in an eternal black hole background in \cite{Hartman:2013qma}.  }
It is well-known that the SdS background is static under a time evolution that runs along the same direction dictated by the Killing vector $\partial_t$ running along the stretched horizons.
More precisely, due to the boost symmetry associated with the time isometry in the bulk, the conjectured dual state in the boundary theory is invariant under the shift
\beq
\begin{cases}
    t_L \rightarrow t_L + \Delta t \, , \qquad
t_R \rightarrow t_R - \Delta t  & \mathrm{cases} \,\,\, 1,2,4,5 \\
t_L \rightarrow t_L + \Delta t \, , \qquad
t_R \rightarrow t_R + \Delta t & \mathrm{case} \,\,\, 3 \, ,
\end{cases}
\label{eq:time_shift_symmetry}
\eeq
where the different sign of case 3 compared to the other configurations is a consequence of the orientation of the Killing vector $\partial_t$ on the two stretched horizons.

The symmetry \eqref{eq:time_shift_symmetry} implies that we need to perform a different evolution of the boundary times in case 3 (compared to the other cases) in order to achieve a time-dependent setting.
In other words, this observation justifies the following rule

\begin{deff}
\label{rule}
   A non-trivial time evolution of the SdS background is achieved by evolving the times of the putative boundary theory upward along the stretched horizons, except on the left stretched horizon in case 3, where the time $t_L$ evolves downward.
\end{deff}

In principle, the time coordinates $t_L, t_R$ are independent, but one can synchronize them by considering an extremal codimension-one surface connecting the two stretched horizons. This geometric construction connects the time slice at time $t_R$ on the right stretched horizon with the one at time $t_L= \pm t_R$ (as specified below) on the left stretched horizon.
Throughout this work, we will refer to the special case of a symmetric (antisymmetric in case 3) time evolution whenever the following condition is met:\footnote{At the cost of being pedantic, we stress that $t$ is a boundary time, related to the bulk time $t_{\rm bulk}$ via the identification of the time coordinate on the stretched horizons dictated by the Killing vector $\p_t$, see discussion below eq.~\eqref{eq:identification_times_umax}.  }
\beq
\begin{cases}
  \frac{t}{2} \equiv t_L = t_R  \, ,   & \mathrm{cases} \,\,\, 1,2,4,5 \\
\frac{t}{2} \equiv -t_L = t_R  \, , & \mathrm{case} \,\,\, 3 \, .
\end{cases}
\label{eq:symmetric_times}
\eeq
Using the symmetry \eqref{eq:time_shift_symmetry} of the extended black hole background, this case can always be achieved for the holographic complexity investigations, and it is not restrictive.\footnote{ In principle, the time coordinates are independent among each copy of SdS. However, requiring a smooth junction between the various copies implies that the coordinates and the Killing vector $\partial_t$ are continuous. Therefore, one can synchronize the clocks of the surfaces belonging to different copies of SdS, too. }

\section{Codimension-zero proposals}
\label{sec:Cod0 complexity}

We begin our journey across the holographic proposals in extended SdS geometry by considering the codimension-zero observables.
After defining in section \ref{ssec:WDW_patch} the main properties of the WDW patch, we explore in sections \ref{ssec:CV20} and \ref{ssec:CA} the time evolution of CV2.0 and CA conjectures, respectively.

\subsection{WDW patch}
\label{ssec:WDW_patch}

We build the WDW patch, \ie the bulk domain of dependence of a spacelike surface anchored at the stretched horizons, according to the list of proposals presented in section \ref{ssec:stretched_horizons}.
In order to avoid divergences of the spacetime volume or of the gravitational action when approaching timelike infinities $\mathcal{I}^{\pm}$, we introduce a cutoff surface located at $r=r_{\rm max}$.
Although the region close to the black hole singularity does not bring any divergence, we will also introduce by analogy a cutoff surface located at $r=r_{\rm min}$.
At the end of the computation, we will send $r_{\rm max} \rightarrow \infty$ and $r_{\rm min} \rightarrow 0$. 

In all the cases described below, the WDW patch is conveniently described by the null coordinates introduced in eq.~\eqref{eq:general_null_coordinates}.
In particular, the null boundaries delimiting the shape of the WDW patch are identified by the conditions that either the $u$ or $v$ coordinate is constant, as we will specify case by case below.

\subsubsection*{Case 1}

\begin{figure}[t!]
\centering
\includegraphics[width=0.6\textwidth]{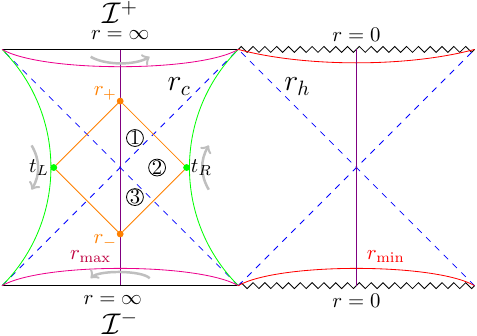}
\caption{General configuration of the WDW patch in case 1. Pink and red curves near $\mathcal{I}^\pm$ and the singularities indicate cutoff surfaces $r_{\rm max}$ for the cosmological and $r_{\rm min}$ for the black hole patches, respectively. The boundaries of the WDW patch are denoted in orange.
The vertical purple line denotes a locus of constant $t=0$.
The gray arrows indicate the orientation of the Killing vector $\partial_t$, and the circled numbers indicate the regions of integration in (\ref{eq:decomposition_CV20_general}).
}
\label{fig:WDW_cosmo_stretched}
\end{figure}

The time dependence of the WDW patch was studied in various dimensions for empty dS space (with and without perturbations) \cite{Jorstad:2022mls,Anegawa:2023wrk,Anegawa:2023dad} and for the SdS black hole in the presence of a shock wave \cite{Baiguera:2023tpt}.
We depict the relevant geometric configuration in Fig.~\ref{fig:WDW_cosmo_stretched}, focusing on the case with symmetric time evolution \eqref{eq:symmetric_times} and symmetric location of the stretched horizons, \ie $r_{\rm st} \equiv r_{\rm st}^L = r_{\rm st}^R$.

The top and bottom joints $r_{\pm}$ of the WDW patch are related to the position of the stretched horizons as
\beq
u_{\rm max} = \frac{t}{2} - r^*(r_{\rm st}) = -r^*(r_+) \, , \qquad
v_{\rm min} = \frac{t}{2} + r^*(r_{\rm st}) = r^*(r_-) \, , \qquad
r_{\pm} \geq r_c \, ,
\label{eq:umax_vmin_WDWpatch}
\eeq
where we used the symmetry of the configuration to infer that the joints are located along the vertical line at $t=0$ in the Penrose diagram.
The time dependence of the joints can be computed by taking the time derivative of the previous identities, leading to
\beq
\frac{d r_{+}}{dt} = - \frac{1}{2} f(r_{+}) \, , \qquad
\frac{d r_{-}}{dt} = \frac{1}{2} f(r_{-}) \, .
\label{eq:rates_rprmcase1}
\eeq
Due to the isometries of the spacetime, the time evolution is symmetric with respect to the boundary time $t=0.$
The critical time $t_{\infty}$ (and its opposite $-t_{\infty}$) is identified by the condition that the top (bottom) joint of the WDW patch reaches future (past) timelike infinity.
More precisely, the critical time $t_{c1}$ when the top joint crosses the future cutoff surface is given by
\beq
t_{c1} = 2 \le  r^*(r_{\rm st}) - r^*(r_{\rm max}) \ri \, , \qquad
t_{\infty} = \lim_{r_{\rm max} \rightarrow \infty} t_{c1} \, .
\label{eq:critical_time_tc1}
\eeq  

\subsubsection*{Case 2}

\begin{figure}[t!]
\centering
\includegraphics[width=0.6\textwidth]{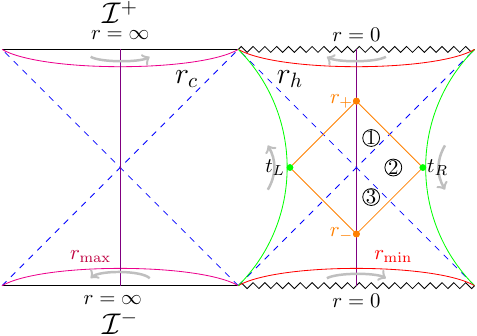}
\caption{General configuration of the WDW patch in case 2. The same coloring scheme as Fig. \ref{fig:WDW_cosmo_stretched} is used, with the circled regions indicating the evaluation of (\ref{eq:CV20_case2_int_regime}).}
\label{fig:WDW_cosmo_stretched_case2}
\end{figure}

This setting, depicted in Fig.~\ref{fig:WDW_cosmo_stretched_case2}, has several analogies with computations of holographic complexity performed in asymptotically AdS spacetimes (\eg see \cite{Chapman:2016hwi,Carmi:2017jqz}) because the WDW patch covers a region behind the black hole horizon.
The main differences with these standard calculations are the specific blackening factor and the presence of the stretched horizons, which cut part of the geometry and forbid the WDW patch to reach any region close to $r=\infty$.
As a consequence, the corresponding holographic complexity will not be UV divergent, nevertheless, we expect the rate of growth to present similar features compared to the AdS case.

Next, we assume without loss of generality that the boundary times satisfy eq.~\eqref{eq:symmetric_times} and that the stretched horizons are symmetric, \ie $r_{\rm st} \equiv r_{\rm st}^L = r_{\rm st}^R$.
In this configuration, we define the special positions of the WDW patch as 
\beq
u_{\rm min} = -\frac{t}{2} - r^*(r_{\rm st}) = -r^*(r_+) \, , \qquad
v_{\rm max} = -\frac{t}{2} + r^*(r_{\rm st}) = r^*(r_-) \, , \qquad
r_{\pm} \leq r_h \, .
\label{eq:umax_vmin_WDWpatch_case2}
\eeq
The difference with eq.~\eqref{eq:umax_vmin_WDWpatch} is that $r_{\pm}$ lie in a different region, and the Killing vector corresponding to the time directions flows in the opposite way along the stretched horizons.
As a consequence, the derivatives of the joints now satisfy the different relations
\beq
\frac{d r_{+}}{dt} =  \frac{1}{2} f(r_{+}) \, , \qquad
\frac{d r_{-}}{dt} = - \frac{1}{2} f(r_{-}) \, .
\label{eq:rates_rprmcase2}
\eeq
The time evolution is still symmetric. The instant $t_0$ (and its opposite $-t_0$) when the top (bottom) joint of the WDW patch reaches the future (past) singularity can be obtained from a limit of the general critical time $t_{c2}$ when the top joint cross the future cutoff surface in the black hole region, \ie
\beq
t_{c2} = 2 \le  r^*(r_{\rm min}) - r^*(r_{\rm st})  \ri \, , \qquad
t_{0} = \lim_{r_{\rm min} \rightarrow 0} t_{c2} \, .
\label{eq:critical_time_tc2}
\eeq  

\subsubsection*{Case 3}

\begin{figure}[t!]
\centering
\includegraphics[width=0.6\textwidth]{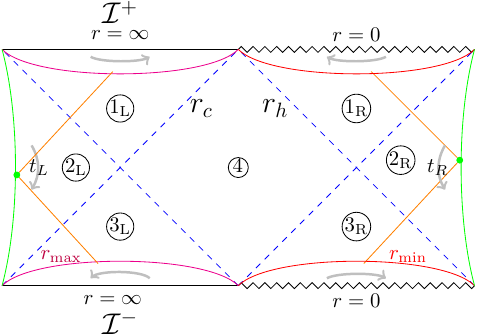}
\caption{General configuration of the WDW patch in case 3. The same coloring scheme as Fig. \ref{fig:WDW_cosmo_stretched} is used, with the circled regions indicating the evaluation of (\ref{eq:Case3_CV20}).} 
\label{fig:WDW_BH_stretched}
\end{figure}

Let us now consider the setting depicted in Fig.~\ref{fig:WDW_BH_stretched}.
We will keep $t_L, t_R$ and the locations of the stretched horizons $r_{\rm st}^L, r_{\rm st}^R$ arbitrary in the following considerations.
The joints of the WDW patch are always located behind the cutoff surfaces (either $r_{\rm max}$ or $r_{\rm min}$), even in the limit where they approach timelike infinity $\mathcal{I}^{\pm}$ and the singularity.
Regarding the time evolution of the diagram, it should be noted that the Killing vector $\p_t$ is downward-oriented along both stretched horizons, implying that the time coordinate increases towards the bottom of the Penrose diagram.
Therefore, from the perspective of a dual boundary theory, we should denote the times on the stretched horizons as $t_{\rm bulk}= -t_L$ and $t_{\rm bulk} = -t_R$.

\subsubsection*{Cases 4-5}

\begin{figure}[t!]
\centering
\subfloat[]{\label{fig:WDW_case4} \includegraphics[width=0.9\textwidth]{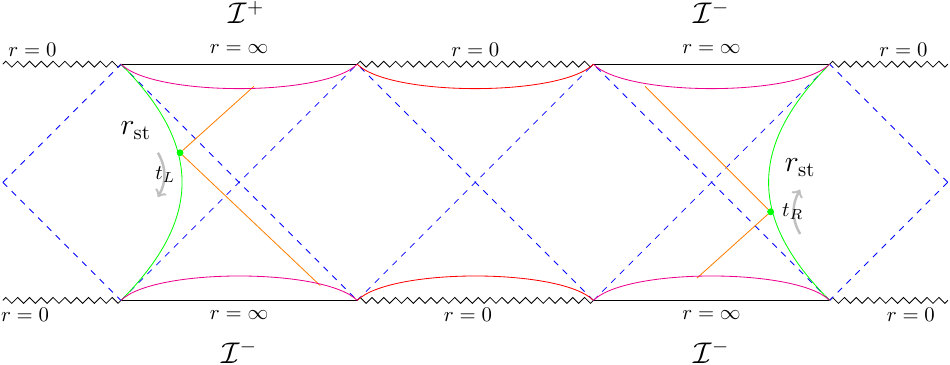}}\vspace{0.5cm}\\ 
\subfloat[]{\label{fig:WDW_case5}\includegraphics[width=0.9\textwidth]{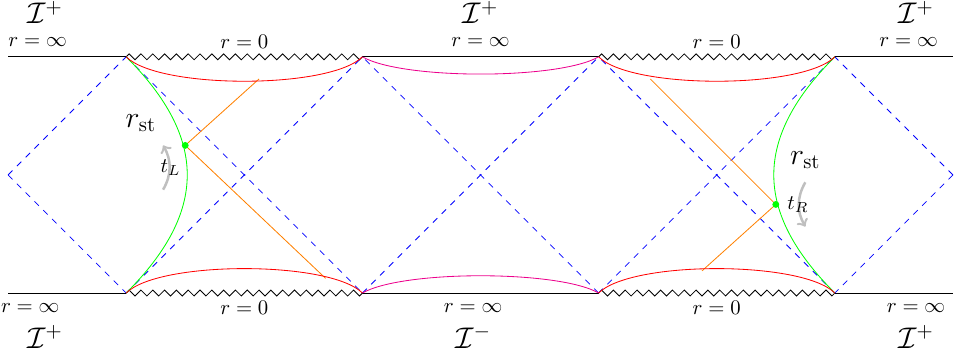}}
\caption{General configuration of the WDW patch in cases 4 (Fig.~a) and 5 (Fig.~b).}
\label{fig:WDW_case_45}
\end{figure}

The relevant shapes of the WDW patch in these cases are depicted in Fig.~\ref{fig:WDW_case_45}.
We observe that the diagrams are similar to case 3 in Fig.~\ref{fig:WDW_BH_stretched}, except that there is an additional inflating patch (case 4) or black hole patch (case 5).
The main feature of these configurations is that the top and bottom joints of the WDW patch always lie behind timelike infinities or singularities, independently of the precise location of the stretched horizons and of the value of the boundary times.

\subsection{Complexity=volume 2.0}
\label{ssec:CV20}

As mentioned in section \ref{sec:Intro}, the CV2.0 conjecture states that holographic complexity is proportional to the spacetime volume of the WDW patch. Given the spherical symmetry of the black hole solution \eqref{eq:asympt_dS}, we get the expression
\beq 
\mathcal{C}_{2.0V} = \frac{V_{\rm WDW}}{G_N L^2} = 
\frac{\Omega_{d-1}}{G_N L^2} \int dr dt \, r^{d-1} \mathcal{F}(r) \, , 
\label{eq:general_CV20}
\eeq
where $\Omega_{d-1} = 2 \pi^{d/2}/\Gamma(d/2)$ is the volume of a $(d-1)$--dimensional unit sphere, and 
$\mathcal{F}$ is an appropriate integrand which depends on the radial coordinate only.

Next, let us analyze the time dependence of the quantity \eqref{eq:general_CV20} in the settings defined in section \ref{ssec:stretched_horizons}.
For convenience, we will always split the evaluation of CV2.0 as
\beq
\mathcal{C}_{2.0V} = \sum_{k} \mathcal{J}_k \, ,
\label{eq:decomposition_CV20_general}
\eeq
where the index $k$ refers to a subregion of the WDW patch that we specify case by case in the configurations discussed below, and $\mathcal{J}_k$ is the complexity \eqref{eq:general_CV20} computed in such subregion.

\subsubsection*{Case 1}

The shape of the WDW patch during the time evolution is determined by the sign of the critical time $t_{\infty}$ defined in eq.~\eqref{eq:critical_time_tc1}.
One can check that increasing the value of $\rho$ defined in eq.~\eqref{eq:stretched_horizon_SdS4} along the interval $[0,1]$, there is a transition from negative to positive values of $t_{\infty}$.
Since static patch holography proposes that the stretched horizon should be located close to the cosmological one ($\rho \rightarrow 1$), we will focus on this case and assume $t_{\infty} >0$. 
We will comment on the opposite scenario at the end of this paragraph.

The configuration of the WDW patch represented in Fig.~\ref{fig:WDW_cosmo_stretched} is the prototype for investigations of CV2.0 in this regime.
According to the decomposition into subregions reported in the picture, we compute the quantities entering eq.~\eqref{eq:decomposition_CV20_general} as
\begin{subequations}
\label{eq:CV20_case1_int_regime}
\beq
\mathcal{J}_{1L}=\mathcal{J}_{1R} = \frac{\Omega_{d-1}}{G_{N} L^2} \int_{r_c}^{r_+} dr  \,  r^{d-1} \le  \frac{t}{2} +r^*(r) -r^*(r_{\rm st})  \ri \, ,
\label{eq:CV20_case1_int_regime_p1}
\eeq
\beq
\mathcal{J}_{2L} =
\mathcal{J}_{2R}=
\frac{\Omega_{d-1}}{G_{N} L^2} \int_{r_{\rm st}}^{r_c} dr  \,  r^{d-1} \le  2 r^*(r) - 2 r^*(r_{\rm st})  \ri \, , 
\label{eq:CV20_case1_int_regime_p2}
\eeq
\beq
\mathcal{J}_{3L} =
\mathcal{J}_{3R} =  \frac{\Omega_{d-1}}{G_{N} L^2} \int_{r_c}^{r_-} dr  \,  r^{d-1} \le - \frac{t}{2} +r^*(r) - r^*(r_{\rm st})  \ri \, ,
\label{eq:CV20_case1_int_regime_p3}
\eeq
\end{subequations}
where $r_{\pm}$ were implicitly defined in eq.~\eqref{eq:umax_vmin_WDWpatch}.
The total complexity is obtained by combining the previous terms according to eq.~\eqref{eq:decomposition_CV20_general}.

\begin{figure}[t!]
    \centering
    \subfloat[]{
\includegraphics[width=0.3\textwidth]{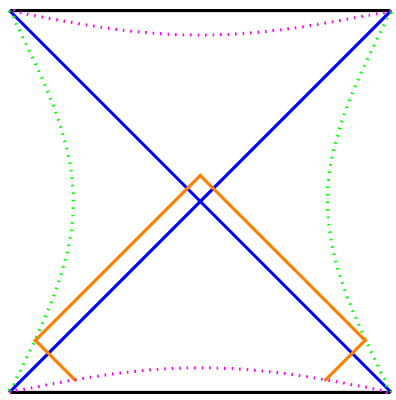} \label{subfig:WDW_evo1_regime1}}
\hfill \subfloat[]{\label{subfig:WDW_evo1_regime2} 
\includegraphics[width=0.3\textwidth]{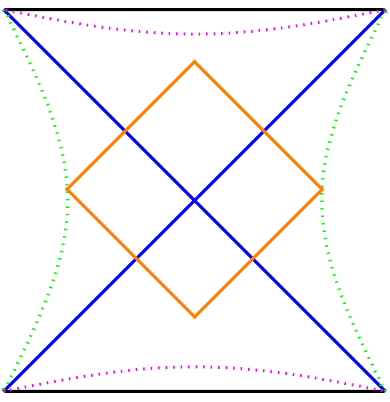}}
\hfill 
\subfloat[]{ \label{subfig:WDW_evo1_regime3}
\includegraphics[width=0.3\textwidth]{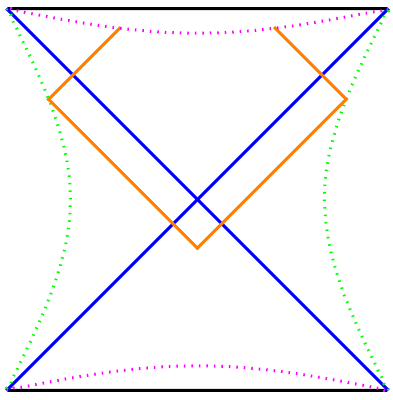}}
    \caption{Possible time evolution of the WDW patch. In the picture, we only report the part of the Penrose diagram included between the stretched horizons. In case 1 (2), the blue lines represent the cosmological (black hole) horizon, while the magenta cutoff surfaces are located at $r=r_{\rm max}$ ($r=r_{\rm min}$). }
    \label{fig:WDW_evo1}
\end{figure}

The evolution of the WDW patch is composed by the three regimes summarized in Fig.~\ref{fig:WDW_evo1}.
At early times $t < -t_{\infty}$, the bottom joint sits behind past timelike infinity $\mathcal{I}^-$, while the top joint is located outside the cosmological horizon $r_c$.
CV2.0 is determined by eq~\eqref{eq:CV20_case1_int_regime} with the replacement $r_{-} \rightarrow r_{\rm max}$.
The intermediate regime $-t_{\infty} \leq t \leq t_{\infty}$ is described by the configuration in Fig.~\ref{subfig:WDW_evo1_regime2}, whose complexity is given by the sum of the terms in eq.~\eqref{eq:CV20_case1_int_regime}.
Finally, at late times $t > t_{\infty}$ the top joint moves behind future timelike infinity $\mathcal{I}^+$, and CV2.0 is obtained by replacing $r_+ \rightarrow r_{\rm max}$ in eq.~\eqref{eq:CV20_case1_int_regime}.

\begin{figure}[t!]
    \centering
    \subfloat[]{ \label{subfig:CV20rate_case1}
\includegraphics[width=0.475\textwidth]{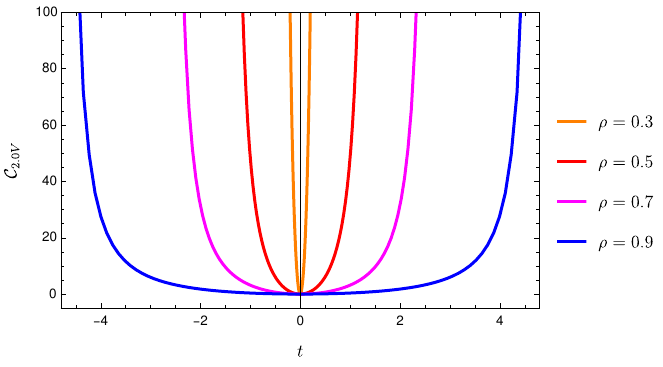}}
\hfill \subfloat[]{\label{subfigb:CV20rate_case1} 
\includegraphics[width=0.475\textwidth]{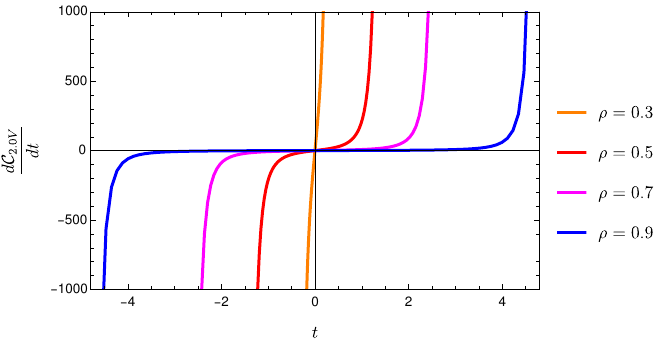}}
    \caption{Time dependence of (a) CV2.0 and (b) its rate, computed according to case 1 in eq.~\eqref{eq:rateCV20_case1} during the interval $[-t_{\infty}, t_{\infty}]$ with critical time defined in eq.~\eqref{eq:critical_time_tc1}. We set $G_N L=1, d=3, \mu=0.14$, and we vary the value of $\rho$ in eq.~\eqref{eq:stretched_horizon_SdS4}, but always keeping $t_{\infty} >0$.   }
    \label{fig:CV20case1}
\end{figure}

While there is no closed expression for the above integrals in generic dimension $d$, we can get the complexity rate by applying the fundamental theorem of integral calculus, together with eq.~\eqref{eq:rates_rprmcase1}:
\beq
\frac{d \mathcal{C}_{2.0V}}{dt} =  \frac{\Omega_{d-1}}{d \, G_{N} L^2} \times  
\begin{cases}
  r_+^d   &  \mathrm{if} \,\, t < -t_{\infty} \\
 \le r_+^d - r_-^d \ri  & \mathrm{if} \,\, -t_{\infty} \leq t \leq t_{\infty} \\
  (-r_-^d)   &  \mathrm{if} \,\, t > t_{\infty} \\
\end{cases}
\label{eq:rateCV20_case1}
\eeq
Let us focus on the intermediate regime.
At the critical time $t_{\infty} (-t_{\infty})$ defined in eq.~\eqref{eq:critical_time_tc1}, the complexity rate clearly becomes positively (negatively) divergent, since $r_+ (r_-) \rightarrow \infty$ and $r_- (r_+)$ is finite. 
This is the hyperfast growth characteristic of asymptotically dS spacetimes \cite{Susskind:2021esx}.
The plots of the time evolution of CV2.0 and its time derivative during the interval $[-t_{\infty}, t_{\infty}]$ are reported in Fig.~\ref{fig:CV20case1}, when $d=3$ and for various choices of $\rho$.
As expected, the qualitative behaviour is similar to the case of empty dS space, see Fig.~5 in \cite{Jorstad:2022mls}.
One can also show that by keeping the regulator $r_{\rm max}$ finite, CV2.0 approaches a linear behaviour at early and late times, in the same fashion as Fig.~7 of \cite{Jorstad:2022mls}.
This is also consistent with reference \cite{Baiguera:2023tpt} (see Fig.~24), in the limit of vanishing shock wave.

\begin{figure}[t!]
    \centering
    \subfloat[]{
\includegraphics[width=0.3\textwidth]{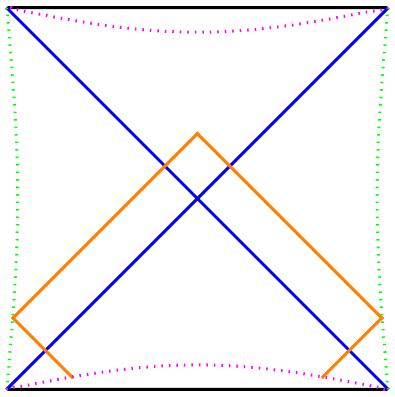} \label{subfig:WDW_evo2_regime1}}
\hfill \subfloat[]{\label{subfig:WDW_evo2_regime2} 
\includegraphics[width=0.3\textwidth]{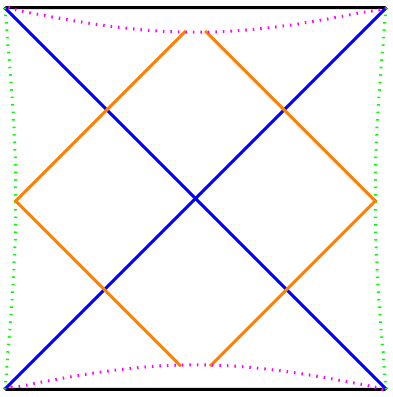}}
\hfill 
\subfloat[]{ \label{subfig:WDW_evo2_regime3}
\includegraphics[width=0.3\textwidth]{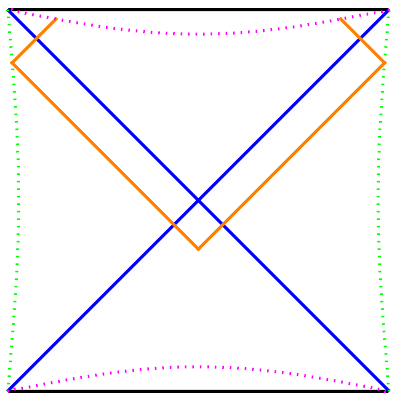}}
    \caption{Alternative time evolution of the WDW patch. Same colour scheme as in Fig.~\ref{fig:WDW_evo1}. In the intermediate regime, both joints of the WDW patch lie behind the cutoff surfaces. }
    \label{fig:WDW_evo2}
\end{figure}

Let us briefly comment on the case (depicted in Fig.~\ref{fig:WDW_evo2}) when $\rho \rightarrow 0$ in eq.~\eqref{eq:stretched_horizon_SdS4}, \ie the stretched horizon approaches the black hole one and the critical time satisfies $t_{\infty} < 0$. 
When $t_{\infty} \leq t \leq -t_{\infty}$,
the top and bottom joints of the WDW patch both sit behind the respective region $\mathcal{I}^{\pm}$.
As a consequence, by direct computation one finds that the rate vanishes during the intermediate regime.
Furthermore, since at least one of the joint always reaches timelike infinity, CV2.0 is always infinite.
These results are inconsistent with the intuition coming from circuit models and from the geometry of dS space \cite{Susskind:2021esx,Shaghoulian:2022fop,Jorstad:2022mls}. 
Therefore, the case with $\rho \rightarrow 0$ does not seem the relevant regime to capture the features expected by a holographic realization of computational complexity.

\subsubsection*{Case 2}

We refer to Fig.~\ref{fig:WDW_cosmo_stretched_case2} for the splitting of the WDW patch into the subregions corresponding to the integrals
\begin{subequations}
\label{eq:CV20_case2_int_regime}
\beq
\mathcal{J}_{1L}=\mathcal{J}_{1R} = \frac{\Omega_{d-1}}{G_{N} L^2} \int_{r_+}^{r_h} dr  \,  r^{d-1} \le  \frac{t}{2} -r^*(r) +r^*(r_{\rm st})  \ri \, ,
\label{eq:CV20_case2_int_regime_p1}
\eeq
\beq
\mathcal{J}_{2L} =
\mathcal{J}_{2R}=
\frac{\Omega_{d-1}}{G_{N} L^2} \int_{r_h}^{r_{\rm st}} dr  \,  r^{d-1} \le  2 r^*(r_{\rm st}) - 2 r^*(r)  \ri \, , 
\label{eq:CV20_case2_int_regime_p2}
\eeq
\beq
\mathcal{J}_{3L} =
\mathcal{J}_{3R} =  \frac{\Omega_{d-1}}{G_{N} L^2} \int_{r_-}^{r_h} dr  \,  r^{d-1} \le - \frac{t}{2} +r^*(r_{\rm st}) - r^*(r)  \ri \, ,
\label{eq:CV20_case2_int_regime_p3}
\eeq
\end{subequations}
where $r_{\pm}$ were defined in eq.~\eqref{eq:umax_vmin_WDWpatch_case2}.
All the other possible configurations of the WDW patch that can arise in case 2, together with the corresponding expressions for CV2.0, can be obtained by a limiting procedure from the above results.
In the following discussion, we will directly perform the limit $r_{\rm min} \rightarrow 0$ since it leads to regular results.

Famously, holographic complexity computed in a black hole background in asymptotically AdS spacetime admits linear growth at early (late) times.
The geometric origin for this behaviour is that the (bottom) top joint of the WDW patch moves behind the singularity, while the top (bottom) joint approaches a constant radial coordinate which coincides with the location of the horizon \cite{Carmi:2017jqz}.
In order to capture the linear growth, we need to study the early and late time regimes in the evolution of CV2.0.

\begin{figure}[t!]
    \centering
    \subfloat[]{ \label{subfig:CV20rate_case2}
\includegraphics[width=0.475\textwidth]{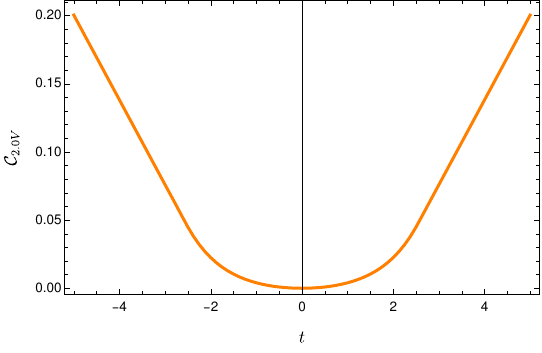}}
\hfill \subfloat[]{\label{subfigb:CV20rate_case2} 
\includegraphics[width=0.475\textwidth]{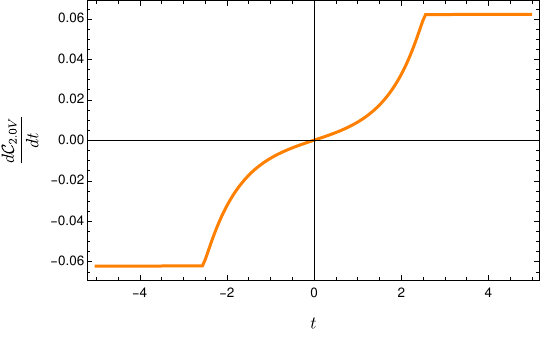}}
    \caption{Time dependence of (a) CV2.0 and (b) its rate, computed according to eq.~\eqref{eq:rateCV20_case2_p1}. We set $G_N L=1, d=3, \mu=0.14, \rho=0.01$. The intermediate interval that connects the regimes with linear growth corresponds to the times $t \in [-t_0, t_0]$, in terms of the critical value in eq.~\eqref{eq:critical_time_tc2}.  }
    \label{fig:CV20case2}
\end{figure}

The WDW patch can evolve in two different ways, distinguished by the sign of $t_0$ defined in eq.~\eqref{eq:critical_time_tc2}.
If $t_0>0$, or equivalently for a stretched horizon \eqref{eq:stretched_horizon_SdS4} close to the black hole horizon ($\rho \rightarrow 0$), the time evolution is characterized by the three regimes depicted in Fig.~\ref{fig:WDW_evo1}.
At times $t < -t_{0}$ the bottom joints is located behind the past singularity and CV2.0 is given by eq.~\eqref{eq:CV20_case2_int_regime} with the replacement $r_- \rightarrow 0$.
Next, we have an intermediate regime where both joints lie inside the black hole horizon.
When $t > t_{0}$, the top joint moves behind the future singularity and CV2.0 is obtained by replacing $r_+ \rightarrow 0$ in eq.~\eqref{eq:CV20_case2_int_regime}. 
Using the time derivatives \eqref{eq:rates_rprmcase2}, we find the rates
\beq
\frac{d \mathcal{C}_{2.0V}}{dt} = 
\frac{\Omega_{d-1}}{d \, G_{N} L^2}  \times 
 \begin{cases}
 \le - r_+^d \ri  &  \mathrm{if} \,\, t < -t_{0} \\
 \le r_-^d - r_+^d \ri  & \mathrm{if} \,\, -t_{0} \leq t \leq t_0 \\
  r_-^d   &  \mathrm{if} \,\, t > t_{0} \\
\end{cases}
\label{eq:rateCV20_case2_p1}
\eeq
In particular, the late (early) time behaviour can be found analytically, since it corresponds to $r_- \rightarrow r_h$ ($r_+ \rightarrow r_h$).
This gives
\beq
\frac{d \mathcal{C}_{2.0V}}{dt} \,\, \underset{t \rightarrow \pm t_{0}}{{\longrightarrow}} \,\,
\pm \frac{\Omega_{d-1}}{d \, G_{N} L^2} \, r_h^d \, ,
\label{eq:late_times_CV20_case2}
\eeq
which is the characteristic linear growth of complexity.
The time dependence of complexity and its rate of growth are depicted in Fig.~\ref{fig:CV20case2}.
The plots show that there is an intermediate regime with non-trivial time evolution that interpolates between early and late times when the advocated linear growth is achieved.

\begin{figure}[t!]
    \centering
    \subfloat[]{ \label{subfig:CV20rate_case2r05}
\includegraphics[width=0.475\textwidth]{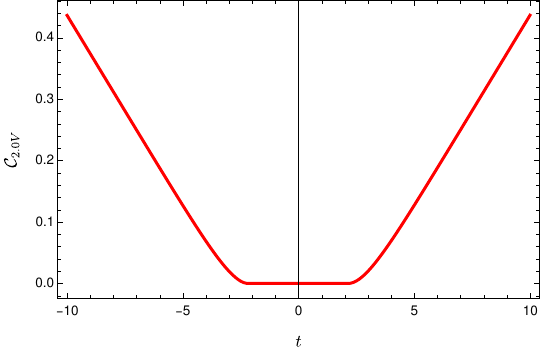}}
\hfill \subfloat[]{\label{subfigb:CV20rate_case2r05} 
\includegraphics[width=0.475\textwidth]{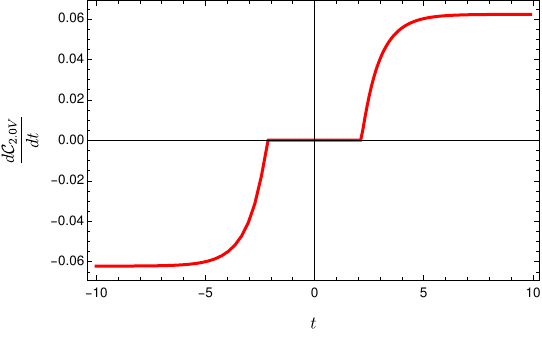}}
    \caption{Time dependence of (a) CV2.0 and (b) its rate, computed according to eq.~\eqref{eq:rateCV20_case2_p2}. We set $G_N L=1, d=3, \mu=0.14, \rho=0.5$. The intermediate regime when complexity is constant corresponds to the interval $t \in [t_0, -t_0]$ in terms of the critical time in eq.~\eqref{eq:critical_time_tc2}. }
    \label{fig:CV20case2r05}
\end{figure}

The other possible evolution of the WDW patch (reported in Fig.~\ref{fig:WDW_evo2}) shows up when the critical time satisfies $t_0 \leq 0$, which always happens when $\rho \rightarrow 1$.
In this case, the WDW patch starts at times $t < t_0$ with the bottom joint behind the past singularity, while the top joint is located behind the black hole horizon.
Afterwards, both joints sit behind the respective singularities, 
and after $t > t_0$ the bottom joints moves after the past singularity and stays in the interior of the black hole.
The rate of CV2.0 reads
\beq
\frac{d \mathcal{C}_{2.0V}}{dt} = 
\frac{\Omega_{d-1}}{d \, G_{N} L^2}  \times 
 \begin{cases}
 \le - r_+^d \ri  &  \mathrm{if} \,\, t < t_{0} \\
0  & \mathrm{if} \,\, t_{0} \leq t \leq -t_0 \\
  r_-^d   &  \mathrm{if} \,\, t > -t_{0} \\
\end{cases}
\label{eq:rateCV20_case2_p2}
\eeq
Again, at late (early) times the bottom (top) joint of the WDW patch approaches the black hole horizon, leading to the constant rate computed in eq.~\eqref{eq:late_times_CV20_case2}.
This scenario is depicted in Fig.~\ref{fig:CV20case2r05}.
The result is very similar to the case of black holes in asymptotically AdS spacetime, where complexity is constant in the intermediate regime between the critical times \cite{Carmi:2017jqz}, while at late and early times approaches a constant rate.
Therefore, this analysis suggests that the expected interior growth of a black hole in asymptotically dS spacetime is recovered when the stretched horizon is close to the cosmological one.
Notably, this choice is the same that reproduced the hyperfast growth for the inflationary patch of spacetime in case 1.

\subsubsection*{Case 3}

The configuration of the WDW patch is given by the plot depicted in fig.~\ref{fig:WDW_BH_stretched}.
Let us split the integration region into various parts, according to the labels used in the picture.
We begin by studying the right side of the Penrose diagram.
The corresponding terms in the sum \eqref{eq:decomposition_CV20_general} are defined as follows:
\begin{subequations}
\label{eq:Case3_CV20}
\beq
\mathcal{J}_{1R} = 
\frac{\Omega_{d-1}}{d G_{N} L^2} \le r_h^d - r_{\rm min}^d \ri \int_{-t_R - r^*(r_{\rm st}^R)}^{+\infty} du    \, ,
\eeq
\beq
\mathcal{J}_{2R}=
\frac{\Omega_{d-1}}{G_{N} L^2} \int_{r_h}^{r_{\rm st}^R} dr  \,  r^{d-1} \le  2 r^*(r_{\rm st}^R) - 2 r^*(r)  \ri = \mathrm{const.} 
\eeq
\beq
\mathcal{J}_{3R} = \frac{\Omega_{d-1}}{d G_{N} L^2} \le r_h^d - r_{\rm min}^d \ri \int_{-\infty}^{-t_R + r^*(r^R_{\rm st})} dv     \, , 
\eeq
\beq
\mathcal{J}_{4} = \frac{\Omega_{d-1}}{d G_{N} L^2} \le r_c^d - r_{h}^d \ri \int_{-\infty}^{\infty} du = \mathrm{const.} 
\eeq
\end{subequations}
In the previous analysis, we denoted with ``$\mathrm{const.}$'' the time-independent terms.
For convenience, the integrals referring to regions inside the black hole are performed by first integrating along the radial coordinate $r$, and then along one of the null directions $u,v$.
While some expressions are formally divergent and can be explicitly evaluated using an appropriate regularization, here we will only focus on the time dependence of the problem.
In particular, we notice that
\beq
\frac{d \mathcal{J}_{1R} }{dt_R} = - \frac{d \mathcal{J}_{3R}}{d t_R} =  \frac{\Omega_{d-1}}{d G_{N} L^2} \le r_h^d - r_{\rm min}^d \ri  \quad \Rightarrow \quad  
\frac{d}{dt_R} \le \mathcal{J}_{1R} + \mathcal{J}_{3R}  \ri = 0 \, .
\eeq
A similar computation shows that
\beq
\frac{d \mathcal{J}_{1L} }{dt_L} = - \frac{d \mathcal{J}_{3L}}{d t_L} = - \frac{\Omega_{d-1}}{d G_{N} L^2} \le r_{\rm max}^d - r_{c}^d \ri  \quad \Rightarrow \quad  
\frac{d}{dt_L} \le \mathcal{J}_{1L} + \mathcal{J}_{3L}  \ri = 0 \, .
\eeq
In conclusion, we trivially find 
\beq
\frac{d \mathcal{C}_{V2.0}}{d t_R} = 
\frac{d \mathcal{C}_{V2.0}}{d t_L} = 0 \, . 
\label{eq:rate_CV20_case3}
\eeq
Remarkably, the spacetime volumes located on the left and right sides of the WDW patch are separately time-independent.
Therefore, any choice of the relative evolution between the left and right boundary times $t_L, t_R$ leads to a vanishing complexity rate!
This behaviour is very different compared to the previous cases and reflects the geometric feature that the portion of WDW patches that emerge from the past cutoff surfaces of the Penrose diagram is compensated by other regions of the same size that move behind the future cutoff surfaces.
If we aim to give a dual interpretation to the spacetime volume inside the WDW patch, we would conclude that the putative dual state has always the same computational complexity.
In particular, it is not able to give us any information neither about the growth of black holes or of the space beyond the cosmological horizon.

\subsubsection*{Cases 4-5}

Decomposing the WDW patch in a similar way as case 3, we find that CV2.0 is time-independent.


\subsection{Complexity=action}
\label{ssec:CA}

According to the CA proposal, holographic complexity is proportional to the on-shell gravitational action evaluated in the WDW patch, \ie
\beq
\mathcal{C}_A = \frac{I_{\rm WDW}}{\pi} \, , \qquad
I_{\rm WDW} = \sum_{\mathcal{X}}  I_{\mathcal{X}} \, , \qquad
\mathcal{X} \in \lbrace \mathcal{B}, \rm GHY, \mathcal{N}, \mathcal{J} ,\rm ct  \rbrace \, .
\label{eq:tot_grav_action}
\eeq
The terms contributing to the list labelled by $\mathcal{X}$ are collected in \cite{Lehner:2016vdi} or in appendix A of \cite{Carmi:2016wjl} (see also \cite{Parattu:2015gga,Chakraborty:2016yna}).
We briefly review their expressions:
\begin{itemize}
    \item The bulk term is the Einstein-Hilbert action
    \beq
I_{\mathcal{B}} = \frac{1}{16 \pi G_N} \int_{\rm WDW} d^{d+1} x \, \sqrt{-g}	 \,  \le  R - 2 \Lambda  \ri \, ,
\label{eq:general_bulk_action}
\eeq
where $R$ is the Ricci scalar, and $\Lambda$ the cosmological constant.
\item The Gibbons-Hawking-York (GHY) term is evaluated on timelike or spacelike codimension-one boundary surfaces
\beq
I_{\rm GHY} = \frac{\varepsilon_{t,s}}{8 \pi G} \int_{\mathcal{B}_{t,s}} d^d x \, \sqrt{|h|} \, K \, ,
\label{eq:GHY_term}
\eeq
where $h$ is the determinant of the induced metric and $K$ is the trace of the extrinsic curvature.
The parameter $\varepsilon_{t,s}=\pm 1$ distinguishes if the surface of interest ${\mathcal{B}_{t,s}} $ is timelike or spacelike, respectively.
\item The codimension-one null boundaries admit the contribution 
\beq
I_{\mathcal{N}} = \frac{\varepsilon_n}{8 \pi G} 
\int_{\mathcal{B}_n} d\lambda d^{d-1} x \, \, \sqrt{|\gamma|} \, \kappa (\lambda) \, ,
\label{eq:null_codim1_term}
\eeq
where $\varepsilon_n = \pm 1$ depends on the orientation of the null normal to the surface, $\lambda$ is a parameter along the congruence of geodesics, 
$\gamma$ is the induced metric along the other orthogonal directions and $\kappa(\lambda)$ is the acceleration of the null geodesics.
In the case of an affine parametrization, $\kappa=0$.
\item Codimension-two joints arise from the intersection of codimension-one surfaces. In this work, the relevant joints will always arise from intersections involving at least one null surface. Their expression reads
 \beq
I_{\mathcal{J}} =  \frac{\varepsilon_{\mathfrak{a}}}{8 \pi G} \int_{\mathcal{J}} d^{d-1} x \, 
 \sqrt{|\gamma|} \, \mathfrak{a}  \, ,
 \label{eq:joints_action}
\eeq
where the coefficient $ \varepsilon_{\mathfrak{a}} = \pm 1$ depends on the orientations of the joint, and $\gamma$ is the determinant of the induced metric along the codimension-two joint.
The integrand $\mathfrak{a}$ is 
\beq
\mathfrak{a} = \begin{cases}
\log \left| \mathbf{t} \cdot \mathbf{k}  \right|  & \mathrm{if} \,\, \mathbf{t} \,\, \mathrm{timelike} \,\, \mathrm{and} \,\, \mathbf{k} \,\, \mathrm{null}   \\
\log \left| \mathbf{n} \cdot \mathbf{k}  \right|  & \mathrm{if} \,\, \mathbf{n} \,\, \mathrm{spacelike} \,\, \mathrm{and} \,\, \mathbf{k} \,\, \mathrm{null} \\
\log \left| \frac{1}{2} \mathbf{k}_L \cdot \mathbf{k}_R  \right|  & \mathrm{if} \,\, \mathbf{k}_L, \mathbf{k}_R \,\, \mathrm{null}  \\
\end{cases}
\label{eq:integrand_a_actionjoints}
\eeq
where the argument of the logarithms is the scalar product between the outward-directed one-forms normal to the codimension-one surfaces intersecting at the joint.  
\item Since the gravitational action is not reparametrization-invariant as it stands, we need to add a counterterm on codimension-one null boundaries
\beq
I_{\rm ct} = \frac{1}{8 \pi G} \int_{\mathcal{B}_n} 
d\lambda d^{d-1} x \, \sqrt{\gamma} \,  \Theta \, \log |\ell_{\rm ct} \Theta| \, ,
\label{eq:counterterm_action}
\eeq 
where $\Theta$ is the expansion of the congruence of null geodesics and $\ell_{\rm ct}$ is an arbitrary length scale.
\end{itemize}

The asymptotically dS spacetimes with metric \eqref{eq:asympt_dS} have constant Ricci scalar, thus the bulk term is simply proportional to the CV2.0 conjecture:
\beq
I_{\mathcal{B}} = 
\frac{d}{8 \pi G_N L^2} V_{\rm WDW} = \frac{d}{8 \pi} \, \mathcal{C}_{2.0V} \, .
\label{eq:bulk_term_CA}
\eeq
Therefore, the bulk contribution satisfies the same properties discovered for the CV2.0 computation performed in section \ref{ssec:CV20}.
In the following, we will denote the combination of boundary terms with 
\beq
I_{\rm bdy} = I_{\rm GHY} + I_{\mathcal{N}} + I_{\mathcal{J}} + I_{\rm ct} \, .
\label{eq:bdy_action}
\eeq
We report here the main results, while we refer the reader to appendix~\ref{app:details_CA} for more details.

\subsubsection*{Case 1}

\begin{figure}[t!]
    \centering
    \subfloat[]{ \label{subfig:CArate_case1}
\includegraphics[width=0.475\textwidth]{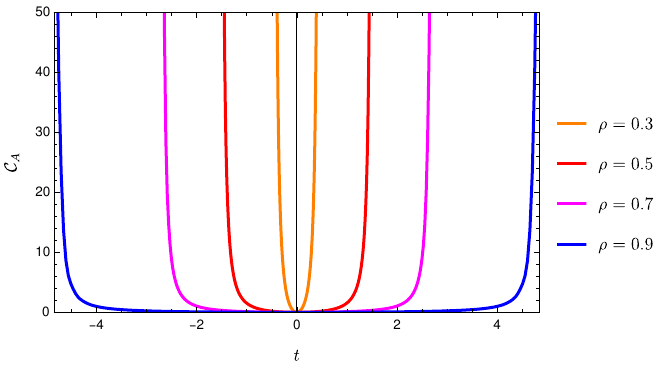}}
\hfill \subfloat[]{\label{subfigb:CArate_case1} 
\includegraphics[width=0.475\textwidth]{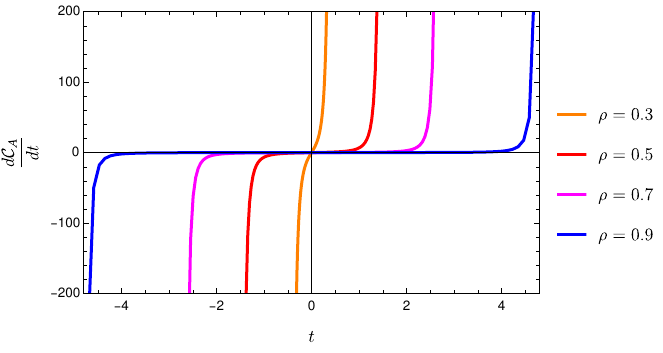}}
    \caption{Time dependence of (a) CA and (b) its rate, computed according to case 1 in eq.~\eqref{eq:rate_CA_case1} during the interval $[-t_{\infty}, t_{\infty}]$ with critical time defined in eq.~\eqref{eq:critical_time_tc1}. We set $G_N L=1, d=3, \mu=0.14, \ell_{\rm ct}=1/3$, and we vary $\rho$ in eq.~\eqref{eq:stretched_horizon_SdS4}, while keeping $t_{\infty} >0$.  }
    \label{fig:CAcase1}
\end{figure}

We study a symmetric configuration of both the time evolution \eqref{eq:symmetric_times} and the location of the stretched horizons $r_{\rm st}^R=r_{\rm st}^L$.
When the critical time satisfies $t_{\infty} >0$, the evolution of the WDW patch is described by Fig.~\ref{fig:WDW_evo1}.
During the intermediate time regime $-t_{\infty} \leq t \leq t_{\infty}$, the CA conjecture is obtained by summing the bulk term \eqref{eq:bulk_term_CA} and the boundary contribution evaluated in eq.~\eqref{eq:bdy_action_case1}.
It is more insightful to study the rate of growth of CA, obtained by applying the time derivatives \eqref{eq:rates_rprmcase1}:
\beq
\begin{aligned}
\frac{d \mathcal{C}_A}{dt} = \frac{\Omega_{d-1}}{8 \pi^2 G_N} & \left[  \frac{(r_+)^d}{L^2}  + \frac{d-1}{2} (r_+)^{d-2} f(r_+) \log \left| \frac{\ell_{\rm ct}^2 (d-1)^2 f(r_+) }{(r_+)^2} \right| + (r_+)^{d-1} \, \frac{f'(r_+)}{2}  \right. \\
&  \left.  - \frac{(r_-)^d}{L^2}  - \frac{d-1}{2} (r_-)^{d-2} f(r_-) \log \left| \frac{\ell_{\rm ct}^2 (d-1)^2 f(r_-) }{(r_-)^2} \right| - (r_-)^{d-1} \, \frac{f'(r_-)}{2}  \right] \, ,
\end{aligned}
\label{eq:rate_CA_case1}
\eeq
where $r_{\pm}$ were defined in eq.~\eqref{eq:umax_vmin_WDWpatch}.
We already established by inspection of eq.~\eqref{eq:rateCV20_case1} that CV2.0 displays a hyperfast growth when approaching the critical times $\pm t_{\infty}$.
To check whether the rate of CA is also divergent when the top joint of the WDW patch approaches $\mathcal{I}^+$, we perform a series expansion around $r_{+} = \infty$ of eq.~\eqref{eq:rate_CA_case1}.
Using the explicit blackening factor in eq.~\eqref{eq:asympt_dS}, we get
\beq
 \frac{d \mathcal{C}_A}{dt} \,\, \underset{t \rightarrow t_{\infty}}{{\longrightarrow}} \,\,\frac{\Omega_{d-1}}{16 \pi^2 G_N L^2} (d-1)    (r_+)^{d}  \log \left| \frac{L^2}{\ell_{\rm ct}^2 (d-1)^2} \right| \, .
\label{eq:approx_rate_CA_case1}
\eeq
The complexity rate (and therefore CA itself) is positively divergent if the counterterm length scale satisfies
\beq
\ell_{\rm ct} < \frac{L}{d-1} \, .
\label{eq:positive_CA_case1}
\eeq
This requirement is the same that fixes the positivity of complexity in empty dS space \cite{Jorstad:2022mls}.
In a similar way, when $t \rightarrow -t_{\infty}$, a series expansion around $r_-=\infty$ gives
\beq
 \frac{d \mathcal{C}_A}{dt} \,\, \underset{t \rightarrow -t_{\infty}}{{\longrightarrow}} \,\, - \frac{\Omega_{d-1}}{16 \pi^2 G_N L^2} (d-1)    (r_-)^{d}  \log \left| \frac{L^2}{\ell_{\rm ct}^2 (d-1)^2} \right| \, ,
\label{eq:approx_rate_CA_case1_past}
\eeq
and now the condition \eqref{eq:positive_CA_case1} implies that the rate is negatively divergent instead.
The time evolution of CA during the intermediate regime is depicted in Fig.~\ref{fig:CAcase1}.
In this plot, we consider various possibilities for the parameter $\rho$ defining the location of the stretched horizon, but we always keep $\ell_{\rm ct}=1/3$ to satisfy \eqref{eq:positive_CA_case1}, so that complexity is positive (negative) at late (early) times. 

The time evolution of the WDW patch when $t< - t_{\infty}$ or $t> t_{\infty}$ is different compared to the intermediate regime that we studied above because GHY terms are now non-vanishing.
The most relevant and characteristic feature of asymptotically dS spacetimes is the hyperfast growth, which we already capture with the expansion in eq.~\eqref{eq:approx_rate_CA_case1}.
At early and late times, the divergent growth of complexity can be regularized by keeping the radial coordinate $r=r_{\rm max}$ of the cutoff surface finite.
The analysis and results in the latter case are analogous to empty dS space (see Fig.~8 in \cite{Jorstad:2022mls}), and we will not report them here.

When the critical time satisfies $t_{\infty} <0$, the WDW patch follows the evolution depicted in Fig.~\ref{fig:WDW_evo2}, and  CA is always trivially divergent.
Therefore, this case does not seem to not capture the expected behaviour of complexity for a putative dual state.

\subsubsection*{Case 2}

\begin{figure}[t!]
    \centering
    \subfloat[]{ \label{subfig:CArate_case2}
\includegraphics[width=0.475\textwidth]{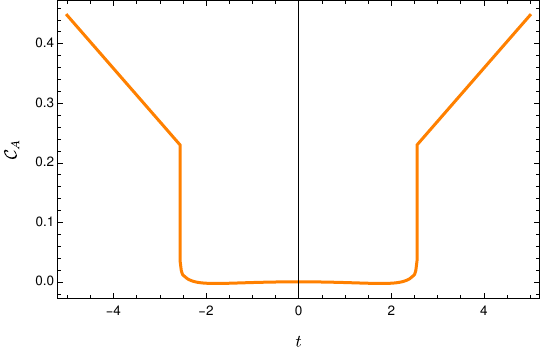}}
\hfill \subfloat[]{\label{subfigb:CArate_case2} 
\includegraphics[width=0.475\textwidth]{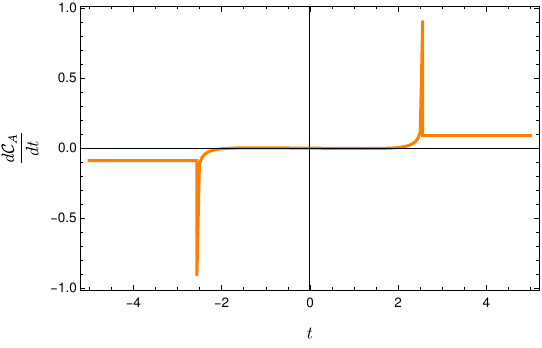}}
    \caption{Time dependence of (a) CA and (b) its rate, computed according to eq.~\eqref{eq:rateCA_case2_p1}. We set $G_N L=1, d=3, \mu=0.14, \rho=0.01, \ell_{\rm ct}=1/3$. }
    \label{fig:CAcase2}
\end{figure}

In the following, we focus on a symmetric configuration of the boundary times \eqref{eq:symmetric_times} and of the stretched horizons $r_{\rm st}^R=r_{\rm st}^L$.
Let us first assume that the critical time satisfies $t_0>0$, when the evolution of the WDW patch is governed by the diagrams in Fig.~\ref{fig:WDW_evo1}.
By applying the time derivatives \eqref{eq:rates_rprmcase2} to the CA conjecture \eqref{eq:tot_grav_action}, we obtain
\small
\beq
\frac{8 \pi^2 G_N}{\Omega_{d-1}}  \frac{d \mathcal{C}_A}{dt} =  \begin{cases}
   - \mu d - \frac{(r_+)^d}{L^2} - \frac{d-1}{2} (r_+)^{d-2} f(r_+) \log \left| \frac{\ell_{\rm ct}^2 (d-1)^2 f(r_+) }{(r_+)^2} \right| - (r_+)^{d-1} \, \frac{f'(r_+)}{2}       &  \mathrm{if}  \,\, t < -t_{0} \\ 
 \Big(  - \frac{(r_+)^d}{L^2}  - \frac{d-1}{2} (r_+)^{d-2} f(r_+) \log \left| \frac{\ell_{\rm ct}^2 (d-1)^2 f(r_+) }{(r_+)^2} \right| - (r_+)^{d-1} \, \frac{f'(r_+)}{2}   & \\
    + \frac{(r_-)^d}{L^2}  + \frac{d-1}{2} (r_-)^{d-2} f(r_-) \log \left| \frac{\ell_{\rm ct}^2 (d-1)^2 f(r_-) }{(r_-)^2} \right| + (r_-)^{d-1} \, \frac{f'(r_-)}{2}  \Big) & \mathrm{if} \,\, t \in [-t_0, t_0] \\
 \mu d  + \frac{(r_-)^d}{L^2}  + \frac{d-1}{2} (r_-)^{d-2} f(r_-) \log \left| \frac{\ell_{\rm ct}^2 (d-1)^2 f(r_-) }{(r_-)^2} \right| + (r_-)^{d-1} \, \frac{f'(r_-)}{2}   &  \mathrm{if} \,\, t > t_{0} \\
\end{cases}
\label{eq:rateCA_case2_p1}
\eeq
\normalsize
where the results for the boundary terms were taken from appendix \ref{app:ssec:case2}.
We represent the numerical evolution of CA and of its rate in Fig.~\ref{fig:CAcase2}.
It is evident that while the complexity is continuous, instead its time derivative presents a discontinuity at the critical times.
The value of the jump at $t= t_0 $ is simply obtained by comparing the rate in the last regime with the rate at intermediate times in the limit $r_{+} \rightarrow 0$, giving
\beq
\frac{d \mathcal{C}_A}{dt}\Big|_{t= t_0^{+}} - \frac{d \mathcal{C}_A}{dt}\Big|_{t=t_0^-} = \frac{\Omega_{d-1}}{8 \pi^2 G_N} \mu d \, .
\label{eq:disc_rateCA_case2p1}
\eeq
The GHY term is fully responsible for this jump since all the terms on null boundaries (either codimension-one or codimension-two) evaluated at the singularity vanish.
At $t=-t_0,$ the same conclusion (except for an overall minus sign) holds.
It is straightforward to determine the late (early) time limits, which correspond to the bottom (top) joint of the WDW patch approaching the horizon radius $r_h$.
The rate reads
\beq
\lim_{t \rightarrow \pm \infty} \frac{d \mathcal{C}_A}{dt} = \pm \frac{\Omega_{d-1}}{8 \pi^2 G_N} \left[ \mu d  + \frac{(r_h)^d}{L^2}  + (r_h)^{d-1} \, \frac{f'(r_h)}{2}  \right] \, ,
\label{eq:rate_CA_case2_latelimit}
\eeq
which is clearly a constant.
This shows that CA asymptotically reproduces the celebrated linear trend of complexity.

\begin{figure}[t!]
    \centering
    \subfloat[]{ \label{subfig:CArate_case2r05}
\includegraphics[width=0.475\textwidth]{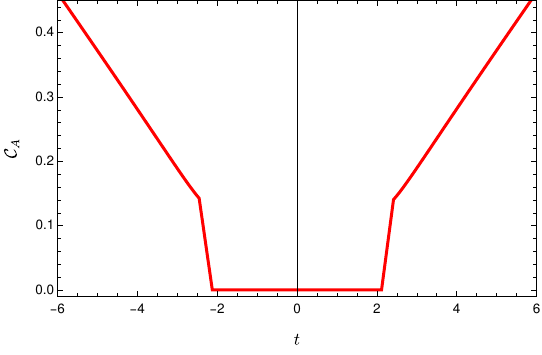}}
\hfill \subfloat[]{\label{subfigb:CArate_case2r05} 
\includegraphics[width=0.475\textwidth]{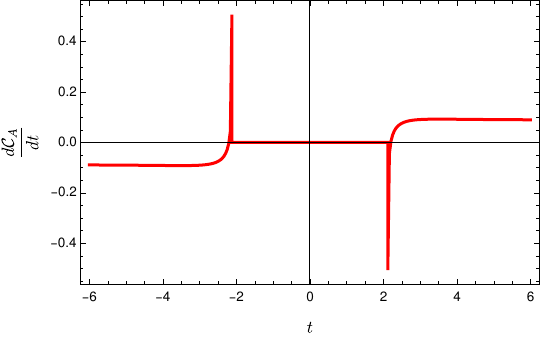}}
    \caption{Time dependence of (a) CA and (b) its rate, computed according to eq.~\eqref{eq:rateCA_case2_p2}. We set $G_N L=1, d=3, \mu=0.14, \rho=0.5, \ell_{\rm ct} = 1/3$. }
    \label{fig:CAcase2r05}
\end{figure}

Next, let us assume that the critical time is instead negative, \ie $t_0 <0$.
In this case the evolution of the WDW patch is depicted in Fig.~\ref{fig:WDW_evo2}, and similar computations lead to the rate
\small
\beq
\frac{8 \pi^2 G_N}{\Omega_{d-1}} \frac{d \mathcal{C}_A}{dt} = \begin{cases}
    - \mu d - \frac{(r_+)^d}{L^2} - \frac{d-1}{2} (r_+)^{d-2} f(r_+) \log \left| \frac{\ell_{\rm ct}^2 (d-1)^2 f(r_+) }{(r_+)^2} \right| - (r_+)^{d-1} \, \frac{f'(r_+)}{2}     &  \mathrm{if}  \,\, t < t_{0} \\
 0  & \mathrm{otherwise} \\
 \mu d  + \frac{(r_-)^d}{L^2}  + \frac{d-1}{2} (r_-)^{d-2} f(r_-) \log \left| \frac{\ell_{\rm ct}^2 (d-1)^2 f(r_-) }{(r_-)^2} \right| + (r_-)^{d-1} \, \frac{f'(r_-)}{2}    &  \mathrm{if} \,\, t > -t_{0} \\
\end{cases}
\label{eq:rateCA_case2_p2}
\eeq
\normalsize
whose main difference with the previous scenario is the time-independence of the intermediate regime.\footnote{The time-independence of the bulk term is a trivial consequence of the intermediate regime of CV2.0 in eq.~\eqref{eq:rateCV20_case2_p2}. The vanishing rate of the boundary terms is found in eq.~\eqref{eq:rate_bdy_app_case2}.   }
We present a numerical evaluation of CA and of its rate in Fig.~\ref{fig:CAcase2r05}.
First of all, we observe again that the time derivative of CA is discontinuous at the critical time.
One can evaluate the jump at $t=-t_0$ by performing the limit $r_- \rightarrow 0$ to get
\beq
\frac{d \mathcal{C}_A}{dt}\Big|_{t= -t_0^{+}} - \frac{d \mathcal{C}_A}{dt}\Big|_{t= -t_0^-} = \frac{\Omega_{d-1}}{8 \pi^2 G_N} \mu d \, ,
\label{eq:disc_rateCA_case2p2}
\eeq
which is the same result obtained in eq.~\eqref{eq:disc_rateCA_case2p1}.
An analogous computation shows that the same conclusion holds for the other critical time.
The early and late time limits are also given by the same expression \eqref{eq:rate_CA_case2_latelimit} determined above.
Therefore, we conclude that the location of the stretched horizons in case 2 mainly affects the intermediate regime.
When taking $\rho \rightarrow 1$, complexity shows an evolution very similar to the case of black holes in AdS \cite{Carmi:2017jqz}.

\subsubsection*{Case 3}

The boundary terms of the gravitational action in the setting described by Fig.~\ref{fig:WDW_BH_stretched} are evaluated in appendix \ref{app:ssec:case3}.
The main observation is that the rates in eqs.~\eqref{eq:rate_CV20_case3} and \eqref{eq:rate_CA_bdy_case3} vanish for any choice of the boundary times $t_L, t_R$, therefore implying that
\beq
\frac{d \mathcal{C}_A}{d t_L} = \frac{d \mathcal{C}_A}{d t_R} = 0 \, .
\label{eq:rate_CA_case3}
\eeq
In particular, the on-shell action inside the WDW patch evaluated on the right and left sides of the Penrose diagram are separately time-independent.

\subsubsection*{Cases 4-5}

In analogy with case 3, one can show that CA conjecture in cases 4 and 5 is also time-independent.
The precise steps are similar to the ones reported in appendix \ref{app:ssec:case3}.

\section{Codimension-one proposals}
\label{sec:Cod1 complexity}

We continue the analysis of holographic complexity in the extended SdS background with the codimension-one proposals: the volume (section \ref{ssec:CV}), and a special class of the complexity=anything conjectures (section \ref{sec:CAny CMC}).

\subsection{Complexity=volume}
\label{ssec:CV}

The CV conjecture in asymptotically dS space states that complexity is associated with the maximal volume $V$ of a slice anchored at the stretched horizons.
Assuming that the codimension-one surface preserves the spherical symmetry of the background, we describe it in terms of a parameter $\sigma$ such that 
\begin{equation}\label{eq:vol computation}
    \begin{aligned}
    \mathcal{C}_V = \frac{V}{G_N L} &=   \frac{\Omega_{d-1}}{G_N L} \int r^{d-1} \sqrt{-f(r) \dot{u}^2 - 2 \dot{u} \dot{r}} \, d\sigma \\
    &= \frac{\Omega_{d-1}}{G_N L} \int r^{d-1} \sqrt{-f(r) \dot{v}^2 + 2 \dot{v} \dot{r}} \, d\sigma \, ,
\end{aligned}
\end{equation}
where we introduced the null coordinates \eqref{eq:general_null_coordinates} and used the notation $\cdot \equiv d/d\sigma$.
In the following, we will focus on the computation of $V$.
Since the volume functional does not depend on $u (v)$, there is a corresponding conserved momentum
\begin{equation}
    P_u \equiv \frac{-f(r) \dot{u} - \dot{r}}{\sqrt{-f(r) \dot{u}^2 - 2 \dot{u} \dot{r}}} \, r^{d-1} =  \frac{-f(r) \dot{v} + \dot{r}}{\sqrt{-f(r) \dot{v}^2 + 2 \dot{v} \dot{r}}} \, r^{d-1} \equiv P_v \, .
    \label{eq:conserved_momentum_CV}
\end{equation}
From now on, we define $P_u = P_v \equiv P \, L^{d-1}$.
The functional \eqref{eq:vol computation} is invariant under reparametrizations.
This allows us to fix the gauge
\begin{equation}
    \sqrt{-f(r) \dot{u}^2 - 2 \dot{u} \dot{r}} = \sqrt{-f(r) \dot{v}^2 + 2 \dot{v} \dot{r}} = \left( \frac{r}{L} \right)^{d-1} \, ,
    \label{gauge}
\end{equation}
under which choice the volume reads
\begin{equation}
    V = \Omega_{d-1} \int \frac{r^{2(d-1)}}{L^{d-1}} d\sigma \, .
\end{equation}
With the gauge choice in eq.~\eqref{gauge}, the conserved momentum \eqref{eq:conserved_momentum_CV} becomes
\begin{equation}\label{eq:P u dot}
    P = -f(r) \dot{u} - \dot{r} = -f(r) \dot{v} + \dot{r} \, .
\end{equation}
Solving for the derivatives of the null coordinates and plugging the results into eq.~\eqref{gauge}, we get
\begin{equation}\label{eq:dot u dot v}
    \dot{r}_{\pm} = \pm \sqrt{P^2 + f(r) \qty(\frac{r}{L})^{2(d-1)}} \, , \qquad
    \dot{u}_{\pm} = \frac{-P - \dot{r}_{\pm}}{f(r)} \, , 
    \qquad \dot{v}_{\pm} = \frac{-P + \dot{r}_{\pm}}{f(r)} \, .
\end{equation}
For a fixed orientation of the parameter $\sigma$ along the maximal slice, when $r$ increases in the same direction as $\sigma$ the solution is given by the $+$ branch, otherwise we must look at the $-$ branch. 
We stress that the sign of $P$ for a given maximal slice depends on the orientation of $\sigma$. This can be understood by noting from eq.~\eqref{eq:P u dot} that
\begin{equation}
    P = -f(r) \dot{t}  \, .
\end{equation}
Outside the cosmological horizon and inside the black hole we have $f(r) < 0$, so ${\rm sign}(P) ={\rm sign}(\dot{t})$. 
In other words, in these regions $P>0$ if $\sigma$ is oriented as the Killing vector $\partial_t$, otherwise $P<0$.\footnote{Of course, in the external region $f(r) > 0$, so ${\rm sign}(P) =- {\rm sign}(\dot{t})$ and opposite conclusions follow.} From Fig. \ref{fig:WDW_BH_stretched}, we conclude that if $\sigma$ runs from the right stretched horizon to the left one, $P>0$ in the future black hole interior and $P<0$ in the future inflating region. 
This is the convention we will employ in the rest of the work.

The first equation in \eqref{eq:dot u dot v} can be written in the form
\begin{equation}
    \dot{r}_{\pm}^2 + U_{\rm eff}(r) = P^2 \, , \qquad
    U_{\rm eff}(r) = - f(r) \, \le \frac{r}{L} \ri^{2(d-1)} \, , 
    \label{eq:classical_motion}
\end{equation}
which describes the motion of a classical non-relativistic particle with energy $P^2$ in a potential $U_{\rm eff}(r)$ \cite{Carmi:2017jqz,Chapman:2018dem,Chapman:2018lsv,Belin:2021bga}.
The turning points $r_t$ of the extremal codimension-one surface satisfy $\dot{r}_{\pm}=0$, and are thus the solutions to the equation
\begin{equation}
 P^2 = U_{\rm eff}(r_t) = 
 -\left( \frac{r_t}{L} \right)^{2(d-1)} + \frac{2 \mu}{L^{2(d-1)}} \, r_t^d + \left( \frac{r_t}{L} \right)^{2d}  \, ,
    \label{eq:turning}
\end{equation}
where the explicit form of the blackening factor \eqref{eq:asympt_dS} has been used in the last step.
The qualitative behaviour of the function $U_{\rm eff}(r)$ can be easily inferred from the properties of the blackening factor $f(r)$. In particular, $U_{\rm eff}(r)$ vanishes for $r= 0, r_h, r_c$ and has an opposite sign compared to $f(r)$. In other words, $U_{\rm eff}(r) \geq 0$ for $0 \leq r \leq r_h$ and $r \geq r_c$, whereas $U_{\rm eff}(r) < 0$ for $r_h < r < r_c$. Hence, we deduce that $U_{\rm eff}(r)$ has at least a (local) maximum in the interval $(0, r_h)$ and at least a minimum in the range $(r_h, r_c)$, while it diverges for $r \rightarrow +\infty$. A plot of the function $U_{\rm eff}(r)$ is displayed in Fig. \ref{subfig:Ueff_CV_case1}, from which it can be seen that there exist only one local maximum and one local minimum. 
In the parameter space we have numerically explored, the function $U_{\rm eff}(r)$ keeps the same qualitative behaviour. So, in what follows we will assume the existence of a single local maximum in $U_{\rm eff}(r)$.

The shape of the effective potential $U_{\rm eff}(r)$ has crucial consequences on the time evolution of the maximal codimension-one slice. 
For a fixed value of the conserved momentum $P$, from eq.~\eqref{eq:classical_motion} it is clear that the corresponding maximal surface can only explore spacetime regions where $P^2 \geq U_{\rm eff}(r)$. When the condition is saturated, \ie $P^2 = U_{\rm eff}(r_t)$, a potential barrier is met and the maximal surface presents a turning point $r=r_t$. The following observations thus follow:
\begin{itemize}
    \item If the maximal surface extends into the inflating region $r \geq r_c$, the potential barrier prevents the surface from reaching timelike infinity $\mathcal{I}^{\pm}$. Strictly speaking, timelike infinity can only be reached by the maximal surfaces with $P \rightarrow \pm \infty$.
    \item If the maximal surface enters the black hole $r \leq r_h$, it ends up in the singularity $r=0$ unless a potential barrier is met. Therefore, the singularity is avoided when the corresponding $P^2$ is lower than the maximal value of the effective potential inside the black hole.
\end{itemize}    

These observations have important consequences for the evolution of holographic complexity.
First, the CV prescription requires to compute the maximal volume of a connected surface anchored at the stretched horizons. 
Therefore, any extremal surface starting from one stretched horizon and falling into the singularity before reaching the other stretched horizon will be discarded in our analysis.\footnote{ \label{footnote:disconnected_solutions}  In principle, one can generalize the CV conjecture to include the possibility that an extremal surface is composed of two parts ending at $r=0$, and connected through the singularity. This case is considered in section 2.2.2 of \cite{Belin:2022xmt} for asymptotically AdS space, and a similar prescription is employed in empty dS space in section 5 of \cite{Jorstad:2022mls}. However, these configurations are either extremal but not maximal, or they appear after some critical time. 
We will not consider them in this work.  }
Second, hyperfast growth is achieved whenever the extremal surface reaches timelike infinity $\mathcal{I}^{\pm}$. 
Given the discussion above, we conclude that as long as the conserved momentum is bounded $|P| < \infty$, hyperfast growth is avoided.
In what follows, we perform a detailed description of the time evolution of maximal codimension-one slices for the various choices of stretched horizons introduced in section \ref{ssec:stretched_horizons}.

\subsubsection*{Case 1}

\begin{figure}[t!]
  	\centering
 \subfloat[]{ \label{subfig:extremal_surfaces_CV_case1} \includegraphics[width=0.4\textwidth]{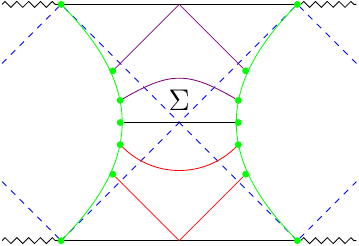} }
    \qquad
  \subfloat[]{ \label{subfig:Ueff_CV_case1} \includegraphics[width=0.5\textwidth]{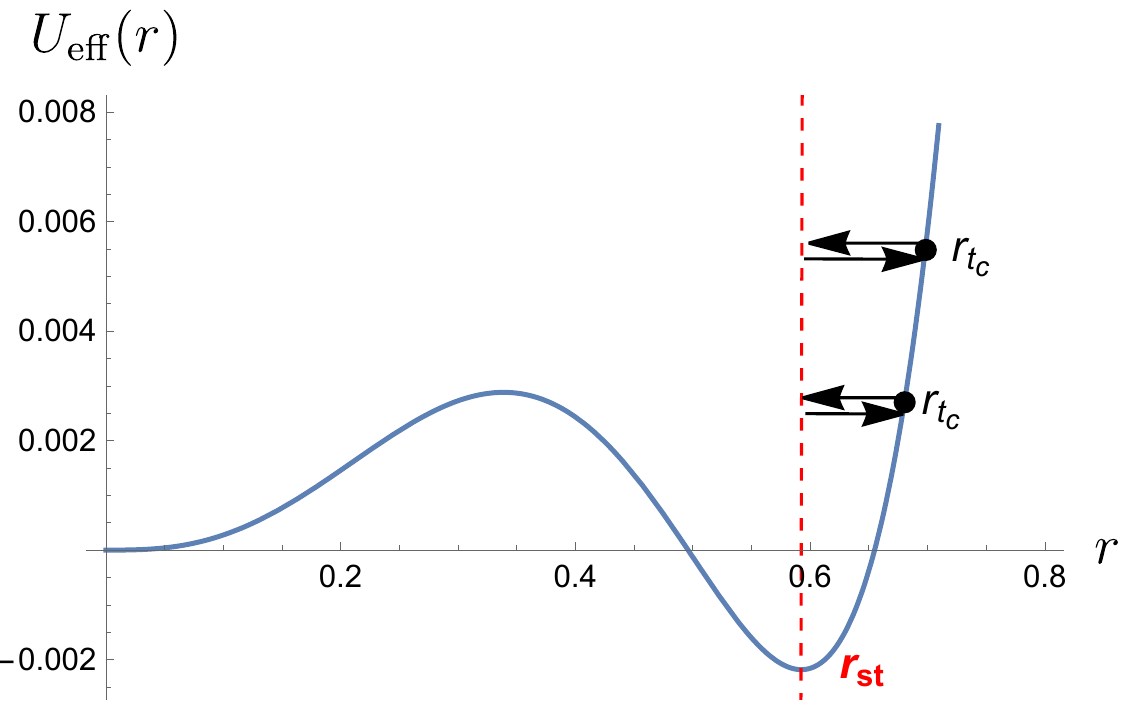} }
  	\caption{(a) The left and right stretched horizons (in green) are placed in consecutive static patches. In this prescription, the codimension-one extremal surface $\Sigma$, anchored to the stretched horizon (green dots), explores the region outside the cosmological horizon. The extremal surfaces are characterized by $P<0$ ($P>0$) in the upper (lower) half of the Penrose diagram, see the purple (red) curves. The black curve has $P=0$.
    (b) Qualitative behavior of the effective potential $U_{\rm eff}(r)$ in eq.~\eqref{eq:turning}. For every value of $P^2$, the surface connects the two stretched horizons after going through the turning point $r_{t_c}$.
    We have set $d=3$, $\mu= 0.187$, and $L=1$.
    }
  	\label{fig:Ueff_1}
  \end{figure}

\begin{figure}[t!]
    \centering
    \subfloat[]{ \label{subfig:CV_case1}
\includegraphics[width=0.45\textwidth]{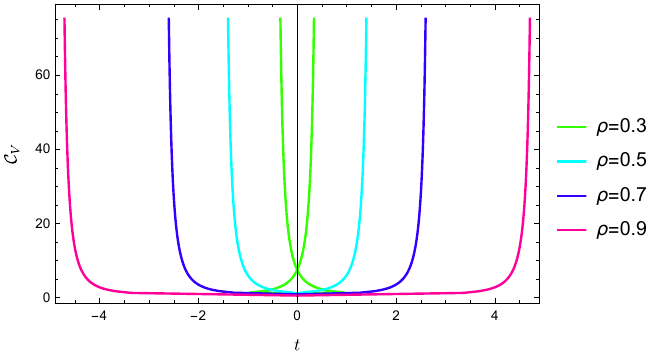}}
\hfill \subfloat[]{\label{subfigb:CVrate_case1} 
\includegraphics[width=0.48\textwidth]{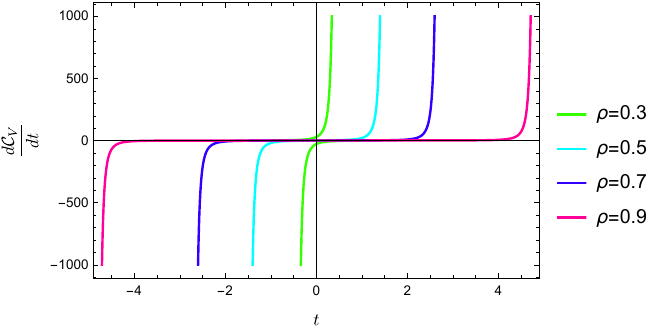}}
    \caption{Time dependence of (a) CV and (b) its rate for the prescription in Fig. \ref{fig:Ueff_1}. Different curves correspond to different values of the stretched horizon parameter $\rho$.
    We have set $d=3$, $\mu = 0.14$, and $G_N L=1$.
    When multiple extremal surfaces exist (see the green curve), it is understood that complexity corresponds to the maximal value of the volume at each fixed time $t$.  }
    \label{fig:volume_cosmo}
\end{figure}

We place the left and right stretched horizons $r_{\rm st} \equiv r_{\rm st}^L = r_{\rm st}^R$ into two consecutive static patches, see Fig.~\ref{subfig:extremal_surfaces_CV_case1}, and we attach the codimension-one surface to the left and right stretched horizon at times $-t_L$ and $t_R$, respectively. 
Due to the time-translation symmetry of the configuration, it is not restrictive to consider the symmetric case \eqref{eq:symmetric_times}.  
After leaving the right stretched horizon $r_{\rm st}$, the maximal surface moves in the direction of increasing $r$ and enters the inflating region. As it can be seen in Fig. \ref{subfig:extremal_surfaces_CV_case1}, here it passes through a turning point and proceeds in the direction of decreasing $r$ until it reaches the left stretched horizon $r_{\rm st}$. 
In summary, the maximal surface admits a single turning point $r_{t_c}$ satisfying $r_{t_c} \geq r_c$. 
Therefore, the extremal volume and the boundary time at the stretched horizon $r_{\rm st}$ are respectively given by
\begin{equation}
    \frac{V(P)}{\Omega_{d-1}} = \frac{2}{L^{d-1}} \int_{r_{\rm st}}^{r_{t_c}} \frac{r^{2(d-1)}}{\sqrt{P^2 + f(r) (r/L)^{2(d-1)}}} dr \, ,
    \label{eq:V_CV_1}
\end{equation}
\begin{equation}
    t(P) =   2 \int_{r_{\rm st}}^{r_{t_c}} \frac{P}{f(r) \sqrt{P^2 + f(r) (r/L)^{2(d-1)}}} dr \, ,
    \label{eq:t_CV_1}
\end{equation}
where the second expression is obtained by adding the left and right-hand sides of
\begin{align}
    - r^*(r_{t_c}) -\frac{t}{2} + r^*(r_{\rm st})  &= \int_{r_{\rm st}}^{r_{t_c}} \frac{\dot{u}_+}{\dot{r}_+} dr =\int_{r_{\rm st}}^{r_{t_c}} \frac{-P - \sqrt{P^2 + f(r) (r/L)^{2(d-1)}}}{f(r) \sqrt{P^2 + f(r) (r/L)^{2(d-1)}}} dr \, , \label{eq:equi1}\\
 -\frac{t}{2} + r^*(r_{\rm st}) - r^*(r_{t_c}) &= \int_{r_{t_c}}^{r_{\rm st}} \frac{\dot{v}_-}{\dot{r}_-} dr = \int_{r_{t_c}}^{r_{\rm st}} \frac{P + \sqrt{P^2 + f(r) (r/L)^{2(d-1)}}}{f(r) \sqrt{P^2 + f(r) (r/L)^{2(d-1)}}} dr \, , \label{eq:equi2}
\end{align}
in which we set the time at the turning point $t_{t_c}=0$ by symmetry.
In the previous equations, we exploited the fact that
\begin{equation}
    \int_a^b \frac{dr}{f(r)} = r^*(b) - r^*(a) \, .
\end{equation}
In Fig.~\ref{fig:volume_cosmo} we show numerical plots of volume complexity and its growth rate as functions of time for several values of the stretched horizon parameter $\rho$ for $d=3$. For different spacetime dimensions, we qualitatively get the same behaviours.  
In particular, the volume diverges at a finite critical time, which increases both with the spacetime dimension $d$ and with the value of $\rho$.
Restricting to $t>0$, for small values of $\rho$ there are two main regimes (see the green curve in Fig. \ref{fig:volume_cosmo}): at small times there are three extremal surfaces anchored at the same boundary time $t$, whereas at large times there are two. 
Such configurations are symmetric under time-reflection $t \rightarrow -t$.
According to the CV prescription, complexity is determined by the maximal volume among these possibilities. 
An analog case with multiple extremal surfaces happened in \cite{Chapman:2021eyy,Jorstad:2022mls,Auzzi:2023qbm}.
Following the same arguments presented in \cite{Jorstad:2022mls,Auzzi:2023qbm}, we deduce that, for fixed time $t$, the surface with larger volume is the one with larger $\left| P \right|$.
For bigger values of $\rho$, there is instead only one extremal surface, and no further maximization is required. 

The volume diverges when the turning point reaches $r_{t_c} = r_{\rm max} \rightarrow +\infty$. From eq.~\eqref{eq:turning}, we note that this happens for 
\begin{equation}
    P^2 \sim \left( \frac{r_{\rm max}}{L} \right)^{2d} \rightarrow +\infty \, . 
\end{equation}
Consequently, from eq. \eqref{eq:t_CV_1} the critical time (for $P \rightarrow -\infty$) can be expressed as
\begin{equation}
    t_{c1} = -2 \int_{r_{\rm st}}^{r_{\rm max}} \frac{dr}{f(r)} = 2 (r^*(r_{\rm st})-r^*(r_{\rm max})) \, ,
    \qquad 
    t_{\infty} = \lim_{r_{\rm max} \rightarrow \infty} t_{c1} \, ,
\end{equation}
in accordance with the result obtained for codimension-zero proposals in eq.~\eqref{eq:critical_time_tc1}.
Since the contribution from empty dS space in the blackening factor $f(r)\sim-(r/L)^2$ is the dominant term in the previous integrals, the critical time obtained in empty dS (discussed in section 5 of \cite{Jorstad:2022mls}) is valid for this case as well. Therefore, we conclude that in the SdS background we get the following dominant contribution around the critical time
\begin{equation}
    t - t_{\infty} \sim \frac{L^2}{r_{\rm max}} \, .
    \label{t_critical_case1}
\end{equation}

\paragraph{Growth rate.}
In order to analyze the growth rate of volume complexity, it is convenient to observe that
\begin{equation}
  \frac{(r/L)^{2(d-1)}}{\sqrt{P^2 + f(r) (r/L)^{2(d-1)}}} = 
  - \frac{P^2}{f(r) \sqrt{P^2 + f(r) (r/L)^{2(d-1)}}}
  +\frac{\sqrt{P^2 + f(r) (r/L)^{2(d-1)}}}{f(r)}  \, .
  \label{eq:rewriting_CV}
\end{equation}
We can thus re-express the volume as
\begin{equation}
    \frac{V}{\Omega_{d-1} L^{d-1}} = 
   - P \, t +
    2 \int_{r_{\rm st}}^{r_{t_c}} \frac{\sqrt{P^2 + f(r) (r/L)^{2(d-1)}}}{f(r)} dr  \, .
\end{equation}
Taking the derivative with respect to the boundary time $t$, we obtain
\beq
\begin{aligned}
    \frac{1}{\Omega_{d-1} L^{d-1}} \frac{dV}{dt} &= - P - \frac{dP}{dt} \left( t - 2 \int_{r_{\rm st}}^{r_{t_c}} \frac{P}{f(r) \sqrt{P^2 + f(r) (r/L)^{2(d-1)}}} dr \right) \\ 
    &+ 2 \frac{dr_{t_c}}{dt} \frac{\sqrt{P^2 + f(r_{t_c}) (r_{t_c}/L)^{2(d-1)}}}{f(r_{t_c})} \, .
\end{aligned}
\label{eq:trick_CV_case1}
\eeq
The second term vanishes because of eq. \eqref{eq:t_CV_1}, while the third term vanishes due to the definition of turning point in eq.~\eqref{eq:turning}. 
Therefore, we get 
\begin{equation}
   \frac{dV}{dt} =-\Omega_{d-1} L^{d-1} P \, .
   \label{eq:V_rate}
\end{equation}
In particular, at the critical time, we find
\begin{equation}
    \lim_{t \to t_{\infty}} \frac{dV}{dt} = 
    \Omega_{d-1} \frac{r_{\rm max}^d}{L} 
    \sim \frac{1}{(t_{\infty} -t)^d} \, ,
    \label{eq:ratelate_CV_case1}
\end{equation}
where eq.~\eqref{t_critical_case1} has been used.
This result describes the hyperfast growth of CV in asymptotically dS geometries.

\subsubsection*{Case 2}
\begin{figure}[t!]
  	\centering
  	\subfloat[]{\label{subfig:Penrose_CV_case2}  \includegraphics[width=0.4\textwidth]{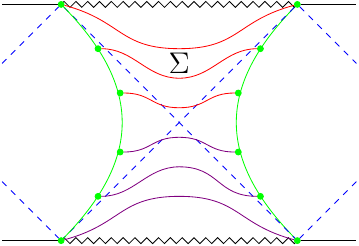}}
    \qquad
 \subfloat[]{ \label{subfig:potential_CV_case2} \includegraphics[width=0.5\textwidth]{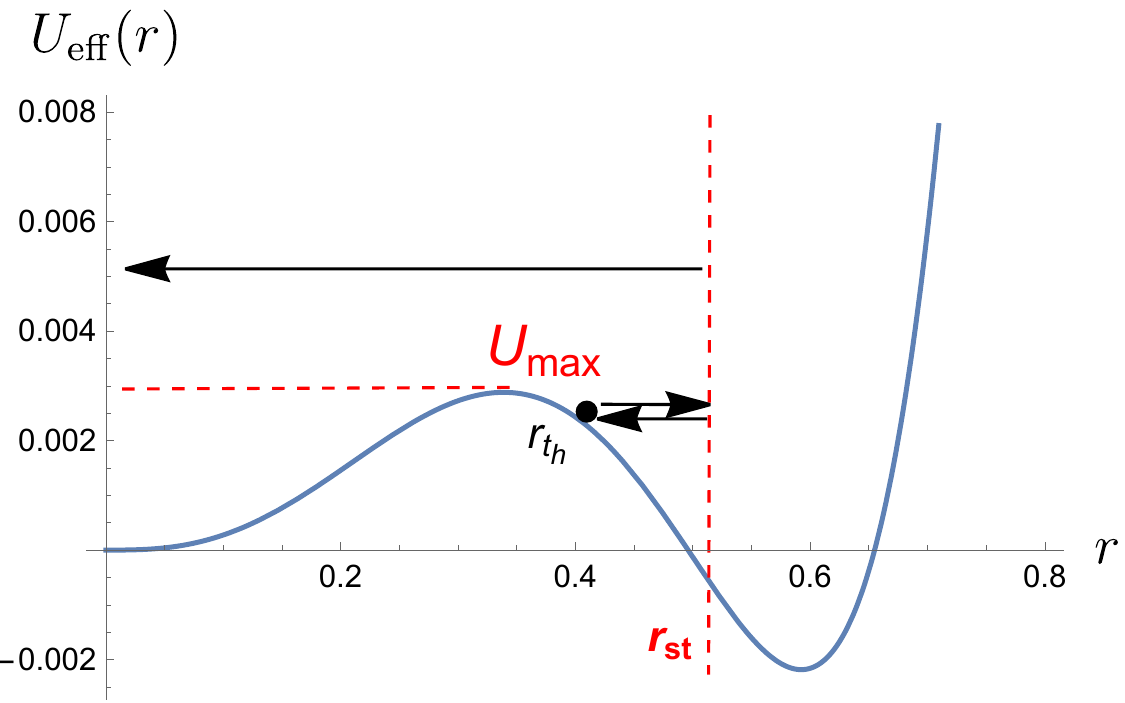}}
  	\caption{(a) The left and right stretched horizons are placed in consecutive static patches. In this prescription, the codimension-one extremal surface $\Sigma$ explores the region behind the black hole horizon.
    The extremal surfaces have $P>0$ ($P<0$) in the upper (lower) half of the Penrose diagram, see the red (purple) curves.
    (b) Qualitative behavior of the effective potential $U_{\rm eff}(r)$. For large $P^2$, the surface falls into the black hole singularity. For small enough $P^2$, the surface connects the two stretched horizons after going through the turning point $r_{t_h}$.
    We have set $d=3$, $\mu= 0.187$, and $L=1$.
    }
  	\label{fig:Ueff_2}
  \end{figure}

\begin{figure}[t!]
    \centering
    \subfloat[]{ \label{subfig:CV_case2}
\includegraphics[width=0.45\textwidth]{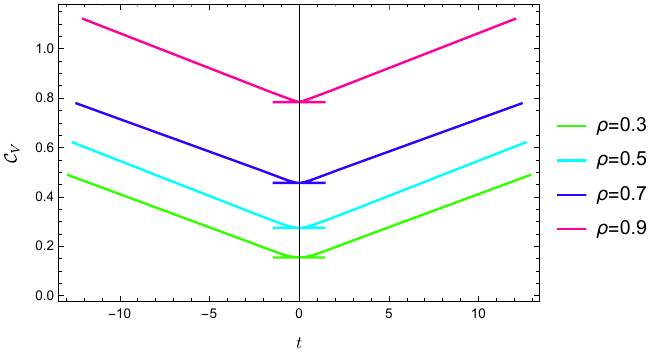}}
\hfill \subfloat[]{\label{subfigb:CVrate_case2} 
\includegraphics[width=0.48\textwidth]{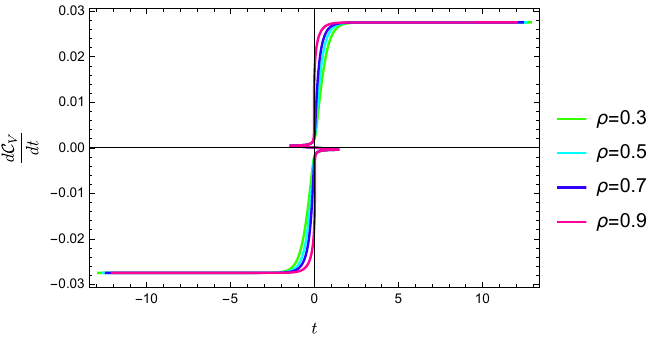}}
    \caption{Time dependence of (a) CV and (b) its rate for the prescription in Fig. \ref{fig:Ueff_2}. Different curves correspond to different values of the stretched horizon parameter.
    We have set $d=3$, $\mu = 0.14$, and $G_N L=1$. For any fixed time $t$, holographic complexity corresponds to the maximum of each coloured curve.  }
    \label{fig:volume_BH}
\end{figure}

One can proceed in a similar way with the configuration depicted in Fig.~\ref{subfig:Penrose_CV_case2}, where we anchor the maximal volume surfaces to a pair of black hole stretched horizons $r_{\rm st} \equiv r_{\rm st}^L = r_{\rm st}^R$. Without loss of generality, we consider the symmetric choice \eqref{eq:symmetric_times} for the boundary times along the stretched horizons.
After leaving the right stretched horizon $r_{\rm st}$, the maximal surface extends in the direction of decreasing $r$ and enters the black hole. 
Contrary to case 1, the value of $|P|$ cannot be arbitrarily large. Let us define $r=r_f$ such that
\begin{equation}
    U'_{\rm eff}(r_f) =0 \, , \qquad U''_{\rm eff}(r_f) \leq 0 \, .
    \label{eq:def_rf_case2CV}
\end{equation}
Namely, $r=r_f$ is the local maximum of the effective potential: $U_{\rm max} \equiv U_{\rm eff} (r_f)$.
As it can be understood from Fig.~\ref{subfig:potential_CV_case2}
, for $P^2 > U_{\rm max}$ the codimension-one slice falls into the black hole singularity $r=0$ before connecting the left and right stretched horizons. 
Instead, for $0 \leq P^2 \leq U_{\rm max}$, the maximal surface meets the potential barrier and passes through a turning point. Then, it moves in the direction of increasing $r$ and reaches the left stretched horizon.
Therefore, $0 \leq P^2 \leq U_{\rm max}$ is the range of the conserved momentum $P^2$ that properly defines an extremal surface according to CV conjecture (see also footnote \ref{footnote:disconnected_solutions}).
In this case, the turning point is located at $r_{t_h} \leq r_h$. Thus, the extremal volume and the boundary time at the stretched horizons $r_{\rm st}$ read
\begin{equation}\label{eq:Vol BB}
    \frac{V(P)}{\Omega_{d-1}} = \frac{2}{L^{d-1}} \int^{r_{\rm st}}_{r_{t_h}} \frac{r^{2(d-1)}}{\sqrt{P^2 + f(r) (r/L)^{2(d-1)}}} dr \, ,
\end{equation}
\begin{equation}
    t(P) =  -2 \int_{r_{t_h}}^{r_{\rm st}} \frac{P}{f(r) \sqrt{P^2 + f(r) (r/L)^{2(d-1)}}} dr \, ,
    \label{eq:t_CV_2}
\end{equation}
respectively.\footnote{\label{fnt:Sign of P}Notice the opposite sign in (\ref{eq:t_CV_2}) with respect to (\ref{eq:t_CV_1}). The reason for this difference is that $P$, as defined in (\ref{eq:conserved_momentum_CV}), takes values $P\leq 0$ in the future region of the inflating patch, while $P\geq 0$ in the black hole patch, as the Killing vector corresponding to time translations adopts opposite orientations in these two cases. This determines the sign of the late time rate of growth of the CV proposal, as previously pointed out in \cite{Chapman:2018lsv}.}
The second relation comes from expressing
\begin{align}
    r^*(r_{t_h}) -\frac{t}{2} - r^*(r_{\rm st})  &= \int_{r_{\rm st}}^{r_{t_h}} \frac{\dot{v}_-}{\dot{r}_-} dr =\int_{r_{\rm st}}^{r_{t_h}} \frac{P + \sqrt{P^2 + f(r) (r/L)^{2(d-1)}}}{f(r) \sqrt{P^2 + f(r) (r/L)^{2(d-1)}}} dr \, , \label{eq:integrations_CV_case2_1}\\
    -\frac{t}{2} - r^*(r_{\rm st}) + r^*(r_{t_h}) &= \int_{r_{t_h}}^{r_{\rm st}} \frac{\dot{u}_+}{\dot{r}_+} dr = \int_{r_{t_h}}^{r_{\rm st}} \frac{-P - \sqrt{P^2 + f(r) (r/L)^{2(d-1)}}}{f(r) \sqrt{P^2 + f(r) (r/L)^{2(d-1)}}} dr \, ,
    \label{eq:integrations_CV_case2}
\end{align}
where we set the time at the turning point $t_{t_h}=0$ by symmetry.
In Fig.~\ref{fig:volume_BH} we display the volume and its growth rate as functions of the boundary time $t$ for different values of the stretched horizon parameter $\rho$. For any value of $\rho$, at small times, a window exists in which for fixed $t$ there are three extremal surfaces. Again, the maximal surface is the one with larger $\left| P \right|$.
For larger times, only one extremal surface exists.

With the growth of the anchoring time $t$, the turning point approaches the black hole singularity, and the value of the conserved momentum $P$ increases. 
In the late-time limit $t \rightarrow +\infty$, the value of the conserved momentum is $P^2 = P^2_{\rm max} \equiv U_{\rm max}$, as it can be consistently checked from eq.~\eqref{eq:t_CV_2}. Namely, at $P = P_{\rm max}$ the turning point is $r_{t_h} = r_f$ and
\begin{equation}
    U_{\rm eff}(r) \sim U_{\rm eff}(r_f) + \frac{1}{2} U''_{\rm eff} (r-r_f)^2 + \dots \, ,
\end{equation}
since $U'_{\rm eff}(r_f)=0$. This condition is crucial for the stretched horizon time to diverge. Indeed,
\begin{equation}\label{eq:t_Pmax_case2}
    t(P_{\rm max}) \sim 2 \sqrt{2} \int^{r_f} \frac{P_{\rm max}}{f(r_f) \sqrt{-U''_{\rm eff}(r_f)} (r-r_f)} \, dr \sim \frac{2 \sqrt{2} \, P_{\rm max}}{f(r_f) \sqrt{-U''_{\rm eff}(r_f)}} \, \log|r-r_f| \, ,
\end{equation}
around the turning point.
At late times, the extremal surface hugs the final slice located at constant $r=r_f$.

\paragraph{Growth rate.}
In contrast to case 1, the extremal volume slices do not explore the inflating region. Instead, the same type of late-time growth found for asymptotically AdS black holes trivially follows. Indeed, using the same trick \eqref{eq:trick_CV_case1} as in case 1, we write the volume as
\begin{equation}
    \frac{V}{\Omega_{d-1} L^{d-1}} = 
    P \, t +
    2 \int_{r_{t_h}}^{r_{\rm st}} \frac{\sqrt{P^2 + f(r) (r/L)^{2(d-1)}}}{f(r)} dr  \, .
\end{equation}
Similarly to eq.~\eqref{eq:V_rate}, we find
\begin{equation}\label{eq:dV dt all t}
    \frac{dV}{dt} = \Omega_{d-1} L^{d-1} P \, .
\end{equation}
Therefore, at late times we get
\begin{equation}
    \lim_{t \to +\infty} \frac{dV}{dt} = \Omega_{d-1} L^{d-1} \sqrt{U_{\rm max}} =
    \Omega_{d-1} \sqrt{- f(r_f)} \, r_f^{d-1} \, ,
    \label{eq:ratelate_CV_case2}
\end{equation}
where $r_f$ is defined in eq.~\eqref{eq:def_rf_case2CV}.
We notice that the final result corresponds to the volume measure evaluated on the final slice at constant $r=r_f$. 
This is the characteristic linear growth obtained in asymptotically AdS geometries \cite{Carmi:2017jqz}.

\subsubsection*{Case 3}

 \begin{figure}[t!]
  	\centering
  	\subfloat[]{ \label{subfig:Penrose_CV_case3} \includegraphics[width=0.5\textwidth]{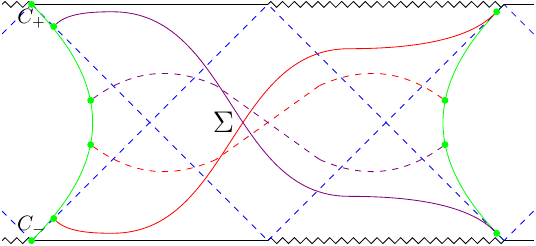}}
    \qquad
   \subfloat[]{ \label{subfig:potential_CV_case3} \includegraphics[width=0.4\textwidth]{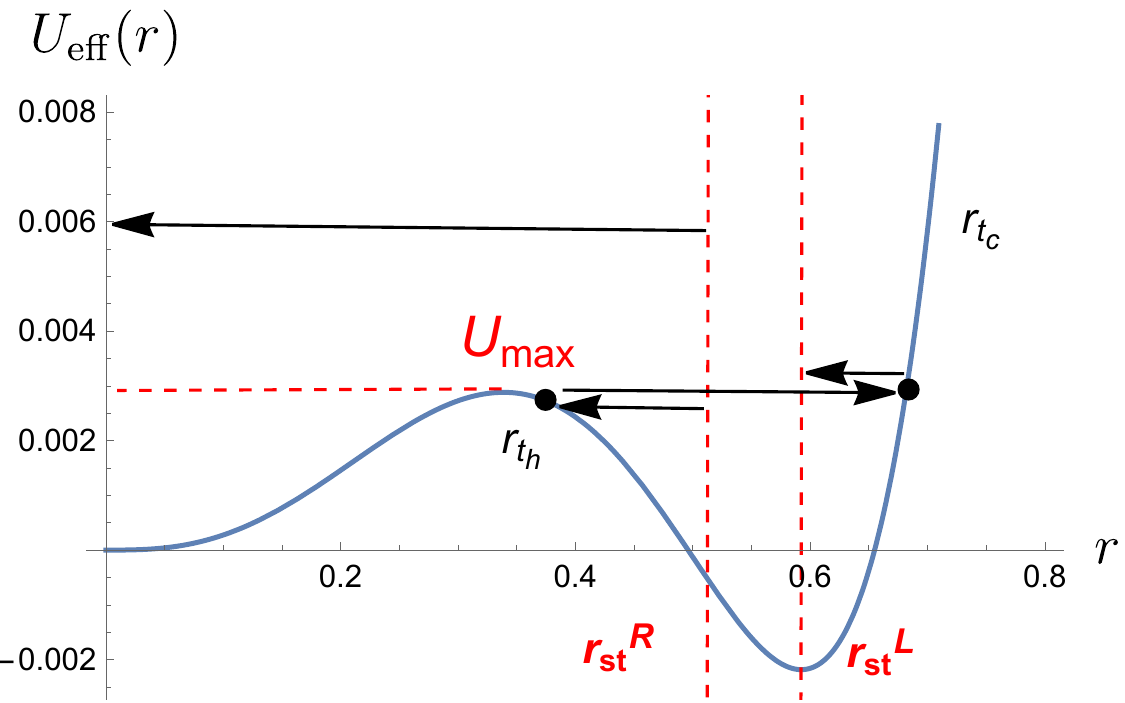}}
  	\caption{(a) The left and right stretched horizons are placed in non-consecutive static patches. According to the prescription \ref{rule}, the maximal volume surfaces $\Sigma$ can evolve to arbitrary late boundary time $t_R=-t_L$. The red (purple) extremal surfaces have $P>0$ ($P<0$). The dashed curves correspond to finite boundary time, while the solid curves are achieved at late (early) times.
    (b) Qualitative behavior of the effective potential $U_{\rm eff}(r)$. For large $P^2$, the surface directly falls into the black hole singularity. For small enough $P^2$, the surface connects the two stretched horizons after going through two turning points $r_{t_c}$ and $r_{t_h}$.
    We have set $d=3$, $\mu= 0.187$, and $L=1$.}
  	\label{fig:Ueff_3}
  \end{figure}

We place the left and right stretched horizons at $r=r_{\rm st}^L$ and $r=r_{\rm st}^R$ into non-consecutive static patches, as in Fig.~\ref{subfig:Penrose_CV_case3}.
Starting from the right stretched horizon $r_{\rm st}^R$, the maximal surface moves towards decreasing $r$ until it enters the black hole. For the surface to reach the left stretched horizon $r_{\rm st}^L$, the value of the conserved momentum $P$ must be properly chosen, see Fig.~\ref{subfig:potential_CV_case3}.
For $P^2 > U_{\rm max}$, the maximal surface does not encounter a potential barrier inside the black hole, so it falls into the singularity.\footnote{Let us remind that $U_{\rm max}$ is the value of the effective potential $U_{\rm eff} (r)$ at the location of its local maximum. } Instead, for $0 \leq P^2 \leq U_{\rm max}$, the surface passes through a turning point $r_{t_h}$ and turns to the direction of increasing $r$.
Next, it explores the inflating region, where it meets another turning point $r_{t_c}$.
Finally, it attaches to the left stretched horizon $r_{\rm st}^L$. 
In summary, for $0 \leq P^2 \leq U_{\rm max}$ there are two turning points, one inside the inflating region and one inside the black hole. 
In this case, the volume reads\footnote{One can straightforwardly generalize the arguments in case of $n$ cosmological and black hole patches upon the replacement $\int_{r_{t_h}}^{r_{t_c}}\rightarrow n\int_{r_{t_h}}^{r_{t_c}}$. However, for clarity, we will limit the discussion to $n=1$ in case 3, and to $n=2$ in cases 4-5. \label{fnt:case3 CV}}
\begin{align}\label{eq:Vol BC}
    \frac{V(P)}{\Omega_{d-1}} \, =& \frac{1}{L^{d-1} } \qty(\int_{r_{t_h}}^{r_{\rm st}^R} +
    \int_{r_{t_h}}^{r_{t_c}}+  \int_{r_{\rm st}^L}^{r_{t_c}})\frac{r^{2(d-1)}}{\sqrt{P^2 + f(r) (r/L)^{2(d-1)}}} dr \, .
\end{align}
To determine the boundary time as a function of the conserved momentum $P$, it is convenient to introduce the function
\begin{equation}
    \tau_{\pm}(r,P) \equiv \frac{-P \pm \sqrt{P^2 + f(r) (r/L)^{2(d-1)}}}{f(r) \sqrt{P^2 + f(r) (r/L)^{2(d-1)}}} \, .
    \label{eq:def_tau}
\end{equation}
For a fixed $P$, the maximal surface explores the future (past) black hole interior and the past (future) inflating region.
We define the anchoring time at the left and right stretched horizon as $-t_L$ and $-t_R$ respectively, according to the convention \ref{conv_times} and the orientation of the Killing vector $\partial_t$ depicted in Fig.~\ref{fig:WDW_BH_stretched}. 
In this way, we have
\begin{subequations}
\label{eq:identities_CV_case3}
\begin{equation}\label{eq:time contribution4}
    t_{t_h} -r^*(r_{t_h}) +t_R + r^*(r_{\rm st}^R) =
   \int^{r_{t_h}}_{r_{\rm st}^R} \frac{\dot{u}_-}{\dot{r}_-} dr =
   -\int^{r_{t_h}}_{r_{\rm st}^R} \tau_+(r,P) dr \, ,
\end{equation}
\begin{equation}
    t_{t_c} +r^*(r_{t_c}) - t_{t_h} - r^*(r_{t_h}) =
   \int^{r_{t_c}}_{r_{t_h}} \frac{\dot{v}_+}{\dot{r}_+} dr =
   \int^{r_{t_c}}_{r_{t_h}} \tau_+(r,P) dr \, ,
   \label{eq:identity2_CV_case3}
\end{equation}
\begin{equation}\label{eq:time contribution1}
   -t_L- r^*(r_{\rm st}^L) - t_{t_c} + r^*(r_{t_c}) =
   \int^{r_{\rm st}^L}_{r_{t_c}} \frac{\dot{u}_-}{\dot{r}_-} dr =
   -\int^{r_{\rm st}^L}_{r_{t_c}} \tau_+(r,P) dr \, ,
\end{equation}
\end{subequations}
where $t_{t_c} (t_{t_h})$ is the value of the time coordinate at the turning point $r_{t_c} (r_{t_h})$.
Notice that from eq.~\eqref{eq:def_tau} we get
\begin{equation}
    \int_a^b \tau_{\pm}(r,P) \, dr = \int_a^b \frac{-P}{f(r) \sqrt{P^2 + f(r) (r/L)^{2(d-1)}}} \, dr \pm r^*(b) \mp r^*(a) \, .
    \label{eq:tau_integral}
\end{equation}
Therefore, by summing the identities~\eqref{eq:identities_CV_case3}, after some algebra we obtain
\begin{align}
\label{eq:time L and R BC}
    t_R-t_L =& -\qty(\int_{r_{t_h}}^{r_{\rm st}^R} + \int_{r_{t_h}}^{r_{t_c}} +\int_{r_{\rm st}^L}^{r_{t_c}})\frac{P}{f(r) \sqrt{P^2 + f(r) (r/L)^{2(d-1)}}} \, dr \, .
\end{align}
We now have two main choices:
\begin{itemize}
    \item $t_R = t_L$, time grows in a symmetric way on the two stretched horizons. 
    In agreement with the time translation symmetry \eqref{eq:time_shift_symmetry}, the volume is time-independent.
    Indeed, all the surfaces with $t_R = t_L$ can be obtained from the $P=0$ solution at $t_R = t_L=0$ by applying the time translation \eqref{eq:time_shift_symmetry}.
    \item $t_R \neq t_L$, the anchoring times are not symmetric.
    The volume is time-dependent. 
    In particular, as argued around eq.~\eqref{eq:symmetric_times}, it is not restrictive to choose the antisymmetric configuration $-t_L=t_R=t/2$.
\end{itemize}

\begin{figure}[t!]
    \centering
    \includegraphics[width=0.6\textwidth]{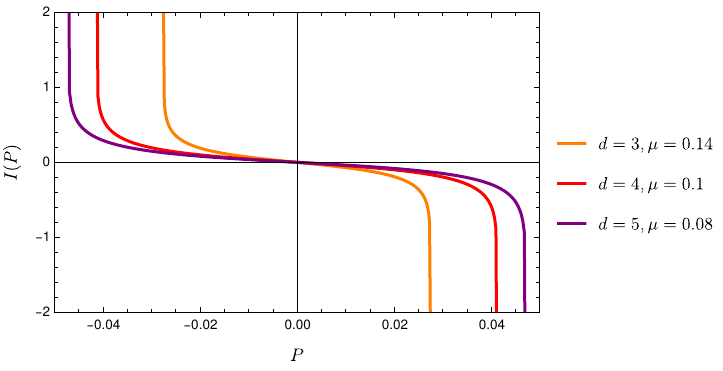}
    \caption{Plot of the integral $I$ in eq.~\eqref{eq:integral_deltat} as a function of the conserved momentum $|P| \leq \sqrt{U_{\rm max}}$, for various choices of the dimension $d$ and the mass parameter $\mu$. We fix $L=1$. }
    \label{fig:plot_deltat_case3}
\end{figure}

Before proceeding, we discuss the consistency of the identities \eqref{eq:identities_CV_case3}.
When $t_L=t_R$ and the stretched horizons are located at the same radial coordinate $r_{\rm st}^L = r_{\rm st}^R$, the extremal surfaces in the Penrose diagram are symmetric, therefore the turning points are located at $t_{t_h}=t_{t_c}=0$.
By plugging these conditions and the identity \eqref{eq:tau_integral} inside eq.~\eqref{eq:identity2_CV_case3}, we find the following constraint
\beq
I(P) = 0 \, , \qquad
I(P) \equiv \int_{r_{t_h}}^{r_{t_c}} \frac{P}{f(r) \sqrt{P^2 + f(r) (r/L)^{2(d-1)}}} dr \, .
\label{eq:integral_deltat}
\eeq
A priori, for any allowed value of $P$, one has to compute the turning points $r_{t_c}, r_{t_h}$, and then check whether the constraint \eqref{eq:integral_deltat} is satisfied.
This step is performed numerically, for various choices of the spacetime dimensions and of the mass parameter, in Fig.~\ref{fig:plot_deltat_case3}. 
It is clear that the constraint only holds when $P=0$.
This result is consistent with the first bullet below eq.~\eqref{eq:time L and R BC}.
In other words, the volume is time-independent, and this case corresponds to the trivial evolution performed with the Killing vectors.

On the contrary, when $t_R=-t_L=t/2$, the extremal surface is \textit{not} symmetric in the Penrose diagram, therefore we are allowed to have $t_{t_c} \ne 0$ and $t_{t_h} \ne 0$.
In this way, there are three unknown variables $t_{t_c}, t_{t_h}, t$ that can be determined by using the three identities \eqref{eq:identities_CV_case3}.
The conserved momentum $P$ is not constrained except for the condition $|P| \leq \sqrt{U_{\rm max}}$, which defines the existence of two turning points.

\begin{figure}[t!]
    \centering
    \includegraphics[width=0.48\textwidth]{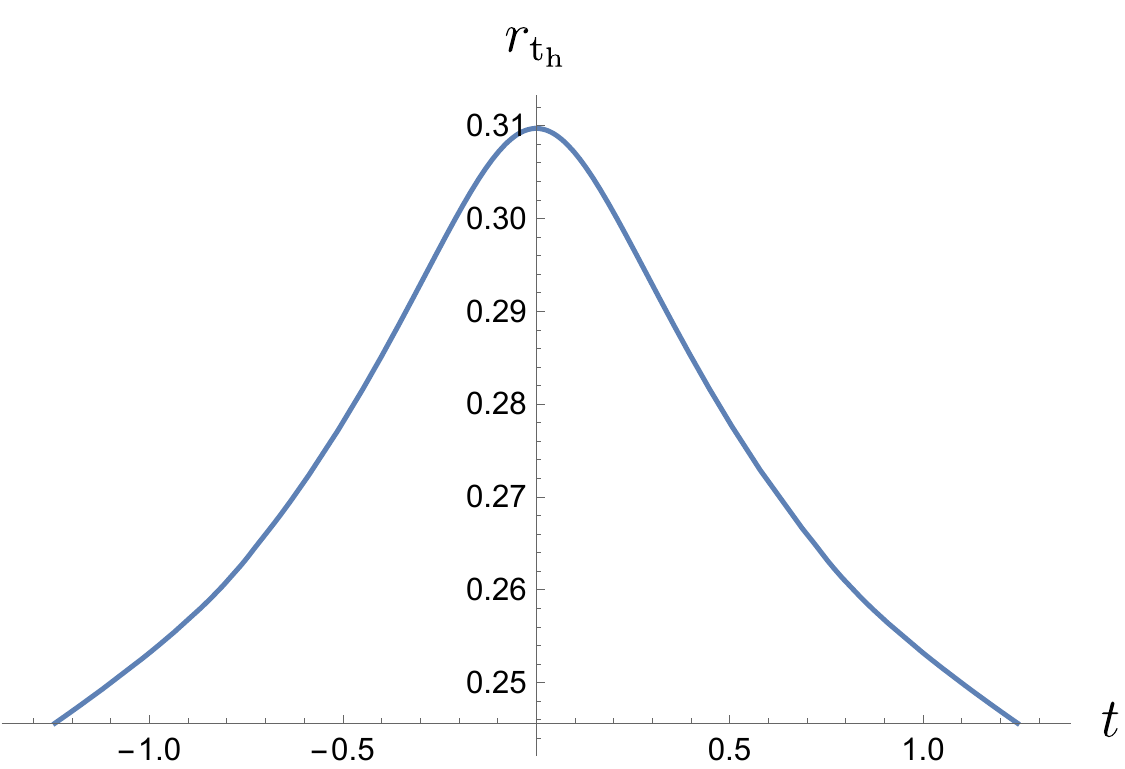}\hfill\includegraphics[width=0.48\textwidth]{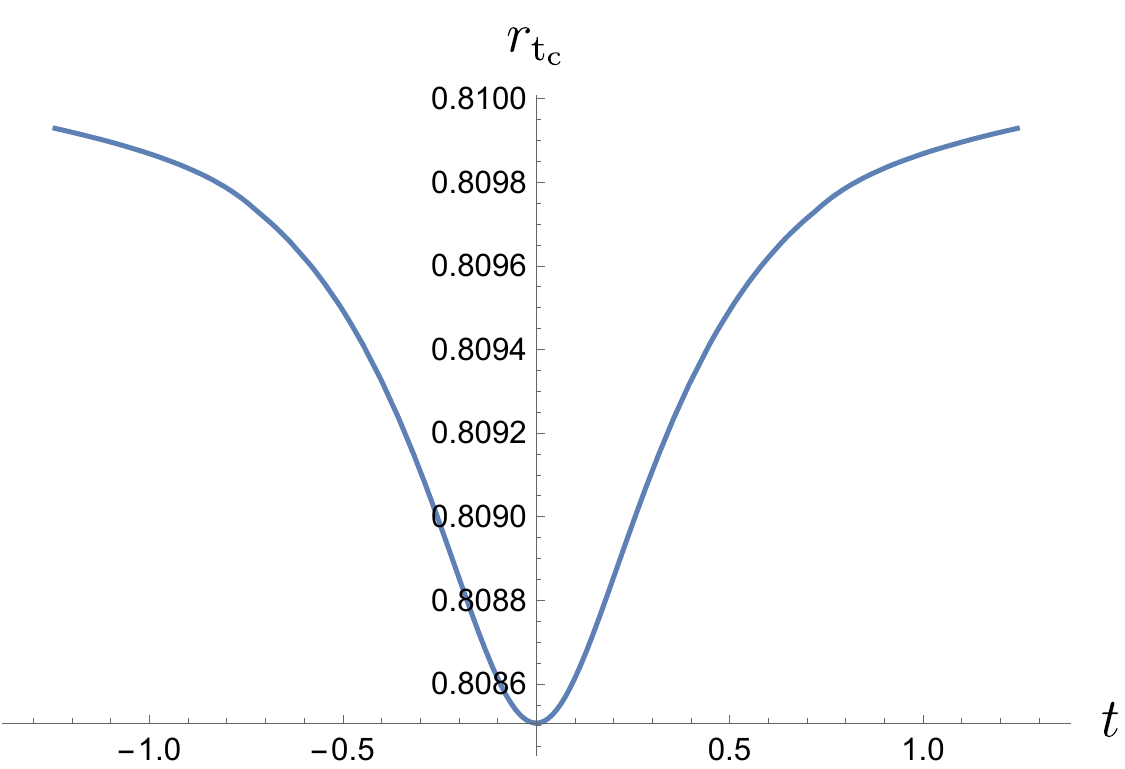}\\
    \includegraphics[width=0.48\textwidth]{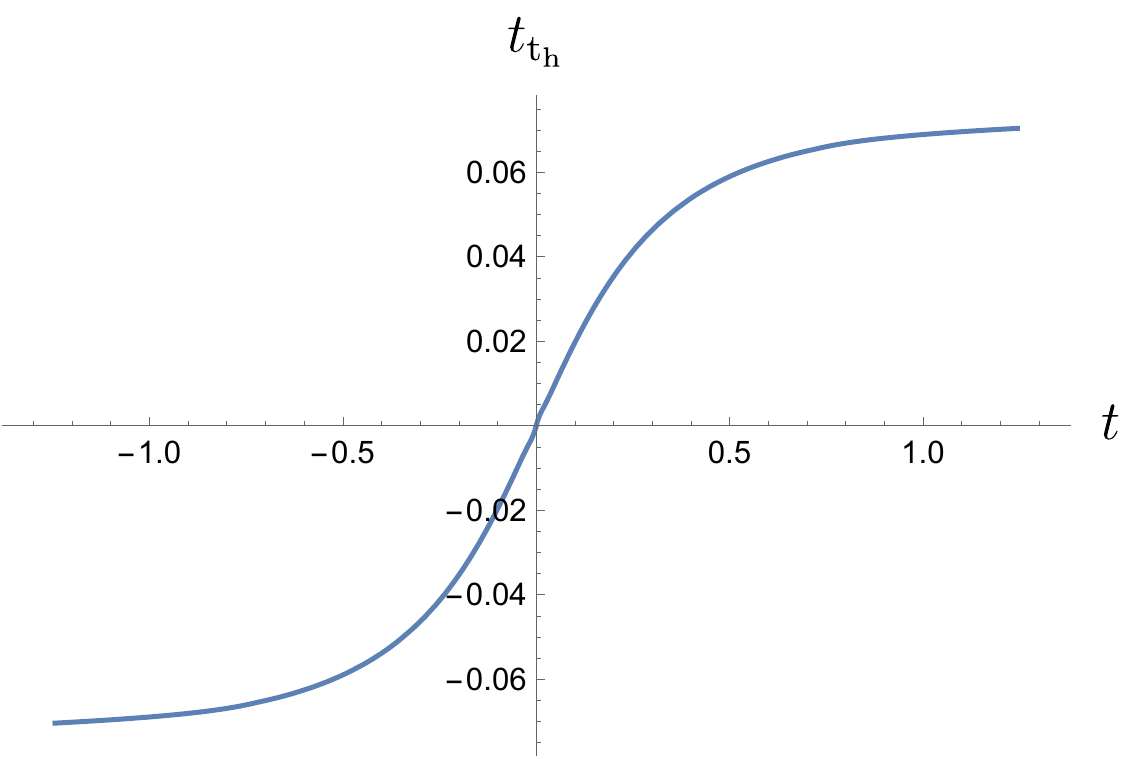}\hfill\includegraphics[width=0.48\textwidth]{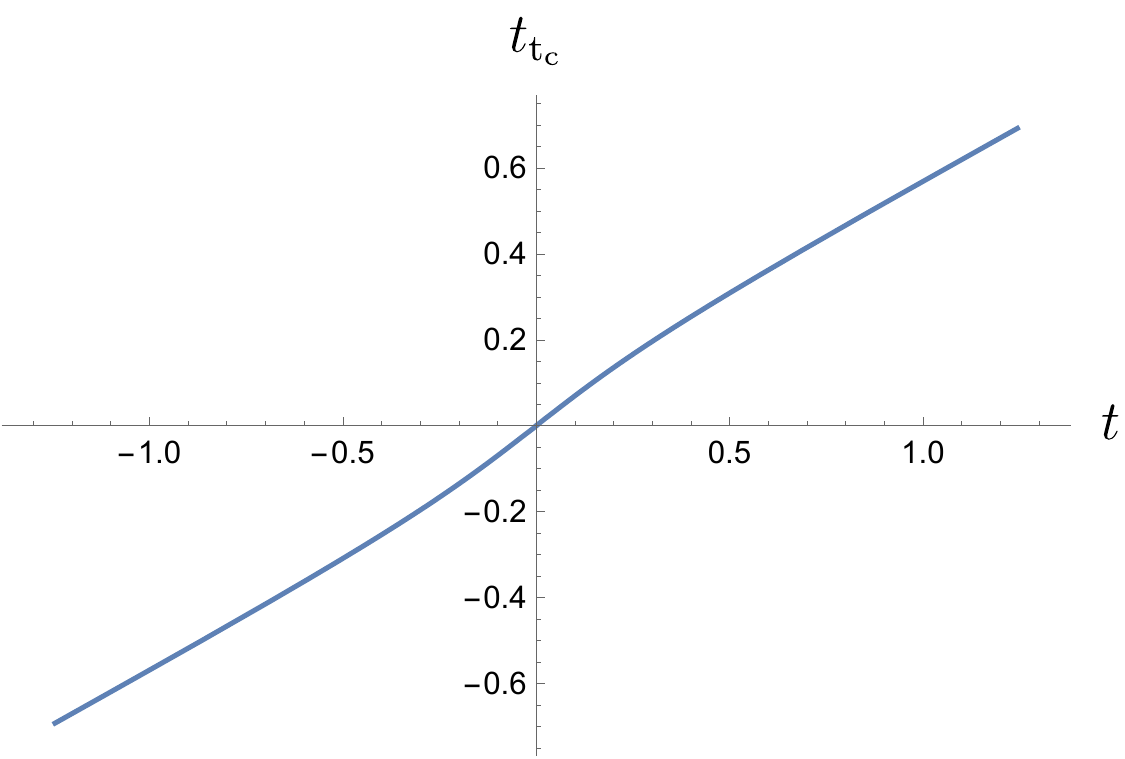}
    \caption{Plots of the radial and time coordinates of the turning point in the black hole and cosmological patches as functions of the boundary time $t$ in case 3. We set $d=3, \mu=0.14$ and $L=1$.}
    \label{fig:turning point in time}
\end{figure}

Let us focus on the time-dependent (antisymmetric) choice.
Similar manipulations as in cases 1-2, applied to eqs.~\eqref{eq:Vol BC} and \eqref{eq:time L and R BC}, lead to
\beq
     \frac{dV}{dt} = \Omega_{d-1} L^{d-1} P~.
     \label{eq:CVrate_case3}
\eeq
We can understand the evolution of the extremal surfaces, schematically depicted in Fig.~\ref{subfig:Penrose_CV_case3}, by studying the positions $(r_{t_c}, r_{t_h})$ and the time coordinates $(t_{t_c}, t_{t_h})$ of the turning points as functions of the boundary time. These plots are shown in Fig.~\ref{fig:turning point in time}.
At times $t>0$, we find that the turning point $r_{t_c}$ moves in the past exterior of the cosmological horizon towards the bottom-left corner $C^-$ of the Penrose diagram where the horizons, the singularity, and past infinity $\mathcal{I}^-$ meet. This corner is characterized by the condition $t_{t_c}=\infty$, while the corresponding radial coordinate is ill-defined.
At late times in the evolution, the turning point formally reaches the corner $C^-$ with finite value $\bar{r}_{t_c} > r_c$ of the radial coordinate, such that $U_{\rm eff}(\bar{r}_{t_c})=U_{\rm max}$.
On the other hand, the turning point $r_{t_h}$ moves towards the past singularity inside the interior of the black hole, until it asymptotically approaches a finite positive time $0<t_{t_h} < \infty$ and a final slice at constant radial coordinate $r_{f_h} < r_h$ such that $U_{\rm eff}'(r_{f_h})=0$.

Starting from eq.~\eqref{eq:time contribution4} and applying manipulations similar to eq.~\eqref{eq:t_Pmax_case2}, we find that the condition $U'(r_{f_h})=0$ implies $t_R (\sqrt{U_{\rm max}}) \rightarrow \infty$.
By imposing the antisymmetry $t_L=-t_R$ in the time evolution, we find that eqs.~\eqref{eq:identity2_CV_case3} and \eqref{eq:time contribution1} are solved when $t_{t_c} \rightarrow \infty$. 
These conditions define the late-time regime of the complexity evolution.
It is difficult to numerically plot the exact shape of the extremal surface at late times. The previous observations suggest that the extremal surface in the limit $t \rightarrow \infty$ becomes disconnected, composed of two parts: the single point at the corner $C^-$, plus a surface at constant radius $r_{f_h}$ in the black hole region.
A priori, it may be possible that the extremal surface remains connected in the strict limit $t \rightarrow \infty$.
However, one can at least rule out that the extremal surface in the inflating patch becomes null and hugs the cosmological horizon, since this would lead to an inconsistent limit of the spacelike surfaces determined by the equations of motion \eqref{eq:classical_motion}. We reserve the precise study of the extremal surface in the late time limit for future studies.  

\begin{figure}[t!]
    \centering
    \subfloat[]{ \label{subfig:CV_case3}
\includegraphics[width=0.48\textwidth]{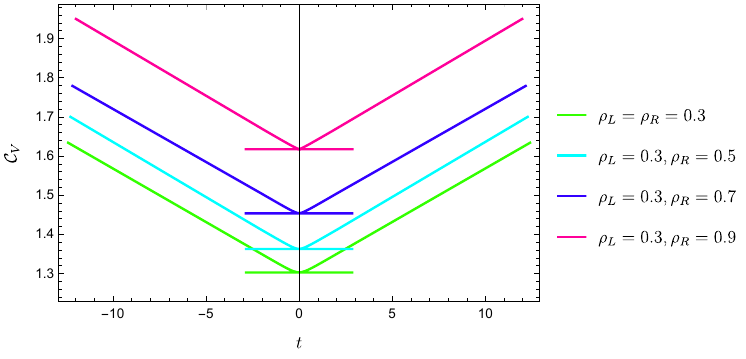}}
\hfill \subfloat[]{\label{subfigb:CVrate_case3} 
\includegraphics[width=0.48\textwidth]{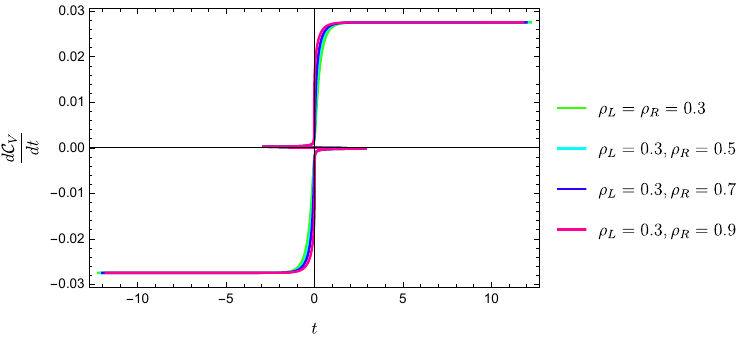}}
    \caption{Time dependence of (a) CV and (b) its rate for the prescription in Fig. \ref{fig:Ueff_3}. Different curves correspond to different values of the left and right stretched horizon parameters.
    It is understood that the complexity function is obtained by taking the maximum of each coloured curve for any fixed time $t$.
    We have set $d=3$, $\mu = 0.14$, and $G_N L=1$. }
    \label{fig:volume_case3}
\end{figure}

Nonetheless, we find the rate of growth of the volume by plugging the value of $P=\sqrt{U_{\rm max}}$ inside eq.~\eqref{eq:CVrate_case3}. The result reads
\beq
\lim_{t\rightarrow\infty} \frac{dV}{dt}= \Omega_{d-1}\sqrt{-f(r_f)}r_f^{d-1}\, .
\label{CV_latetimes_case3}
\eeq
We conclude that CV conjecture approaches a linear increase at late times in case 3, characterized by the volume measure evaluated on the final slice inside the black hole. 
The full time dependence of CV and its rate are plotted in Fig.~\ref{fig:volume_case3}.

\subsubsection*{Case 4}

\begin{figure}[t!]
  	\centering
  	\subfloat[]{ \label{subfig:extremal_surface_CV_case4} \includegraphics[width=0.5\textwidth]{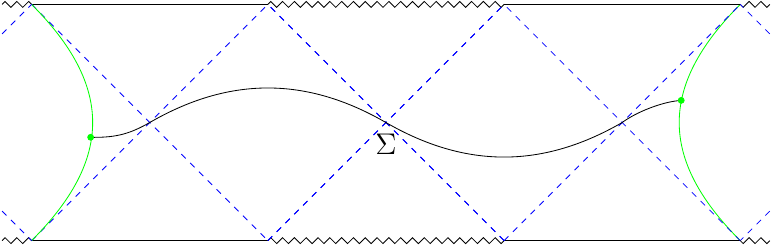}}
    \qquad
  \subfloat[]{ \label{potential_CV_case4} \includegraphics[width=0.4\textwidth]{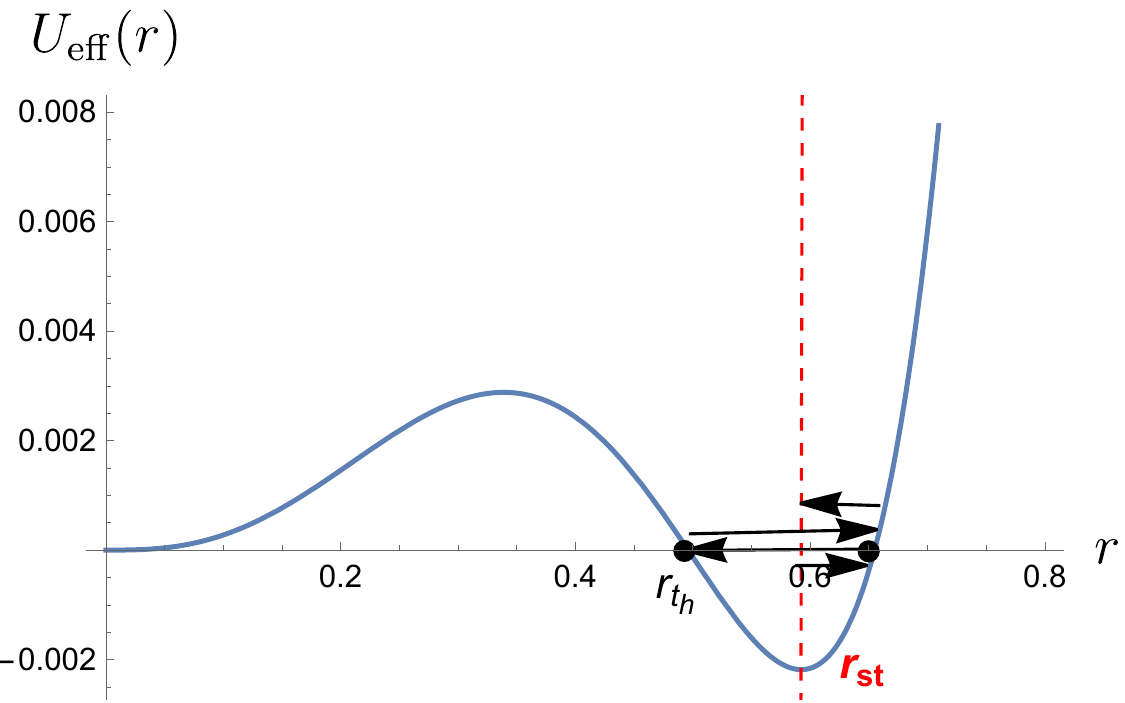}}
  	\caption{(a) The left and right stretched horizons are placed in non-consecutive static patches.
   Given the restriction \eqref{eq:integral_deltat}, only the conserved momentum $P=0$ is allowed for the surface $\Sigma$ depicted in the diagram.
   (b) Qualitative behavior of the effective potential $U_{\rm eff}(r)$. The surface $\Sigma$ connects the two stretched horizons after going through three turning points: $r_{t_c}$, $r_{t_h}$, and $r_{t_c}$ again. 
   In this case, the turning points coincide with the location of the cosmological and black hole horizons, respectively.
    We have set $d=3$, $\mu= 0.187$, and $L=1$.    }
  	\label{fig:Ueff_4}
  \end{figure}

The configuration is illustrated in Fig.~\ref{subfig:extremal_surface_CV_case4}.
We locate the left and right stretched horizons at $r_{\rm st}^L = r_{\rm st}^R \equiv r_{\rm st}$ and, following the orientation of the Killing vector in Fig.~\ref{fig:WDW_case4}, we define the left and right time as $-t_L$ and $t_R$, respectively.
Due to the time translation symmetry \eqref{eq:time_shift_symmetry}, it is not restrictive to consider the symmetric configuration \eqref{eq:symmetric_times}. 
An analysis of the effective potential in Fig.~\ref{potential_CV_case4} implies that the conserved momentum needs to satisfy $0 \leq P^2 \leq U_{\rm max}$ (where $U_{\rm max}$ is the local maximum of the effective potential) in order to obtain a connected extremal surface anchored to the stretched horizons.    
In this regime, the expressions for the volume and boundary time can be found in a similar manner to eqs.~\eqref{eq:Vol BC} and \eqref{eq:time L and R BC}, but in this case there is an additional turning point in the inflating region of the second patch.
The symmetries of the configuration imply that the turning points in the cosmological regions are located at the same radial coordinate $r_{t_{c1}} = r_{t_{c2}} \equiv r_{t_c}$, therefore we find
\begin{equation}\label{eq:Vol CC B}
    \frac{V(P)}{\Omega_{d-1}} = \frac{2}{L^{d-1}} \qty(\int^{r_{t_c}}_{r_{\rm st}}+\int^{r_{t_c}}_{r_{t_h}}) \frac{r^{2(d-1)}}{\sqrt{P^2 + f(r) (r/L)^{2(d-1)}}} dr \, ,
\end{equation}
In performing the previous steps, we employed a smooth coordinate map in the static patch region that connects different copies of SdS space.
For a fixed $P$, the maximal surface explores the future (past) black hole interior and the past (future) inflating regions. 
Next, we study the difference of the null coordinates $u(v)$ on the extremal surface, moving from the right to the left of the Penrose diagram.
The following relations hold
\begin{subequations}
\label{eq:identities_CV_case4}
\begin{equation}\label{eq:time contribution1_case4}
   t_{t_c} -r^*(r_{t_c}) - t_R +r^*(r_{\rm st}^R)  =
   \int_{r_{\rm st}^R}^{r_{t_c}} \frac{\dot{u}_+}{\dot{r}_+} dr =
   -\int^{r_{\rm st}^R}_{r_{t_c}} \tau_-(r,P) dr \, ,
\end{equation}
\begin{equation}
  t_{t_h} +  r^*(r_{t_h}) -t_{t_c} - r^*(r_{t_c}) =
   \int_{r_{t_c}}^{r_{t_h}} \frac{\dot{v}_-}{\dot{r}_-} dr =
   \int_{r_{t_h}}^{r_{t_c}} \tau_-(r,P) dr \, ,
   \label{eq:identity2_CV_case4}
\end{equation}
\begin{equation}
  t_{t_c} -  r^*(r_{t_c}) -t_{t_h} + r^*(r_{t_h}) =
   \int_{r_{t_h}}^{r_{t_c}} \frac{\dot{u}_+}{\dot{r}_+} dr =
   \int_{r_{t_h}}^{r_{t_c}} \tau_-(r,P) dr \, ,
   \label{eq:identity3_CV_case4}
\end{equation}
\begin{equation}\label{eq:time contribution4_case4}
  -t_L + r^*(r_{\rm st}^L) - t_{t_c} + r^*(r_{t_c}) =
   \int^{r_{\rm st}^L}_{r_{t_c}} \frac{\dot{v}_-}{\dot{r}_-} dr =
    \int^{r_{t_c}}_{r_{\rm st}^L} \tau_-(r,P) dr \, ,
\end{equation}
\end{subequations}
where we momentarily kept $t_L, t_R$ arbitrary.
Since the turning points in the two inflating regions are located at the same radial coordinate, we find that the summation of  eqs.~\eqref{eq:identity2_CV_case4}-\eqref{eq:identity3_CV_case4}
leads to the same constraint obtained in eq.~\eqref{eq:integral_deltat}.
This constraint can also be achieved by considering either eq.~\eqref{eq:identity2_CV_case4} or \eqref{eq:identity3_CV_case4} separately, in the symmetric setting \eqref{eq:symmetric_times} where $t_{t_c}=t_{t_h}=0$. 
In analogy with our discussion in case 3 of this section, the vanishing of $I(P)$ forces $P=0$, which means that the maximal volume surface would only exist for a single boundary time in the symmetric configuration, namely $t_L=t_R=0$. A possible way to determine an extremal surface for any boundary time is to evolve its endpoints in the antisymmetric way $t_L=-t_R$. However, this leads to the trivial time evolution generated by the Killing vector $\partial_t$. 
This means that the maximal volume slice will remain with the same constant value
\begin{equation}
    V(0)=\frac{2\Omega_{d-1}}{L^{d-1}}\qty(\int_{r_{\rm st}}^{r_{t_c}}+\int_{r_{t_h}}^{r_{t_c}})\frac{r^{d-1}~\rmd r}{\sqrt{f(r)}}~.
\end{equation}
Thus, the choice $t_L=-t_R$ allows for an evolution (although trivial) of the codimension-one observable.

\subsubsection*{Case 5}
\begin{figure}[t!]
  	\centering
  	\subfloat[]{\includegraphics[width=0.5\textwidth]{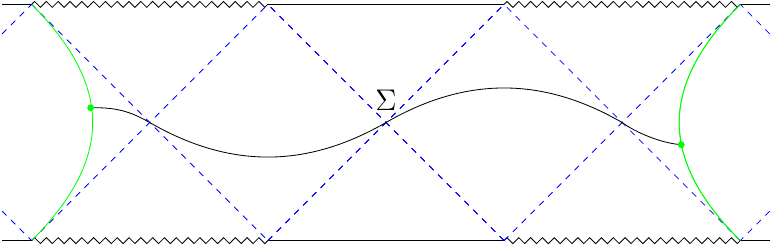}}
    \qquad
 \subfloat[]{\includegraphics[width=0.4\textwidth]{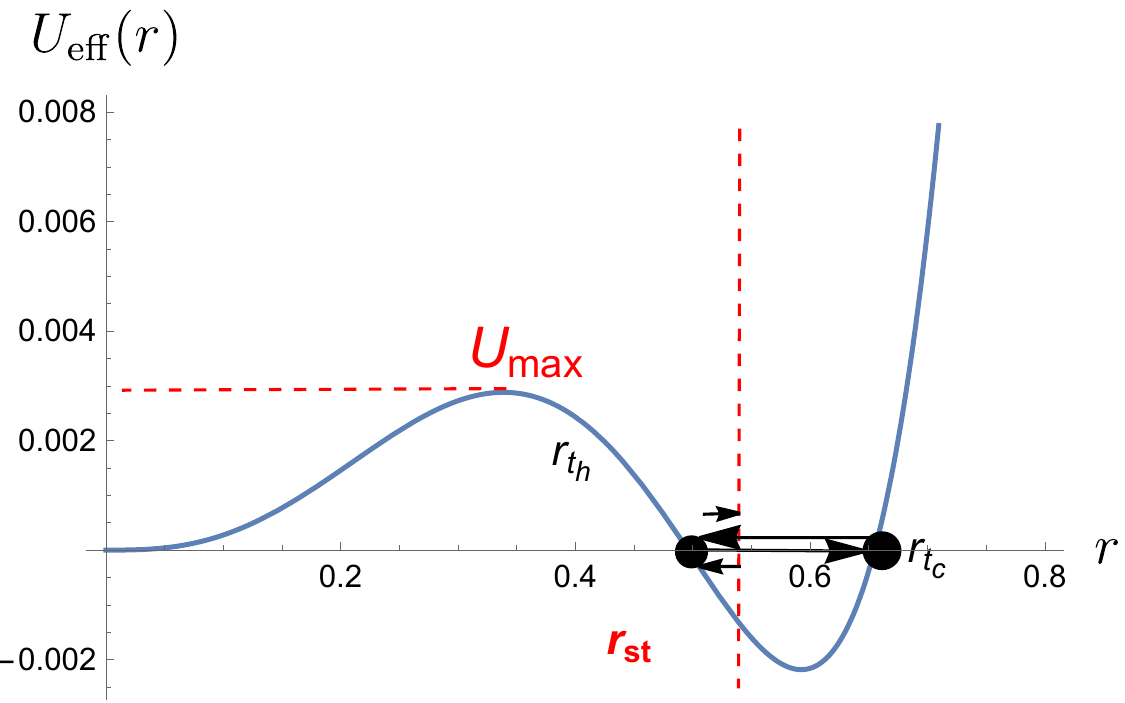}}
  	\caption{(a) The left and right stretched horizons are placed in non-consecutive static patches with a generic boundary time $t_R=-t_L$. 
   Due to the constraint \eqref{eq:integral_deltat}, only the conserved momentum $P=0$ is allowed for the surface $\Sigma$ depicted in the diagram.
    (b) Qualitative behavior of the effective potential $U_{\rm eff}(r)$. The surface $\Sigma$ connects the stretched horizons after going through three turning points: $r_{t_h}$, $r_{t_c}$, and $r_{t_h}$ again. 
   In this case, the turning points coincide with the location of the cosmological and black hole horizons, respectively.
    We have set $d=3$, $\mu= 0.187$, and $L=1$.   }
  	\label{fig:Ueff_5}
  \end{figure}

This configuration is illustrated in Fig.~\ref{fig:Ueff_5}. 
We take $r_{\rm st}^L = r_{\rm st}^R \equiv r_{\rm st}$ and we define the time on the left and right stretched horizons as $t_L$ and $-t_R$, respectively (see Fig.~\ref{fig:WDW_case5}). Again, it is not restrictive to take the symmetric configuration \eqref{eq:symmetric_times}. Similarly to the previous cases, one can express
\begin{equation}\label{eq:Vol BB C}
    \frac{V(P)}{\Omega_{d-1}} = \frac{2}{L^{d-1}} \qty(\int_{r_{t_h}}^{r_{\rm st}}+\int_{r_{t_h}}^{r_{t_c}}) \frac{r^{2(d-1)}}{\sqrt{P^2 + f(r) (r/L)^{2(d-1)}}} dr \, .
\end{equation}
Furthermore, we obtain several identities by computing the null coordinates at various points along the extremal surface
\begin{subequations}
\label{eq:identities_CV_case5}
\begin{equation}\label{eq:time contribution1_case5}
   t_{t_h} -r^*(r_{t_h}) + t_R +r^*(r_{\rm st}^R)  =
   \int_{r_{\rm st}^R}^{r_{t_h}} \frac{\dot{u}_-}{\dot{r}_-} dr =
   \int^{r_{\rm st}^R}_{r_{t_h}} \tau_+(r,P) dr \, ,
\end{equation}
\begin{equation}
  t_{t_c} +  r^*(r_{t_c}) -t_{t_h} - r^*(r_{t_h}) =
   \int_{r_{t_h}}^{r_{t_c}} \frac{\dot{v}_+}{\dot{r}_+} dr =
   \int_{r_{t_h}}^{r_{t_c}} \tau_+(r,P) dr \, ,
   \label{eq:identity2_CV_case5}
\end{equation}
\begin{equation}
  t_{t_h} -  r^*(r_{t_h}) -t_{t_c} + r^*(r_{t_c}) =
   \int_{r_{t_c}}^{r_{t_h}} \frac{\dot{u}_-}{\dot{r}_-} dr =
   \int_{r_{t_h}}^{r_{t_c}} \tau_+(r,P) dr \, ,
   \label{eq:identity3_CV_case5}
\end{equation}
\begin{equation}\label{eq:time contribution4_case5}
  t_L + r^*(r_{\rm st}^L) - t_{t_h} + r^*(r_{t_h}) =
   \int^{r_{\rm st}^L}_{r_{t_h}} \frac{\dot{v}_+}{\dot{r}_+} dr =
    \int_{r_{t_h}}^{r_{\rm st}^L} \tau_+(r,P) dr \, ,
\end{equation}
\end{subequations}
where we momentarily allowed $t_L, t_R$ to be arbitrary.
By adding eqs.~\eqref{eq:identity2_CV_case5} and \eqref{eq:identity3_CV_case5}, we find again the constraint \eqref{eq:integral_deltat}. 
This implies that $P=0$, therefore only the solution $t_L=t_R=0$  is allowed in the non-trivial case determined by the prescription \ref{rule}.
At arbitrary boundary time, the only allowed configuration is the antisymmetric one (with $t_L=-t_R$) for which the volume is simply given by eq.~\eqref{eq:Vol BB C} with $P=0$.

\subsection{Complexity=anything}
\label{sec:CAny CMC}
\begin{figure}[t!]
    \centering
    \includegraphics[width=\textwidth]{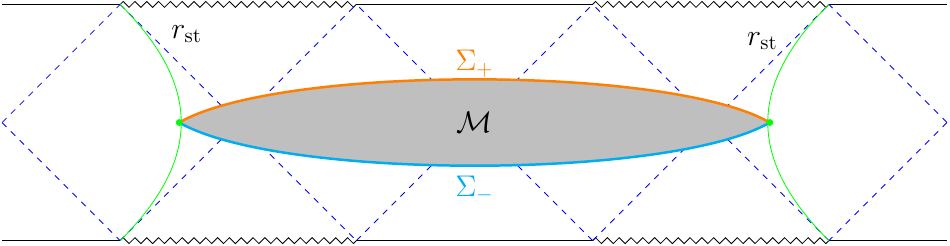}
\caption{Proposal for evaluating the volume of CMC slices for SdS$_{d+1}$ black holes. $\mathcal{M}$ (gray) is the bulk region bounded by the slices $\Sigma_-$ and $\Sigma_+$ (cyan and orange, respectively), where $\mathcal{C}^{-}$ and $\mathcal{C}^{+}$ in eq.~\eqref{eq:Volepsilon} are evaluated. 
In the picture, we anchor the bulk region to the black hole stretched horizons (green lines) at particular locations (green dots).
Similar considerations apply to the cosmological stretched horizons. The precise profile of the $\Sigma_\epsilon$ slices is determined by the extremization of eq.~\eqref{eq:regions CMC}.}
    \label{fig:ProposaldS}
\end{figure}

We are interested in codimension-one observables within the class of the CAny proposal \cite{Belin:2021bga,Belin:2022xmt,Jorstad:2023kmq}.
First, we define a spacetime region by extremizing the following combination of codimension-one and codimension-zero terms with different weights
\begin{equation}\label{eq:regions CMC}
\begin{aligned}
    \mathcal{C}_{\rm CMC}&=\frac{1}{G_N L}\biggl[\alpha_+\int_{\Sigma_+}d^d\sigma\,\sqrt{h}+\alpha_-\int_{\Sigma_-}d^d\sigma\,\sqrt{h}+ \frac{\alpha_B}{L}\int_{\mathcal{M}}d^{d+1}x\sqrt{-g}\biggr] \, ,
\end{aligned}
\end{equation}
where $\mathcal{M}$ is a $(d+1)$--dimensional bulk region with future (past) boundaries $\Sigma_{\pm}$ anchored at the stretched horizons, such that $\partial\mathcal{M}=\Sigma_+\cup\Sigma_-$, as shown in Fig.~\ref{fig:ProposaldS}.\footnote{We will label with $\epsilon=\lbrace +,- \rbrace$ the quantities defined on the codimension-one surfaces $\Sigma_{\pm}$.  }
We denote with $h$ the determinant of the induced metric on $\Sigma_{\pm}$. 
The coefficients $\alpha_{\pm}$ and $\alpha_B$ are dimensionless constants. The extremization of the functional \eqref{eq:regions CMC} defines constant mean curvature (CMC) slices, \ie the extrinsic curvature on these surfaces is given by \cite{MARSDEN1980109,Belin:2022xmt}
\begin{equation}
    K_\epsilon\equiv\eval{K}_{\Sigma_\epsilon} = -\epsilon \, \frac{\alpha_B}{\alpha_\epsilon~L}~,
    \label{eq:K_CMC}
\end{equation}
with the convention that the vectors normal to $\Sigma_{\pm}$ are future-directed.
We then define holographic complexity as the physical observable
\begin{equation}\label{eq:Volepsilon}
    \mathcal{C}^\epsilon \equiv \frac{1}{G_N L}\int_{\Sigma_\epsilon}d^d\sigma\,\sqrt{h}~F[g_{\mu\nu},\,\mathcal{R}_{\mu\nu\rho\sigma},\,\nabla_\mu]~,
\end{equation}
where $F[g_{\mu\nu},\,\mathcal{R}_{\mu\nu\rho\sigma},\,\nabla_\mu]$ is an arbitrary scalar functional composed of $(d+1)$--dimensional bulk curvature invariants built with the metric $g_{\mu\nu}$, the Riemann tensor $\mathcal{R}_{\mu\nu\rho\sigma}$ and the covariant derivative $\nabla_{\mu}$.
The quantity $\mathcal{C}^{\epsilon}$ is defined on either the future (when $\epsilon=-$) or the past (when $\epsilon=+$) codimension-one CMC slices $\Sigma_{\pm}$ determined by the extremization of the functional \eqref{eq:regions CMC} with fixed boundary conditions on the stretched horizons.

To simplify the evaluation of the CMC slices and the associated complexity observable, we employ the EF coordinates \eqref{eq:generic_null_metric_noshock}.
Using the spherical symmetry of the background \eqref{eq:asympt_dS}, we describe the slices $\Sigma_\epsilon$ in terms of a radial parameter $\sigma$, such that the coordinates are expressed as $(v(\sigma)$, $r(\sigma))$. Evaluating eq.~\eqref{eq:regions CMC}  in this coordinate system, we obtain
\beq
\mathcal{C}_{\rm CMC} =\frac{\Omega_{d-1} L^{d-2}}{G_N}\sum_{\epsilon}\alpha_\epsilon\int_{\Sigma_\epsilon}
    d\sigma\,\mathcal{L}_\epsilon~,
    \label{eq:C CMC}
\eeq
where
\begin{equation}
\begin{aligned}
    \mathcal{L}_\epsilon&\equiv \qty(\frac{r}{L})^{d-1}\sqrt{-f(r)\dot{v}^2+2\dot{v}\dot{r}}+\epsilon \, \tfrac{L\abs{K_\epsilon}}{d}\dot{v}\qty(\frac{r}{L})^d\\
    &=\qty(\frac{r}{L})^{d-1}\sqrt{-f(r)\dot{u}^2-2\dot{u}\dot{r}}+\epsilon \, \tfrac{L\abs{K_\epsilon}}{d}\dot{u}\qty(\frac{r}{L})^d \, .
\end{aligned}
\end{equation}
Using the gauge choice (\ref{gauge}), the equations of motion (EOM) associated to eq.~\eqref{eq:C CMC} can be written as 
\begin{equation}
    \dot{r}^2+\mathcal{U}(P_{\epsilon},\,r)=0~,
    \label{eq:EOM_CAny}
\end{equation}
where
\begin{subequations}
    \beq
    \begin{aligned}
    \label{eq:potential arbitrary}
P_{\epsilon}&\equiv\pdv{\mathcal{L}_\epsilon}{\dot{v}}=\dot{r}-\dot{v}\,f(r)+\epsilon \, \frac{L~\abs{K_\epsilon}}{d}\,\le \frac{r}{L}\ri^d\\
&=\pdv{\mathcal{L}_\epsilon}{\dot{u}}=-\dot{r}-\dot{u}\,f(r)+\epsilon \, \frac{L~\abs{K_\epsilon}}{d}\,\le \frac{r}{L}\ri^d~,
\end{aligned}
\eeq
\beq
    \mathcal{U}(P_{\epsilon},\,r) \equiv -f(r) \le \frac{r}{L}\ri^{2(d-1)} - \left(P_{\epsilon} -\epsilon \, \frac{L~\abs{K_\epsilon}}{d}  \le \frac{r}{L}\ri^d\right)^2~.\label{eq:U CAny}
\eeq
\end{subequations}
The differential equation \eqref{eq:EOM_CAny} effectively describes the motion of a particle in a potential $\mathcal{U}$.
The quantity $P_{\epsilon}$ is a conserved momentum corresponding to the fact that the variable $u(v)$ is cyclic for the Lagrangian density $\mathcal{L}_{\epsilon}$.

The codimension-one surfaces solving the variational problem in eq.~\eqref{eq:EOM_CAny} determine the CMC slices
where complexity is calculated according to eq.~\eqref{eq:Volepsilon}.
In EF coordinates, the CAny observable reads
\beq
\begin{aligned}
    \label{eq:C epsilon inter}
    \mathcal{C}^\epsilon &=\frac{\Omega_{d-1} L^{d-2}}{G_N}\int_{\Sigma_\epsilon}d\sigma\,\qty(\frac{r}{L})^{d-1}\sqrt{-f(r)\dot{v}^2+2\dot{v}\dot{r}}\,a(r)\\
    &=\frac{\Omega_{d-1} L^{d-2}}{G_N}\int_{\Sigma_\epsilon}d\sigma\,\qty(\frac{r}{L})^{d-1}\sqrt{-f(r)\dot{u}^2-2\dot{u}\dot{r}}\,a(r)\,,
\end{aligned}
\eeq
where $a(r)$ is a dimensionless scalar function corresponding to the evaluation of the functional $F$ in eq.~\eqref{eq:Volepsilon} in the geometry under consideration.
We remark that the CAny proposal evaluated on CMC slices reduces to the CV conjecture \eqref{eq:vol computation} when $K_{\epsilon}=0$ and $F=1$ in eq.~\eqref{eq:Volepsilon}.\footnote{Equivalently, it follows from eq.~\eqref{eq:K_CMC} that when $K_{\epsilon}=0$, only the $\alpha_+$ or $\alpha_-$ terms in (\ref{eq:regions CMC}) contribute to define the region of integration, which collapses to a single extremal surface.}
Indeed, by imposing the above-mentioned conditions, we find that
the identities \eqref{eq:EOM_CAny}, \eqref{eq:potential arbitrary} and \eqref{eq:U CAny} reduce to eqs.~\eqref{eq:classical_motion} and \eqref{eq:conserved_momentum_CV}, and the observable \eqref{eq:C epsilon inter} reduces to the induced volume on the extremal surface. 

The shape of the effective potential $\mathcal{U}$ is crucial to determine the time evolution of the CMC slices.
As long as $P_{\epsilon}$ lies in a range such that the effective potential admits at least one root, then the corresponding extremal surface will present a turning point $r_t$ defined by $\mathcal{U}(P_{\epsilon}, r_t)=0$.
The major difference with the CV case is that the effective potential \eqref{eq:U CAny} is itself a function of $P_{\epsilon}$, therefore it is more difficult to determine the full time-dependence of complexity.
For this reason, we will mainly focus on the late time behaviour, which provides the most relevant and universal feature. 

In the following, we analyze the CAny observable for the various configurations of the stretched horizons presented in section \ref{ssec:stretched_horizons}.

\subsubsection*{Case 1}
This case has been previously investigated in \cite{Aguilar-Gutierrez:2023zqm}. Here we briefly report the main results, and we include additional information on the time dependence of complexity.  
Without loss of generality, we consider the case of a symmetric time evolution \eqref{eq:symmetric_times} and we locate the stretched horizons at the same radial coordinate $r_{\rm st} = r^L_{\rm st}=r^R_{\rm st}$.
The codimension-one observable in eq.~\eqref{eq:C epsilon inter} can be expressed as
\begin{equation}\label{eq:Anything complexity cod1}
\mathcal{C}^\epsilon=\frac{2\Omega_{d-1} L^{d-2}}{G_N} \int_{{r}_{\rm st}}^{{r}_{t_c}} \frac{a(r)~ {\qty(\frac{r}{L})}^{2(d-1)}  }{\sqrt{-\mathcal{U}(P_{\epsilon},\,{r})}} \, dr \, .
\end{equation}
Meanwhile, the evolution of the boundary time follows from
\begin{equation}\label{eq: t CAny}
\begin{aligned}
t&= 2  \int_{{r}_{\rm st}}^{{r}_{t_c}} \frac{ P_{\epsilon}- \epsilon L\frac{\abs{K_\epsilon}}{d} \qty(\frac{r}{L})^d  }{f(r)\,\sqrt{-\mathcal{U}(P_{\epsilon}, r)}} \, dr \, .
\end{aligned}
\end{equation}
This allows to bring the integral (\ref{eq:Anything complexity cod1}) to the form
\begin{equation}\label{eq:CAny late time C1}
    \begin{aligned}
    {\mathcal{C}^\epsilon}=&\frac{\Omega_{d-1} L^{d-2}}{G_N} \, a(r_{t_c})\sqrt{-f(r_{{t_c}})\qty(\frac{r_{{t_c}}}{L})^{2(d-1)}}~t\\
    &+\frac{2\Omega_{d-1} L^{d-2}}{G_N}\int_{r_{\rm st}}^{r_{{t_c}}}  dr \, \frac{a(r)f(r)\,\qty(\frac{r}{L})^{2(d-1)}-a(r_{t_c})\sqrt{-f(r_{{t_c}})\qty(\frac{r_{\rm t_c}}{L})^{2(d-1)}}\left(P_\epsilon -\frac{\epsilon {L~\abs{K_\epsilon}}}{d} \qty(\frac{r}{L})^{d} \right)}{f(r)\sqrt{-\mathcal{U}(P_\epsilon,\,r)}} ~.
\end{aligned}
\end{equation} 
As discussed in \cite{Aguilar-Gutierrez:2023zqm}, one can perform an additional minimization between $\mathcal{C}^{\pm}$ to define a time-reversal invariant observable in asymptotically dS space that avoids a hyperfast growth. For our purposes, it is sufficient to note that at late (early) times, the knowledge of $\mathcal{C}^{-} (\mathcal{C}^+)$ will be sufficient to determine the complexity, while the other surface might not be defined.

In particular, one can show that the growth rate at late times becomes
\begin{equation}\label{eq:Late time growth}
    \lim_{{t}\rightarrow\infty}\dv{{t}} \mathcal{C}^- =\frac{\Omega_{d-1}}{G_NL}a(r_f) \sqrt{-f({r}_f) {r}_f^{2(d-1)}}
\end{equation}
where $r_f=\lim_{t\rightarrow\infty}r_{t_c}$ is the turning point at late times, which satisfies $r_f \geq r_c$.
This behaviour is achieved whenever $r_f$ is a maximum of the effective potential \eqref{eq:U CAny}, \ie the following conditions hold:
\beq\label{eq: Condition Cany pot}
\mathcal{U}(P_{\epsilon}, r_f) = 0 \, , \qquad
\partial_r \mathcal{U}(P_{\epsilon}, r_f) = 0 \, , \qquad
\partial_r^2 \mathcal{U}(P_{\epsilon}, r_f) \leq  0 \, .
\eeq
The previous constraints can be summarized into the algebraic relation
\begin{equation}\label{eq: rf equation}
    4 {r_f}
   f\left({r_f}\right)
   \left((d-1)
   f'\left({r_f}\right)+{K_\epsilon}^
   2 {r_f}\right)+4
   (d-1)^2
   f\left({r_f}\right){}^2
   +{r_f}^2
   f'\left({r_f}\right){}^
   2=0~.
\end{equation}
When there are no roots to the above equation in the range $r_c \leq r_f < \infty$, eq.~\eqref{eq:Late time growth} does not hold, and complexity presents instead a hyperfast growth in the inflating patch.
Indeed, the latter behaviour happens when $K_{\epsilon}=0$, leading to the CV conjecture studied in section \ref{ssec:CV}.

The novelty carried by the class of holographic proposals \eqref{eq:Volepsilon} is the existence of a range of extrinsic curvatures for the CMC slices  $\abs{K_{\epsilon}}\geq K_{\rm crit}$ such that the identity \eqref{eq: rf equation} admits a turning point located beyond the cosmological horizon, and the hyperfast growth is avoided.
At the special values of the mass parameter $\mu=0$ (pure dS space) and $\mu=\mu_N$ (extremal case), the critical extrinsic curvature is analytically found to be
\begin{align}\label{eq:K crit}
    K_{\rm crit}(\mu=0)=\frac{2\sqrt{d-1}}{L}~, \qquad   
    K_{\rm crit}(\mu=\mu_N)=~\frac{\sqrt{d}}{L}~.
\end{align}
For general $\mu$, there is no closed-form expression for the critical value of $K_\epsilon$, but one can prove that it always exists \cite{Aguilar-Gutierrez:2023zqm}.

\begin{figure}[t!]
\centering
\subfloat[]{\label{subfig:CMC_slices_case1} \includegraphics[height=0.3\textwidth]{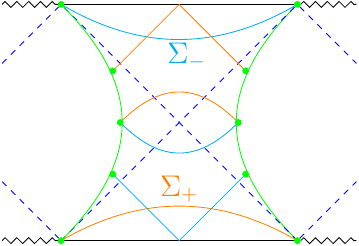}}\hspace{0.75cm}\subfloat[]{\label{subfig:Ueff_CAny_case1}  \includegraphics[height=0.3\textwidth]{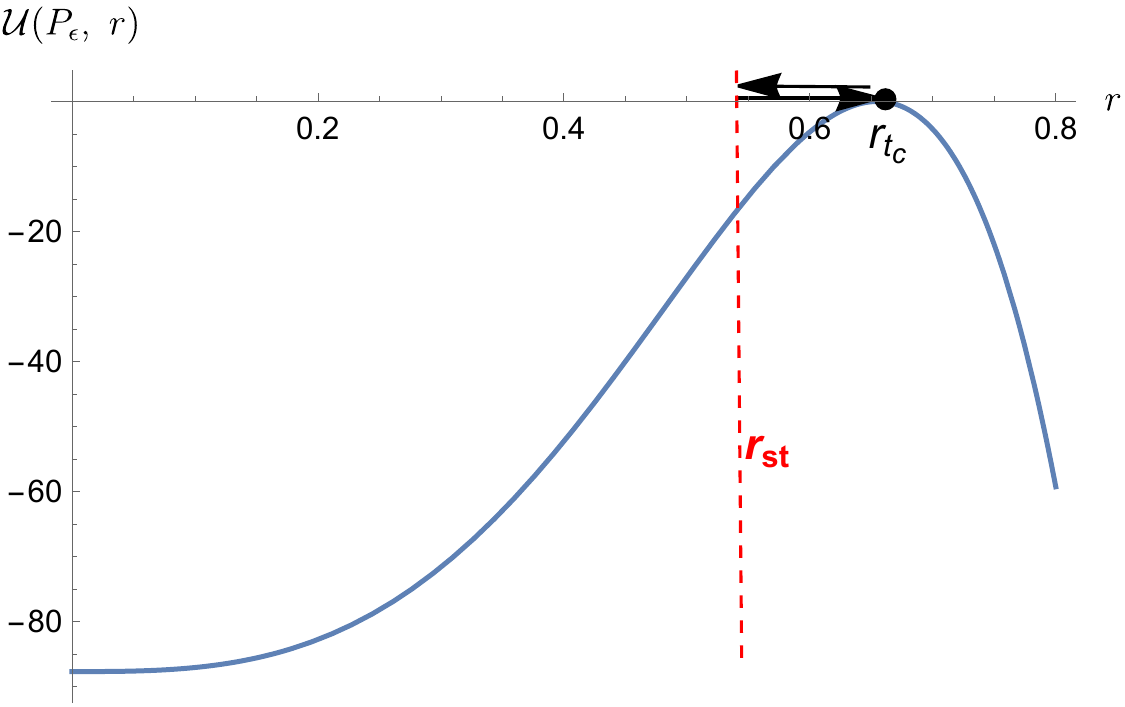}}
    \caption{Case 1: (a) Representation of the profile of the CMC slices $\Sigma_+$ (orange) and $\Sigma_-$ (cyan) when $\abs{K_\epsilon}\geq K_{\rm crit}$ for SdS space. (b) Effective potential (\ref{eq:U CAny}). A particle coming from $r_{\rm st}$ will be reflected at the turning point $r_{t_c}$ if it exists; otherwise it will fall into timelike infinity $\mathcal{I}^{\pm}$. Numerical values are the same as in Fig.~\ref{fig:Ueff_1}, in addition to $P_\epsilon=-9.37$ and $\abs{K_\epsilon}=100$.  }
    \label{fig:CMC1}
\end{figure}

\begin{figure}[t!]
    \centering
   \subfloat[]{\includegraphics[width=0.48\textwidth]{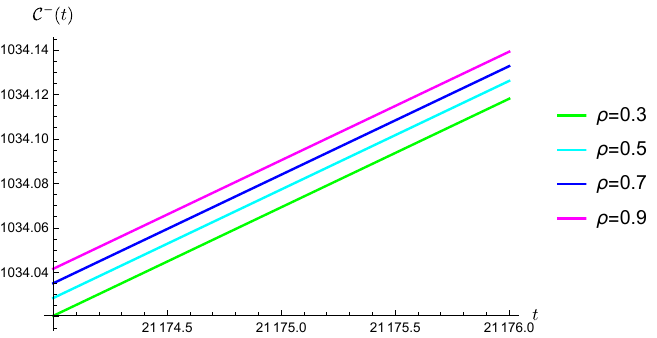}}\hspace{0.25cm}
   \subfloat[]{\includegraphics[width=0.48\textwidth]{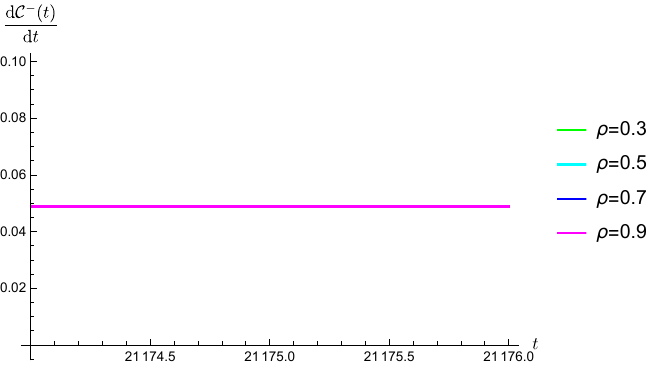}}
    \caption{Case 1: Evaluation of (a) the late-time behavior of $\mathcal{C}^-$ determined by eq.~\eqref{eq:CAny late time C1} and of (b) its rate of growth as a function of time. We take $a(r)=1$, $K_-=100/L$, besides the same parameters as in Fig.~\ref{fig:volume_cosmo}. Notice that all the different choices of $\rho$ lead to the same late-time rate of growth.}
    \label{fig:CAny1}
\end{figure}

The time evolution of the CMC slices in the SdS background is shown in Fig.~\ref{subfig:CMC_slices_case1}. 
We depict the effective potential $\mathcal{U}$ in Fig.~\ref{subfig:Ueff_CAny_case1} for a specific value of the conserved momentum.
This plot shows that at the corresponding boundary time, determined according to eq.~\eqref{eq: t CAny}, there is a single turning point corresponding to the existence of a root of the effective potential.
The late-time behaviour of complexity and its rate of growth are depicted in Fig.~\ref{fig:CAny1}, referring to the case $K_{\epsilon} \geq K_{\rm crit}$ and for a trivial function $F=1$ in eq.~\eqref{eq:Volepsilon}.
As anticipated below eq.~\eqref{eq: rf equation}, the main feature is that CAny displays a linear increase with a rate independent of $\rho$, the value that determines the location of the stretched horizons.

\subsubsection*{Case 2}

We consider again a symmetric configuration of the boundary times \eqref{eq:symmetric_times} and of the stretched horizons $r_{\rm st}^L = r_{\rm st}^R$.
The evaluation of complexity and of the boundary time is very similar to eqs.~\eqref{eq:Anything complexity cod1} and \eqref{eq: t CAny}:
\begin{align}\label{eq:CAny 2}
\mathcal{C}^\epsilon&=\frac{2\Omega_{d-1} L^{d-2}}{G_N} \int^{{r}_{\rm st}}_{{r}_{t_h}} \frac{ {\qty(\frac{r}{L})}^{2(d-1)}}{\sqrt{-\mathcal{U}(P_v^{\epsilon},\,{r})}} \, dr \, ,\\
t&= -2\int^{{r}_{\rm st}}_{{r}_{t_h}}\frac{\left(P_{\epsilon} - \epsilon L\frac{\abs{K_\epsilon}}{d} \qty(\frac{r}{L})^d \right)}{f(r)\,\sqrt{-\mathcal{U}(P_{\epsilon}, r)}} \, dr \, .
\label{eq:t CAny 2}
\end{align}
After similar manipulations, we find that the rate of growth at late times is formally given by the same expression \eqref{eq:Late time growth} with $r_f$ satisfying eq.~\eqref{eq: rf equation}.
However, the difference with the previous case is that the radial coordinate $r_f$ of the turning point generating the late time linear growth now lies inside the interval $[0,~r_{h}]$ (with $r_{\rm st}>r_h$). 
One can now find a solution for $r_f$ in this range even when $K_\epsilon=0$, as obtained in case 2 of section~\ref{ssec:CV}, and in contrast to case 1 for the CAny proposal.
This result is consistent with the behaviour of black holes in asymptotically AdS space \cite{Belin:2022xmt,Jorstad:2023kmq}.

\begin{figure}[t!]
\centering
\subfloat[]{\includegraphics[height=0.3\textwidth]{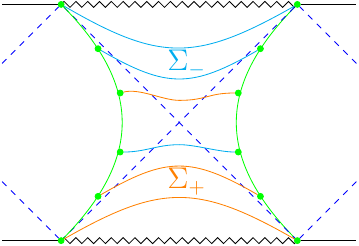}}\hspace{0.75cm}\subfloat[]{\includegraphics[height=0.3\textwidth]{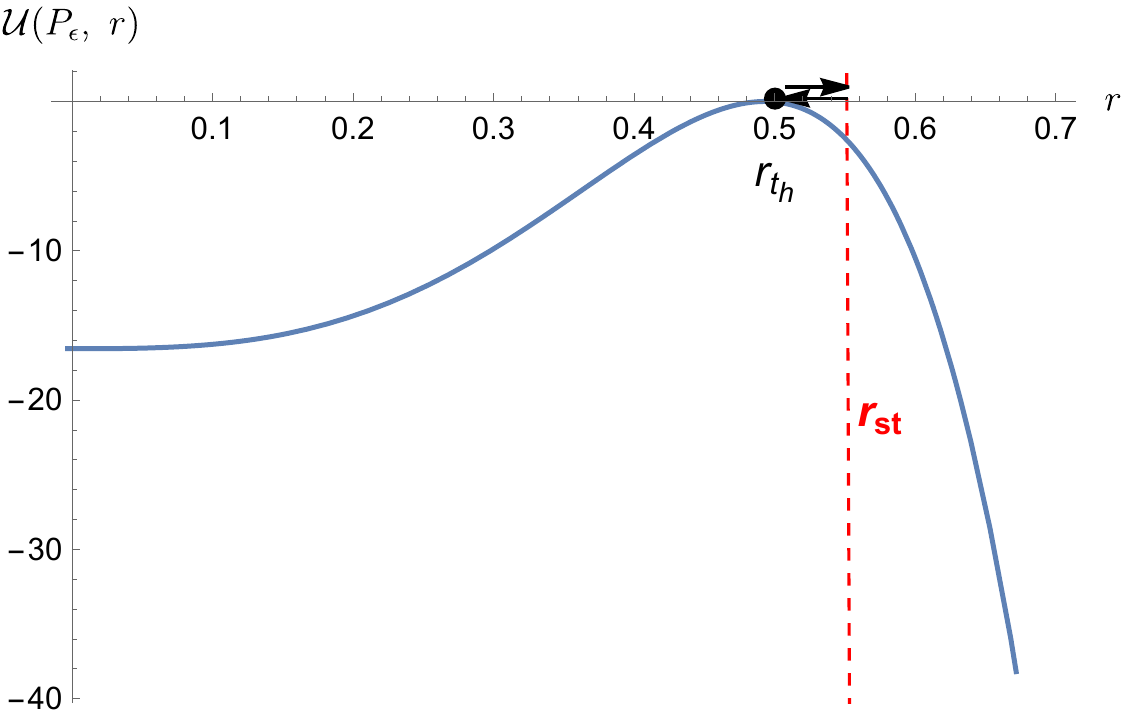}}
    \caption{Case 2: (a) Representation of the profile of the CMC slices for $\Sigma_-$ (cyan) and $\Sigma_+$ (orange) for SdS space with a pair of black hole stretched horizons. (b) Effective potential (\ref{eq:U CAny}). A particle moving in the potential coming from $r_{\rm st}$ will be reflected at the turning point $r_{t_h}$. The numerical parameters are chosen to be the same as in Fig. \ref{fig:CMC1}, except for $P_\epsilon=-4.07$.}
    \label{fig:CMC2}
\end{figure}

\begin{figure}[t!]
    \centering
    \subfloat[]{\includegraphics[width=0.48\textwidth]{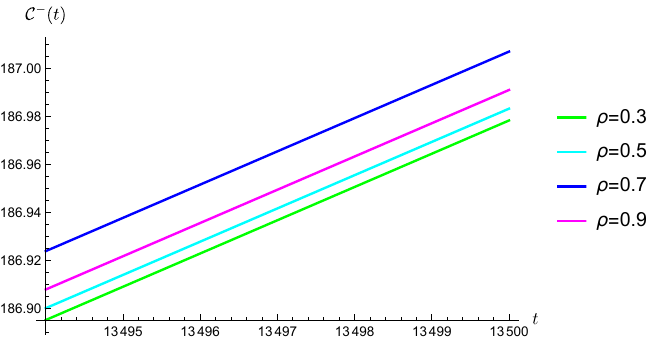}}\hspace{0.2cm}\subfloat[]{\includegraphics[width=0.48\textwidth]{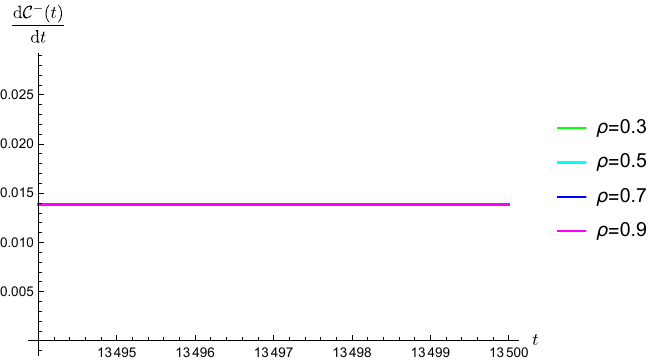}}
    \caption{Case 2: late-time behavior of (a) $\mathcal{C}^\epsilon$ in eq.~\eqref{eq:CAny 2} and (b) its rate of growth. The same parameters as in Fig.~\ref{fig:CAny1} have been used, including $a(r)=1$ and $K_-=100/L$.}
    \label{fig:CAny2}
\end{figure}

The CMC slices in the SdS black hole background and a plot of the effective potential for a specific value of $P_{\epsilon}$ are shown in Fig.~\ref{fig:CMC2}, while the corresponding plot for the complexity growth at late times is shown in Fig.~\ref{fig:CAny2}.


\subsubsection*{Case 3}
We consider the case of a cosmological stretched horizon located at $r_{\rm st}^L$ on the left side, and a black hole stretched horizon at $r_{\rm st}^R$ on the right side of the Penrose diagram. 
The general CAny observable defined in eq.~\eqref{eq:C epsilon inter} becomes
    \begin{align}\label{eq:Cepsilon car2}
    \mathcal{C}^\epsilon&=\frac{\Omega_{d-1} L^{d-2}}{G_N}\qty(\int_{r^{\rm L}_{\rm st}}^{r_{\rm t_c}}+  \int^{r_{\rm t_c}}_{r_{\rm  t_h}}+\int^{r^{R}_{\rm st}}_{r_{\rm t_h}})\frac{\qty(\frac{r}{L})^{2(d-1)}a(r)}{\sqrt{-\mathcal{U}(P_{\epsilon},\,r)}} \, \rmd {r} ~.
\end{align}
Next, we follow similar steps to case 3 in section~\ref{ssec:CV} to derive an expression for the boundary times. 
Namely, we define the anchoring times at the left and right stretched horizons as $-t_L$ and $-t_R$, and we obtain the same eq.~\eqref{eq:identities_CV_case3} {except for} the modification $\tau_\pm\rightarrow\tau_\pm^\epsilon$ in (\ref{eq:def_tau}), where
\begin{equation}
    \tau_{\pm}^\epsilon(P_\epsilon,~r) \equiv \frac{-\qty(P_\epsilon-\epsilon L\frac{\abs{K_\epsilon}}{d}\qty(\frac{r}{L})^d )\pm \sqrt{-\mathcal{U}(P_\epsilon,~r)}}{f(r) \sqrt{-\mathcal{U}(P_\epsilon,~r)}} \, .
    \label{eq:def_taup}
\end{equation}
After this replacement, we can directly add the various contributions to find
\begin{align}
    t_R-t_L &= \qty(\int_{r_{\rm st}^L}^{r_{t_c}} + \int_{r_{\rm t_h}}^{r_{t_c}} + \int_{r_{t_h}}^{r_{\rm st}^R} )\frac{P_{\epsilon} - \epsilon L\frac{\abs{K_\epsilon}}{d} \qty(\frac{r}{L})^d }{f(r) \sqrt{-\mathcal{U}(P_{\epsilon}, r)}} \, \rmd r \, ,\label{eq: t CMC 3}\\
   t_{t_c}-t_{t_h} &=-I(P_\epsilon,~K_\epsilon),\quad I(P_\epsilon,~K_\epsilon)=\int_{r_{t_h}}^{r_{t_c}}\frac{P_{\epsilon} - \epsilon L\frac{\abs{K_\epsilon}}{d} \qty(\frac{r}{L})^d }{f(r) \sqrt{-\mathcal{U}(P_{\epsilon}, r)}} \, \rmd r \, .\label{eq:time dif turning points}
\end{align}
When $\abs{K_\epsilon}\geq K_{\rm crit}$, one can numerically show that there do not exist two turning points $r_{t_c}$ and $r_{t_h}$ at fixed boundary time, thus corresponding to disconnected extremal surfaces $\Sigma_\epsilon$. 
For this reason, we will restrict the following analysis to the case $\abs{K_\epsilon} < K_{\rm crit}$.

Similar to case 3 in Sec. \ref{ssec:CV}, we have two possibilities:
\begin{itemize}
    \item In the symmetric configuration $t_L=t_R$, the only solution to eq.~\eqref{eq: t CMC 3} is characterized by the conditions $P_{\epsilon}=K_{\epsilon}=0$.
Therefore, the CAny proposal is reduced to CV, and the same analysis performed in section \ref{ssec:CV} follows.
In particular, the previous result is consistent with the constraint \eqref{eq:time dif turning points}, and leads to a time-independent complexity observable \eqref{eq:Cepsilon car2}. 
\item In contrast, when we consider (without loss of generality) the relation $t_R=-t_L$ dictated by prescription \ref{rule}, we can have a non-trivial time evolution with non-vanishing $t_{t_c}, t_{t_h}$. We will consider this choice in the remainder of the section.
\end{itemize} 

\begin{figure}[t!]
\centering
\subfloat[]{\includegraphics[height=0.25\textwidth]{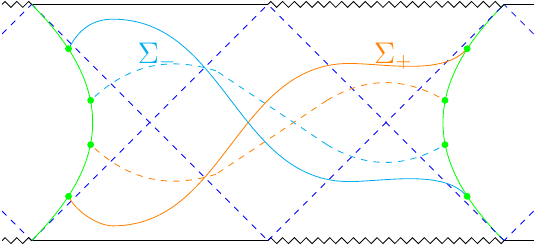}}\hspace{0.2cm}\subfloat[]{\includegraphics[height=0.25\textwidth]{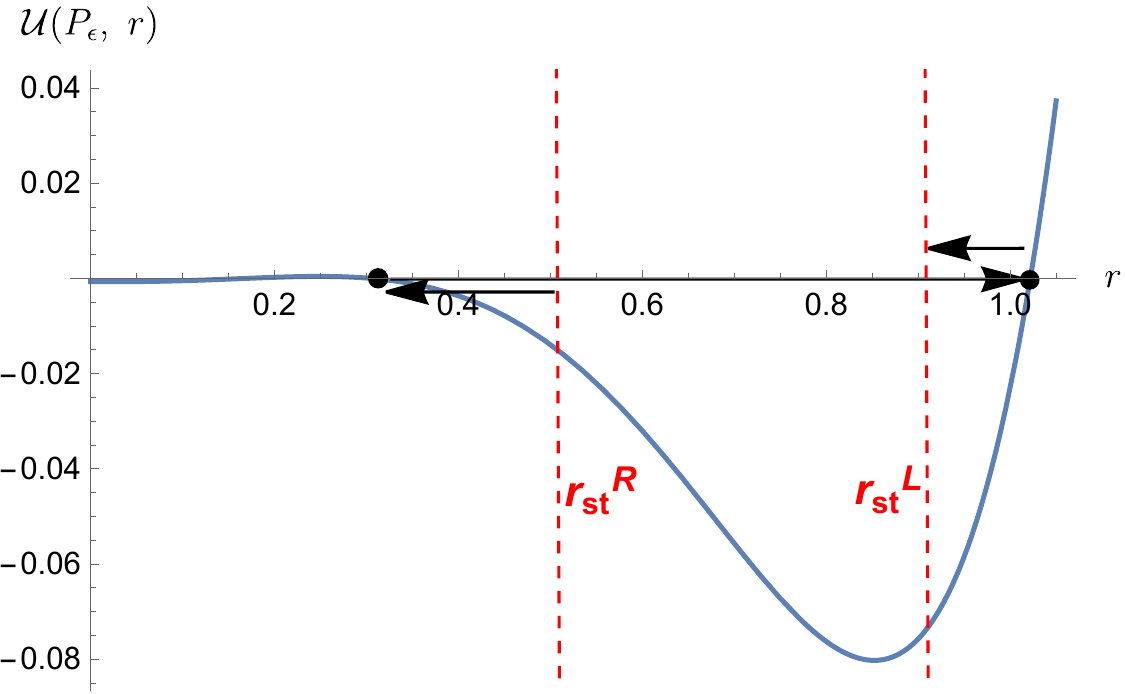}}
    \caption{Case 3: (a) Representation of the profile of the CMC slices for $\Sigma_-$ (cyan) and $\Sigma_+$ (orange) when $\abs{K_\epsilon}\leq K_{\rm crit}$ for a SdS black hole, intersecting the stretched horizons at the green dots. {Dashed and solid lines represent different boundary times along the evolution.} (b) Effective potential (\ref{eq:U CAny}). A particle moving in the potential coming from $r_{\rm st}^R$ will be reflected at the turning points and reach $r_{\rm st}^L$. The numerical parameters are chosen to be the same as in Fig. \ref{fig:CMC2}, except for $\abs{K_\epsilon}=\sqrt{3}$, and $P_\epsilon=-0.027$.}
    \label{fig:CMC3}
\end{figure}

We proceed to evaluate eq.~\eqref{eq:Cepsilon car2} carefully. Since under the gauge fixing \eqref{gauge} we have $P_{\epsilon} - \epsilon L\frac{\abs{K_\epsilon}}{d}=-\dot{r}-f(r)\dot{u}= -f(r) \dot{t}$, the sign of this combination will be determined by the orientation of the Killing vectors, as we remarked in footnote \ref{fnt:Sign of P}. 
The CMC slices corresponding to this regime are shown in Fig.~\ref{fig:CMC3}.

We can now take care of the integrand (\ref{eq:Cepsilon car2}) at the turning points. 
At late times, eq.~\eqref{eq: rf equation} admits a turning point $r_{t_h}$ inside the black hole region but does not admit a turning point $r_{t_c}$ outside the cosmological horizon.
For this reason, we can conveniently re-express the complexity as
\begin{equation}\label{eq:Cepsilon close form BC}
\begin{aligned}
    \frac{G_N~\mathcal{C}^\epsilon}{\Omega_{d-1} L^{d-2}}=&\sqrt{-f(r_{t_h})\qty(\frac{r_{t_h}}{L})^{2(d-1)}}a(r_{t_h})\qty(\int^{r^{\rm R}_{\rm st}}_{r_{\rm t_h}}+  \int^{r_{\rm t_c}}_{r_{\rm  t_h}}+\int^{r_{\rm t_c}}_{r_{\rm st}^{\rm L}})\frac{\left(P_{\epsilon}- \epsilon L\frac{\abs{K_\epsilon}}{d} \qty(\frac{r}{L})^d \right)d r}{f(r)\sqrt{-\mathcal{U}(P_{\epsilon}, r)}}\\
&+\qty(\int^{r^{R}_{\rm st}}_{r_{\rm t_h}} + \int^{r_{\rm t_c}}_{r_{\rm  t_h}}+\int^{r_{\rm t_c}}_{r_{\rm st}^{\rm L}})\tfrac{f(r)\,\qty(\frac{r}{L})^{2(d-1)}a(r)-\sqrt{-f(r_{t_h})\qty(\frac{r_{t_h}}{L})^{2(d-1)}}a(r_{t_h})\left(P_{\epsilon} -\frac{\abs{K_\epsilon}}{d} \qty(\frac{r}{L})^d \right)}{f(r)\sqrt{-\mathcal{U}(P_{\epsilon},\,r)}}\rmd r~.
\end{aligned}
\end{equation}
By denoting with $r_{f_h}$ the local maximum of the effective potential $\mathcal{U}$ approached by the CMC slice at late times, we obtain that the time derivative of eq.~\eqref{eq:Cepsilon close form BC} reads
\begin{equation}\label{eq:t inft dCdt CAny3}
    \lim_{{t}\rightarrow\infty}\dv{\mathcal{C}^\epsilon}{{t}}=\frac{\Omega_{d-1} L^{d-2}}{G_N}\sqrt{-f(r_{f_h})\qty(\frac{r_{f_h}}{L})^{2(d-1)}}a(r_{f_h})~.
\end{equation}
The existence of a final slice at constant radius approached by the CMC slices is responsible for the linear growth at late times.
Notice that the result for $K_\epsilon < K_{\rm crit}$ agrees with the analysis of case 3 in section~\ref{ssec:CV} when  $K_\epsilon=0$ and $a(r)=1$. 

\begin{figure}[t!]
    \centering
    \subfloat[]{\includegraphics[width=0.48\textwidth]{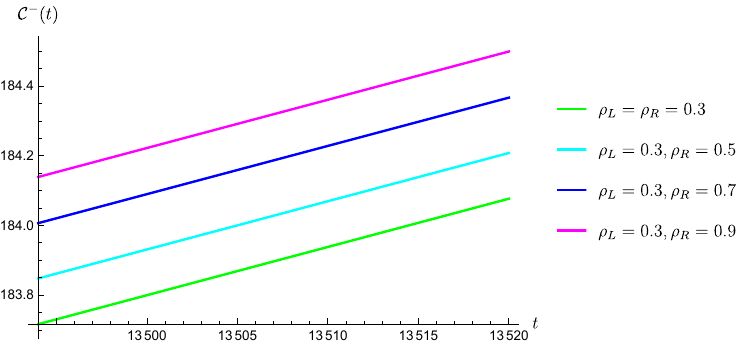}}\hspace{0.2cm}\subfloat[]{\includegraphics[width=0.48\textwidth]{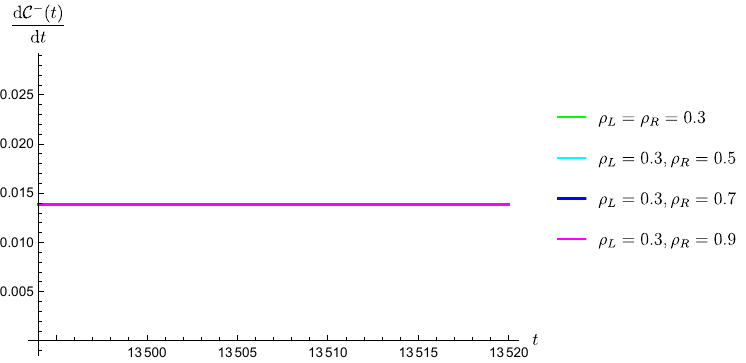}}
    \caption{Case 3: plot of (a) $\mathcal{C}^-$ in eq.~\eqref{eq:Cepsilon close form BC} and (b) its rate of growth. We use the same parameters as in Fig. \ref{fig:CAny1}, in addition to $n=1$ and various choices of $\rho_{\rm L}$ and $\rho_{\rm R}$, as indicated in the legend of the figure.}
    \label{fig:CAny3}
\end{figure}

\subsubsection*{Case 4}

\begin{figure}[t!]
    \centering
    \includegraphics[width=0.7\textwidth]{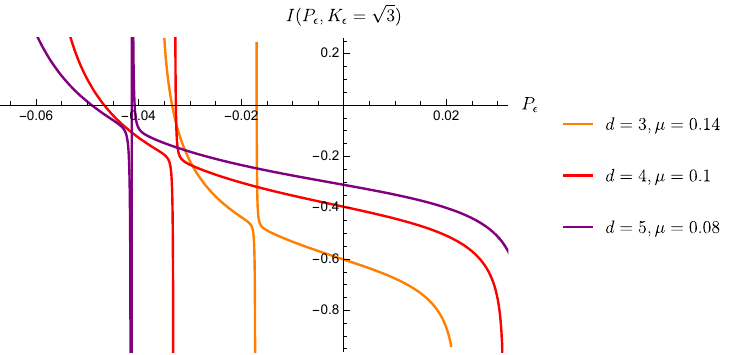}
    \caption{Plot of the integral $I(P_\epsilon,~K)$ in (\ref{eq:time dif turning points}) as a function of the conserved momentum $P_\epsilon$ for fixed $\abs{K_\epsilon}=\sqrt{3} < K_{\rm crit}$ and for various $d, \mu$. We fix $L=1$ and $\epsilon=-1$. The function only vanishes for particular values of $P_\epsilon$.}
    \label{fig:CAny2ndInt}
\end{figure}

We assume that the cosmological stretched horizons are located at the same radial coordinate $r_{\rm st}^L=r_{\rm st}^R \equiv r_{\rm st}$. Without loss of generality, we also consider the symmetric configuration \eqref{eq:symmetric_times} of the boundary times.
In each copy of the extended SdS$^n$ geometry, all connected surfaces $\Sigma_\epsilon$ must at least pass through a turning point behind the black hole horizon and one beyond the cosmological horizon. 
Specializing to the case $n=2$, we have a single turning point $r_{t_h}$ in the interior of the black hole, and two turning points outside the cosmological horizon, located at the same radial coordinate $r_{t_c}$ by symmetry.
When $|K_{\epsilon}| \geq K_{\rm crit}$, there are no connected surfaces.
In the other regime $|K_{\epsilon}| <K_{\rm crit}$, we perform manipulations similar to case 3 to get
\begin{align}\label{eq:Cepsilon car}
    \mathcal{C}^\epsilon&=\frac{2\Omega_{d-1} L^{d-2}}{G_N}\qty(\int_{r_{\rm st}}^{r_{\rm t_c}}+ \int^{r_{t_c}}_{r_{\rm t_h}})\frac{\qty(\frac{r}{L})^{2(d-1)}a(r)}{\sqrt{-\mathcal{U}(P_{\epsilon},\,r)}}d r~,
    \end{align}
\begin{equation}\label{eq:t CMC close form}
    t_R+t_L= 2\qty(\int_{r_{\rm st}}^{r_{\rm t_c}}+ \int^{r_{t_c}}_{r_{\rm t_h}})\frac{\left(P_{\epsilon} - \epsilon L\frac{\abs{K_\epsilon}}{d} \qty(\frac{r}{L})^d \right)d r}{f(r)\,\sqrt{-\mathcal{U}(P_{\epsilon}, r)}} ~,
\end{equation}
together with the constraint (\ref{eq:time dif turning points}). 
Now, there are two main possibilities:
\begin{itemize}
 \item When $t_R=-t_L$, the condition \eqref{eq:time dif turning points}, together with \eqref{eq:t CMC close form}, implies that
 $P_{\epsilon}=K_{\epsilon}=0$, thus leading to the time-independent evolution of CV already analyzed in section \ref{ssec:CV}.  
\item When we consider (without loss of generality) the time-dependent case $t_R= t_L$ proposed by the prescription \ref{rule},  
symmetry arguments imply that the turning points are located at $t_{t_c}=t_{t_h}=0$. In this case, one has to numerically check whether the constraint \eqref{eq:time dif turning points}, with vanishing left-hand-side, is satisfied.
The plots of the integral $I(P_{\epsilon}, K_{\epsilon})$ are depicted in Fig.~\ref{fig:CAny2ndInt} for the case $|K_{\epsilon}| < K_{\rm crit}$.
One can clearly see that the constraint is only satisfied for a specific value of $P_{\epsilon}$. Therefore, holographic complexity would only be defined for a single boundary time.     
\end{itemize}

In summary, CMC surfaces connecting the two stretched horizons only exist at a fixed symmetric boundary time.
Alternatively, we can have solutions defined at arbitrary (but antisymmetric) boundary times by evolving the trivial surface at constant $t=0$ with the time isometry.

\subsubsection*{Case 5}
Analogous to the evaluation of case 5 in section \ref{ssec:CV}, we consider a symmetric location of the black hole stretched horizons $r_{\rm st}^{\rm L}=r_{\rm st}^{\rm R}=r_{\rm st}$. We find that
    \begin{align}\label{eq:Cepsilon car3}
    \mathcal{C}^\epsilon&=\frac{2\Omega_{d-1} L^{d-2}}{G_N}\qty(\int^{r_{\rm st}}_{r_{\rm t_h}}+\int^{r_{\rm t_c}}_{r_{\rm  t_h}})\frac{\qty(\frac{r}{L})^{2(d-1)}a(r)}{\sqrt{-\mathcal{U}(P_{\epsilon},\,r)}}d r~,\\
 t_R+t_L&= -2\qty(\int^{r_{\rm st}}_{r_{\rm t_h}}+\int^{r_{\rm t_c}}_{r_{\rm  t_h}})\frac{\left(P_{\epsilon} - \epsilon L\frac{\abs{K_\epsilon}}{d} \qty(\frac{r}{L})^d \right)d r}{f(r)\,\sqrt{-\mathcal{U}(P_{\epsilon}, r)}} ~,   \label{eq:t CAny case5}
\end{align}
together with the constraint \eqref{eq:time dif turning points}.
The conclusions are the same as case 4: either the CMC slices degenerate to the volume case with a trivial evolution at antisymmetric times $t_L= -t_R$, or they exist at a single boundary time, such that $t_L=t_R$ and the constraint \eqref{eq:time dif turning points} is satisfied.

\section{Conclusions}
\label{sec:Conclusions}

In this work, we studied the time dependence of holographic complexity in the extended SdS black hole geometry for several configurations of the stretched horizons.
The results are summarized in table~\ref{tab:results}.

In case 1, which investigates the cosmological patch of the geometry, complexity is characterized by a hyperfast growth, extending previous studies in empty dS space \cite{Jorstad:2022mls}.
The only exception to this behaviour is represented by a class of codimension-one CAny observables, which present an eternal evolution with a linear increase at late times, if the extrinsic curvature on CMC slices where they are evaluated is larger than a critical value \cite{Aguilar-Gutierrez:2023zqm}.
These results may suggest that the hyperfast growth could be used as a way to discriminate whether a complexity conjecture is well-behaved or not.
In case 2, where only the black hole patch of the geometry is considered, we achieve the characteristic linear growth for all the complexity conjectures, in agreement with the behaviour of AdS black holes (\eg see \cite{Carmi:2017jqz,Belin:2021bga,Belin:2022xmt}).
This seems a robust feature that holds independently of the asymptotics of the spacetime where complexity is evaluated.
Cases 3-5 are novel set-ups that access both the interior of the black hole and the exterior of the cosmological horizon.
In the case of codimension-zero proposals, all the complexity observables are time-independent. This happens because the WDW patch always reaches timelike infinities $\mathcal{I}^{\pm}$ and the singularities. 
In particular, the portion of WDW patch emerging from the past of the Penrose diagram is always compensated by the disappearance of a region of equal size into the future part of the geometry. Remarkably, this feature happens separately on the left and right sides of the diagram, therefore it is not a consequence of any synchronization of the boundary times on the two stretched horizons.

For codimension-one observables, the situation is more peculiar.
Our main result is that in the volume case, \textit{connected extremal surfaces anchored at symmetric times to the stretched horizons only exist when $t_L=t_R=0$}.
In case 3, the isometries of the geometry allow for a non-trivial evolution of complexity that approaches linear growth at late times. 
In cases 4-5, the isometries imply that only the trivial evolution is possible, leading to a time-independent result. 
Similar features appear for the codimension-one CAny proposal.
We conclude that the extended SdS black hole represents an arena where the location of the stretched horizons distinguishes among different behaviours for the holographic proposals.

It is interesting to compare our results on codimension-one extremal surfaces with the recent study of spacelike geodesics connecting static sphere observers in SdS space \cite{Faruk:2023uzs}, in comparison to our case 3.  
When the boundary times are symmetric ($t_L=t_R$), they found that there exists a small window where these spacelike geodesics exist, that connect the future (past) of the inflating patch with the past (future) of the black hole patch.
In empty dS space, the only symmetric spacelike geodesics connecting antipodal static patch observers require $t_{L}=t_R=0$, \eg see \cite{Chapman:2021eyy,Galante:2022nhj}.
These findings resonate with our results, except that we only have the time $t_L=t_R=0$ allowed, instead of a small boundary time interval.
These discrepancies can be probably attributed to the different dimensionality between the geometric objects that we studied, compared to the analysis in the literature.
In the case of geodesics, it is interesting to notice that more general configurations are allowed by considering curves with a complex length \cite{Chapman:2022mqd,Aalsma:2022eru}.
This suggests that we may consider complex maximal volume to find a non-trivial time evolution for symmetric configurations in multiple copies of the SdS geometry.

\subsubsection*{Outlook}

One of the motivations for this work was to relate our results with the coarse-graining of information for general observables in quantum cosmology.
According to static patch holography, we expect that the information
about the geometry in the static patch can be retrieved from observables anchored at the stretched horizon, either near the black hole or the cosmological horizons.
On the other hand, in the quantum cosmology context, one usually considers a meta-observer that lives close to a spacelike surface (such as a reheating surface, see the horizontal purple lines in Fig.~\ref{fig:extSdS}).
One may expect that the observables anchored to the stretched horizons should give the same information to which this meta-observer has access. 
If this is true, then one would conclude that the encoding of information in the extended SdS geometry is redundant.
This is because the meta-observer living on a single spacelike slice close to $\mathcal{I}^{\pm}$ would have access to the information contained in the other copies of the extended SdS background as well.
Our results imply that if the observer were to measure gravitational probes such as those in case 3, they would be able to extract different types of dynamical information with codimension-one observables, while they would not find any time evolution from the codimension-zero observables that we studied.
In general, we expect that other codimension-zero observables in the CAny class \cite{Belin:2022xmt} could allow for non-trivial dynamics, as long as the bulk region where complexity is evaluated does not reach both $\mathcal{I}^\pm$ and the singularities of the black hole. We leave this analysis for future studies.

The covariant objects computed in this work can be interpreted as proper measurements of complexity only if we find a dual quantum state and a corresponding circuit that exhibits the same features. Circuits and gravitational settings based on a double-scaled SYK model were recently built in
\cite{Susskind:2021esx,Blommaert:2023opb,Narovlansky:2023lfz,Blommaert:2023wad,Verlinde:2024znh,Verlinde:2024zrh}.
One may then hope to refine these toy models using Nielsen or Krylov definitions of complexity in quantum mechanics \cite{Nielsen1,Nielsen2,Nielsen3,Parker:2018yvk}.

Next, let us discuss some further developments in the study of gravitational observables.
First, the switchback effect is a crucial ingredient of all the complexity proposals.
While this property has already been analyzed in asymptotically dS space in many instances \cite{Baiguera:2023tpt,Anegawa:2023wrk,Aguilar-Gutierrez:2023pnn},
it would be interesting to extend the analysis to the observables considered in this work.
In particular, a natural step would be to investigate whether the time-independent observables in cases 3-5 would develop non-trivial dynamics after the insertion of a shock wave.
Second, we plan to compute the holographic complexity conjectures in lower-dimensional gravity.
Two-dimensional empty dS space arises from the dimensional reduction of near-extremal SdS black holes in higher dimensions. Explicitly, when $r_{c,h}\rightarrow r_N$ defined in eq.~\eqref{eq:critical_mass_SdS}, the geometry is well-known to become dS$_2\times$S$^{d-1}$ (\eg see \cite{Anninos:2012qw})
    \begin{equation}
        \rmd s^2=-\qty( 1-\tfrac{\rho^2}{L_2^2})\rmd\tau^2+\qty( 1-\tfrac{\rho^2}{L_2^2})^{-1}\rmd \rho^2+r_N^2\rmd\Omega_{d-1}^2~,
    \end{equation}
where $L_2$ is the length scale of dS$_2$, and $\rho$ is a radial coordinate measuring the finite separation between the cosmological and black hole horizons. There are different reasons to be interested in this regime. 
First, let us consider the complexity rate obtained at late times for the CAny observables in cases 1-3 (see table \ref{tab:results} and references therein).
The results always depend on the location $r_{f_i}$ of a final slice which is approached by the extremal surface at late times. In the near-extremal limit, the solution for $r_f$ in eq.~\eqref{eq: rf equation} has been previously found to coincide with the Nariai radius $r_N$ \cite{Aguilar-Gutierrez:2023zqm}.
While in the general case the turning points in the cosmological and black hole region (if they both exist) satisfy $r_{t_c} > r_{t_h}$, the naive near-extremal limit $\mu \rightarrow \mu_N$ at late times would imply that $r_{t_c} \rightarrow r_{t_h} \rightarrow r_N $. According to this limit, we would conclude that the growth rate at late times reads
\begin{equation}
    \lim_{t\rightarrow\infty}\dv{t} \mathcal{C}^\epsilon =0~,
    \label{eq:extremal_limit}
\end{equation}
for cases 1-3, while this equation is already true without taking the near-extremal limit in cases 4-5.
The Lloyd bound on complexity for asymptotically AdS black holes is conjectured to take the form $d\mathcal{C}/dt \sim TS$ \cite{Susskind:2014rva,Brown:2015lvg}. 
The result \eqref{eq:extremal_limit} is usually associated to the vanishing temperature of extremal black holes. 
While the naive Hawking temperature associated with the SdS solution (see eq.~\eqref{eq:hor_cosm_temp_SdS}) vanishes in the limit $r_{c,h} \rightarrow r_N$, a careful analysis of the surface gravity experienced by a static patch observer moving along a worldline shows that the temperature should be non-zero (see footnote \ref{foot:extremal_temp}) \cite{Bousso:1996au,Morvan:2022ybp}. 
If we assume that the Lloyd bound also holds in SdS space, then we expect that a non-vanishing $d\mathcal{C}/dt\sim TS$ should be reproduced using the dS$_2\times$S$^{d-1}$ geometry.

Another reason to be interested in two-dimensional empty dS case is that one might be able to propose modifications of the CAny conjectures based on our higher dimensional results, in a similar fashion as the refinement of CV considered in \cite{Anegawa:2023wrk}.
In the case of the observables in the extended SdS$^n$ with $n>1$, the two-dimensional quantities analog to cases 4-5 are Jackiw-Teitelbom (JT) gravity multiverse models \cite{Aguilar-Gutierrez:2021bns,Levine:2022wos}.
Moreover, in two dimensions it is possible to consider centaur geometries where patches of dS space are glued to asymptotic AdS regions, which have a standard timelike boundary where AdS/CFT duality applies \cite{Anninos:2017hhn,Galante:2023uyf}. It would be interesting to study holographic complexity for multiple copies of this geometry and check whether the time-independence of the codimension-zero proposals in cases 3-5, and codimension-one proposals in 4-5 is lifted. 
In particular, the case of a single copy was studied in \cite{Chapman:2021eyy}, where no hyperfast growth arises. 

Finally, another interesting and challenging task would be to include quantum corrections in the evaluation of holographic complexity, as recently uncovered in JT gravity \cite{Carrasco:2023fcj}. 
A useful test ground for these corrections is provided by double holography, where quantum backreaction can be tracked in an exactly solvable regime \cite{Emparan:2021hyr,Chen:2023tpi}. A brane-world version of the extended SdS black hole was recently proposed in \cite{Aguilar-Gutierrez:2023zoi}, and a SdS$_3$ black hole with quantum corrections has appeared in \cite{Emparan:2022ijy}. 
Therefore, it might be useful to test whether the complexity proposals explored in our work display either a similar or a radically different behavior compared to these models.

\section*{Acknowledgements}
We thank Victor Franken, Thomas Hertog, Rob Myers, and Shan-Ming Ruan for useful discussions, and especially Rotem Berman and Juan Pedraza for their initial collaboration and interesting comments. 
We thank Roberto Auzzi, Shira Chapman and Giuseppe Nardelli for useful comments and discussions on a preliminary version of this work.
We benefited from the workshop ``Gravity meets quantum information" in Würzburg, which allowed this collaboration to start.
SEAG thanks the IFT-UAM/CSIC, the University of Amsterdam, the Delta Institute for Theoretical Physics, and the International Centre for Theoretical Physics for their hospitality and financial support during several phases of the project, and the Research Foundation - Flanders (FWO) for also providing mobility support.
The work of SEAG is partially supported by the FWO Research Project G0H9318N and the inter-university project iBOF/21/084.
The work of SB is supported by the Israel Science Foundation (grant No. 1417/21), the German Research Foundation through a German-Israeli Project Cooperation (DIP) grant “Holography and the Swampland”, Carole and Marcus Weinstein through the BGU Presidential Faculty Recruitment Fund, and the ISF Center of Excellence for theoretical high energy physics. SB is also supported by an Azrieli fellowship awarded by the Azrieli foundation.
The work of NZ is supported by MEXT KAKENHI Grant-in-Aid for Transformative Research Areas A ”Extreme Universe” No. 21H05184.

\appendix

\section{Details on the complexity=action proposal}
\label{app:details_CA}

In this appendix, we present additional technical details on the computation of the boundary action $I_{\rm bdy}$ in eq.~\eqref{eq:bdy_action} for various configurations of the WDW patch.

\subsection{Case 1}
\label{app:ssec:case1}

Assuming that $t_{\infty} > 0$, we focus on the regime at intermediate times $-t_{\infty} \leq t \leq t_{\infty}$ when the WDW patch takes the shape in Fig.~\ref{subfig:WDW_evo1_regime2}.
In this setting all the codimension-one boundaries are null, therefore there are no GHY terms.
In case 1, the special positions of the WDW patch $r_{\pm}$ are defined in eq.~\eqref{eq:umax_vmin_WDWpatch}.

\paragraph{Joint terms.}
The geometry contains four null-null joints: the top and bottom vertices of the WDW patch, plus two joints located at the stretched horizons.
The one-forms orthogonal to the null boundaries of the WDW patch are given by
\beq
\begin{aligned}
&  \mathrm{TL}: \qquad
k_{\mu}^{\rm TL} dx^{\mu} = - \alpha \le du + \frac{2}{f(r)} dr \ri\Big|_{v= -t_L + r^*(r_{\rm st}^L)} \\
&  \mathrm{BL}: \qquad
k_{\mu}^{\rm BL} dx^{\mu} = 
 \beta  du|_{u = -t_L - r^*(r_{\rm st}^L)} 
 \\
&  \mathrm{TR}: \qquad
k_{\mu}^{\rm TR} dx^{\mu} = 
 \beta'  du|_{u = t_R - r^*(r_{\rm st}^R)}  \\
&  \mathrm{BR}: \qquad
k_{\mu}^{\rm BR} dx^{\mu} = - \alpha' \le du + \frac{2}{f(r)} dr \ri\Big|_{v= t_R + r^*(r_{\rm st}^R)}\\
\end{aligned}
\label{eq:list_null_normals_joints}
\eeq
where R denotes right, L left, T top, B bottom.
The corresponding joint contributions read
\begin{subequations}
\beq
I_{\mathcal{J}}^{\rm st,L} = - \frac{\Omega_{d-1}}{8 \pi G_N} (r_{\rm st}^L)^{d-1} \log \left| \frac{\alpha \beta}{f(r_{\rm st}^L)} \right| \, ,
\qquad 
I_{\mathcal{J}}^{\rm st,R} = - \frac{\Omega_{d-1}}{8 \pi G_N} (r_{\rm st}^R)^{d-1} \log \left| \frac{\alpha' \beta'}{f(r_{\rm st}^R)} \right| \, ,
\label{eq:joints_stretched_horizons}
\eeq
\beq
I_{\mathcal{J}}^{\rm T} = \frac{\Omega_{d-1}}{8 \pi G_N} (r_{+})^{d-1} \log \left| \frac{\alpha \beta'}{f(r_+)}  \right| \, , \qquad
I_{\mathcal{J}}^{\rm B} = \frac{\Omega_{d-1}}{8 \pi G_N} (r_{-})^{d-1} \log \left| \frac{\alpha' \beta}{f(r_-)}  \right| \, . 
\eeq
\label{eq:joints_case1}
\end{subequations}

\paragraph{Contributions on null boundaries.}
The parametrization \eqref{eq:list_null_normals_joints} is affine, therefore the codimension-one boundary terms in eq.~\eqref{eq:null_codim1_term} vanish. 
The counterterm \eqref{eq:counterterm_action} is non-trivial.
To evaluate the latter, we compute the expansion parameter of the null geodesics delimiting the WDW patch
\beq
\begin{aligned}
&  \mathrm{TL}: \qquad
\Theta^{\rm TL}  =  - \frac{\alpha (d-1)}{r} \, , \qquad
&  \mathrm{BL}: \qquad
\Theta^{\rm BL}  = - \frac{\beta (d-1)}{r} \, ,  \\
&  \mathrm{TR}: \qquad
\Theta^{\rm TR}  =  - \frac{\beta' (d-1)}{r}  \, , \qquad
&  \mathrm{BR}: \qquad
\Theta^{\rm BR}  = - \frac{\alpha' (d-1)}{r} \, . \\
\end{aligned}
\eeq
Correspondingly, the null counterterm evaluated at each null boundary reads
\begin{subequations}
\beq
I_{\rm ct}^{\rm TL} = -\frac{\Omega_{d-1}}{8\pi G_N} \left\lbrace (r_{+})^{d-1} \left[ \log \left| \frac{\alpha \ell_{\rm ct} (d-1)}{r_{+}} \right| + \frac{1}{d-1} \right] 
- (r_{\rm st}^L)^{d-1} \left[ \log \left| \frac{\alpha \ell_{\rm ct} (d-1)}{r_{\rm st}^L} \right| + \frac{1}{d-1} \right]  \right\rbrace \, ,
\eeq
\beq
I_{\rm ct}^{\rm BL} = -\frac{\Omega_{d-1}}{8\pi G_N} \left\lbrace (r_{-})^{d-1} \left[ \log \left| \frac{\beta \ell_{\rm ct} (d-1)}{r_{-}} \right| + \frac{1}{d-1} \right] 
- (r_{\rm st}^L)^{d-1} \left[ \log \left| \frac{\beta \ell_{\rm ct} (d-1)}{r_{\rm st}^L} \right| + \frac{1}{d-1} \right]  \right\rbrace \, ,
\eeq
\beq
I_{\rm ct}^{\rm TR} = -\frac{\Omega_{d-1}}{8\pi G_N} \left\lbrace (r_{+})^{d-1} \left[ \log \left| \frac{\beta' \ell_{\rm ct} (d-1)}{r_{+}} \right| + \frac{1}{d-1} \right] 
- (r_{\rm st}^R)^{d-1} \left[ \log \left| \frac{\beta' \ell_{\rm ct} (d-1)}{r_{\rm st}^R} \right| + \frac{1}{d-1} \right]  \right\rbrace \, ,
\eeq
\beq
I_{\rm ct}^{\rm BR} = -\frac{\Omega_{d-1}}{8\pi G_N} \left\lbrace (r_{-})^{d-1} \left[ \log \left| \frac{\alpha' \ell_{\rm ct} (d-1)}{r_{-}} \right| + \frac{1}{d-1} \right] 
- (r_{\rm st}^R)^{d-1} \left[ \log \left| \frac{\alpha' \ell_{\rm ct} (d-1)}{r_{\rm st}^R} \right| + \frac{1}{d-1} \right]  \right\rbrace \, .
\eeq
\label{eq:Ict_case1}
\end{subequations}

\paragraph{Total boundary action.}
Finally, the boundary action \eqref{eq:bdy_action} is given by the summation of eqs.~\eqref{eq:joints_case1} and \eqref{eq:Ict_case1}. In the symmetric case $r_{\rm st}^L = r_{\rm st}^R$, the result reads
\beq
\begin{aligned}
   & I_{\rm bdy} (-t_{\infty} \leq t \leq t_{\infty}) = \frac{\Omega_{d-1}}{8 \pi G_N} \left\lbrace  
   - 2 (r_{\rm st})^{d-1} \left[ \log \left| \frac{(r_{\rm st})^2}{f(r_{\rm st}) \ell_{\rm ct}^2 (d-1)^2} \right| - \frac{2}{d-1} \right]
 \right. \\
& \left.  + (r_{+})^{d-1} \left[ \log \left| \frac{(r_{+})^2}{f (r_{+}) \ell_{\rm ct}^2 (d-1)^2} \right| - \frac{2}{d-1} \right] 
+ (r_{-})^{d-1} \left[ \log \left| \frac{(r_{-})^2}{f (r_{-}) \ell_{\rm ct}^2 (d-1)^2} \right| - \frac{2}{d-1} \right]
\right\rbrace \, ,
\end{aligned}
\label{eq:bdy_action_case1}
\eeq
where $r_{\pm}$ were defined in eq.~\eqref{eq:umax_vmin_WDWpatch}, and we assumed that the stretched horizons on the two sides of the geometry are located at the same radial coordinate, \ie $r^L_{\rm st} =r^{R}_{\rm st}$.

\subsection{Case 2}
\label{app:ssec:case2}

When the stretched horizons are determined according to case 2 in section \ref{ssec:stretched_horizons}, we need to study the early and late time regimes to find the characteristic linear growth of complexity.
Depending on the sign of the critical time $t_0$, the time dependence of the WDW patch can follow two different evolutions, represented by the Penrose diagrams in Fig.~\ref{fig:WDW_evo1} or \ref{fig:WDW_evo2}.
The formal expression of CA at early and late times is universal, while it differs at intermediate times between the two cases.

\paragraph{Evolution with critical time $t_0>0$ (see Fig.~\ref{fig:WDW_evo1}).}
The analytic expression of the boundary terms at intermediate times $-t_0 \leq t \leq t_0$ is given by the same analytic formula reported in eq.~\eqref{eq:bdy_action_case1}, with the only difference that the locations of the joints $r_{\pm}$ are now defined by eq.~\eqref{eq:umax_vmin_WDWpatch_case2}.

\vskip 2mm
\noindent
At times $t>t_0$, the configuration of the WDW patch is given by Fig.~\ref{subfig:WDW_evo1_regime3}.
The main novelty is the existence of GHY terms due to the intersection of the WDW patch with the spacelike cutoff surface at $r=r_{\rm min}$.
The corresponding induced metric reads
\beq
\sqrt{h} = r^{d-1} \sqrt{f (r)}\Big|_{r=r_{\rm min}} \, , \qquad
K = \sqrt{f(r)} \p_r \sqrt{h} = \frac{1}{2r \sqrt{f(r)}} \left[ 2(d-1) f(r) + r f'(r) \right]\Big|_{r= r_{\rm min}} \, ,
\label{eq:induced_metric_GHY}
\eeq
which gives the GHY contribution
\beq
\begin{aligned}
I_{\rm GHY} & = - \frac{\Omega_{d-1}}{16 \pi G_N} (r_{\rm min})^{d-2} \le  t_R + t_L + r^*(r_{\rm st}^R) + r^*(r_{\rm st}^L) - 2 r^* (r_{\rm min})  \ri \\
& \times  \left[ 2(d-1) f(r) + r f'(r) \right]\Big|_{r= r_{\rm min}}  \, .
\end{aligned}
\eeq
There are some special choices that simplify the previous result. First, we consider the symmetric time evolution \eqref{eq:symmetric_times}.
Second, we locate the stretched horizons at the same radial coordinate $r^R_{\rm st} = r^L_{\rm st}$.
Finally, we take the limit $r_{\rm min} \rightarrow 0$.
All together, these choices lead to 
\beq
\lim_{r_{\rm min} \rightarrow 0} I_{\rm GHY}  =  \frac{\Omega_{d-1}}{8 \pi G_N} \mu d \le  t + 2 r^*(r_{\rm st}) - 2 r^* (0)  \ri \, ,
\eeq
where $\mu$ is the mass parameter entering the blackening factor in eq.~\eqref{eq:asympt_dS}.

Next, let us evaluate the joints.
The expressions in eq.~\eqref{eq:joints_case1} are also valid in this setting, except for the top vertex $I_{\mathcal{J}}^{\rm T}$, which now moved behind the cutoff surface at $r=r_{\rm min}$.
There are instead two new joints arising from the intersection of the future boundaries of the WDW patch with the cutoff.
The one-form normal to such spacelike surface is given by
\beq
t_{\mu} dx^{\mu} = - \frac{dr}{\sqrt{-f(r)}}\Big|_{r = r_{\rm min}} \, . 
\eeq
This allows to determine the joints
\beq
I_{\mathcal{J}}^{\rm TL} = \frac{\Omega_{d-1}}{8 \pi G_N} (r_{\rm min})^{d-1} \log \left| \frac{\alpha}{f(r_{\rm min})}  \right| \, , \qquad
I_{\mathcal{J}}^{\rm TR} = \frac{\Omega_{d-1}}{8 \pi G_N} (r_{\rm min})^{d-1} \log \left| \frac{\beta'}{f(r_{\rm min})}  \right| \, . 
\label{eq:joints_case2_app}
\eeq
Tthese expressions vanish in the limit $r_{\rm min} \rightarrow 0$.

Finally, the counterterms \eqref{eq:Ict_case1} referring to the top-left and top-right null boundaries of the WDW patch get modified as
\begin{subequations}
\beq
I_{\rm ct}^{\rm TL} = \frac{\Omega_{d-1}}{8\pi G_N} 
(r_{\rm st})^{d-1} \left[ \log \left| \frac{\alpha \ell_{\rm ct} (d-1)}{r_{\rm st}} \right| + \frac{1}{d-1} \right]  \, ,
\eeq
\beq
I_{\rm ct}^{\rm TR} = \frac{\Omega_{d-1}}{8\pi G_N} 
 (r_{\rm st})^{d-1} \left[ \log \left| \frac{\beta' \ell_{\rm ct} (d-1)}{r_{\rm st}} \right| + \frac{1}{d-1} \right] \, ,
\eeq
\label{eq:Ict_case2_late_times}
\end{subequations}
where we already considered the symmetric case $r^R_{\rm st} = r^L_{\rm st}$ and performed the limit $r_{\rm min} \rightarrow 0$.
The contributions arising from the past boundaries in eq.~\eqref{eq:Ict_case1} is unchanged.

Summing all the boundary terms determined above according to eq.~\eqref{eq:bdy_action}, we obtain
\beq
\begin{aligned}
 I_{\rm bdy} (t>t_0) = \frac{\Omega_{d-1}}{8 \pi G_N} & \left\lbrace 
   \mu d \le  t + 2 r^*(r_{\rm st}) - 2 r^* (0)  \ri 
   - 2 (r_{\rm st})^{d-1} \left[ \log \left| \frac{(r_{\rm st})^2}{f_2(r_{\rm st}) \ell_{\rm ct}^2 (d-1)^2} \right| - \frac{2}{d-1} \right]
 \right. \\
& \left.  + (r_{-})^{d-1} \left[ \log \left| \frac{(r_{-})^2}{f (r_{-}) \ell_{\rm ct}^2 (d-1)^2} \right| - \frac{2}{d-1} \right]
\right\rbrace \, .
\end{aligned}
\label{eq:bdy_action_case2_latetimes}
\eeq

\vskip 2mm
\noindent
The computation of CA at times $t< t_0$ is very similar as the case of late times, with the roles of $r_+$ and $r_-$ exchanged.
We directly report the result
\beq
\begin{aligned}
 I_{\rm bdy} (t<-t_0) = \frac{\Omega_{d-1}}{8 \pi G_N} & \left\lbrace  
  \mu d \le  -t + 2 r^*(r_{\rm st}) - 2 r^* (0)  \ri 
  - 2 (r_{\rm st})^{d-1} \left[ \log \left| \frac{(r_{\rm st})^2}{f_2(r_{\rm st}) \ell_{\rm ct}^2 (d-1)^2} \right| - \frac{2}{d-1} \right]
 \right. \\
& \left.  + (r_{+})^{d-1} \left[ \log \left| \frac{(r_{+})^2}{f (r_{+}) \ell_{\rm ct}^2 (d-1)^2} \right| - \frac{2}{d-1} \right] 
\right\rbrace \, .
\end{aligned}
\label{eq:bdy_action_case2_earlytimes}
\eeq

\paragraph{Evolution with critical time $t_0<0$ (see Fig.~\ref{fig:WDW_evo2}).}
Since we already argued that early and late times contribute to CA with the same expressions \eqref{eq:bdy_action_case2_earlytimes} and \eqref{eq:bdy_action_case2_latetimes} computed above, it is sufficient to focus on the intermediate time regime $t_0 \leq t \leq -t_0$.

There are two GHY terms, coming from either the past or the future cutoff surfaces near the singularities.
The sum of these contributions reads
\beq
I_{\rm GHY}^{\rm T} + I_{\rm GHY}^{\rm B}  = \frac{\Omega_{d-1}}{8 \pi G_N} (r_{\rm min})^{d-2}
\le r^*(r_{\rm st}^L) + r^*(r_{\rm st}^R) - 2 r^*(r_{\rm min})   \ri 
\left[ 2(d-1) f(r_{\rm min}) + r f'(r_{\rm min}) \right] \, .
\label{eq:IGHY_case2_app}
\eeq
Next, we consider the joints.
At the stretched horizons we have the usual contributions in eq.~\eqref{eq:joints_stretched_horizons}.
The intersections with the singularities lead to vanishing results when performing the limit $r_{\rm min} \rightarrow 0$, similarly to the analog computation performed in eq.~\eqref{eq:joints_case2_app}.
Finally, the counterterms include contributions similar to eq.~\eqref{eq:Ict_case2_late_times}, and their sum reads
\beq
\begin{aligned}
I_{\rm ct} = \frac{\Omega_{d-1}}{8\pi G_N} & \left\lbrace 
(r_{\rm st}^L)^{d-1} \left[ \log \left| \frac{\alpha \ell_{\rm ct} (d-1)}{r_{\rm st}^L} \right| + \frac{1}{d-1} \right]
+ (r_{\rm st}^R)^{d-1} \left[ \log \left| \frac{\beta' \ell_{\rm ct} (d-1)}{r_{\rm st}^R} \right| + \frac{1}{d-1} \right] \right. \\
& \left.  
+ (r_{\rm st}^L)^{d-1} \left[ \log \left| \frac{\beta \ell_{\rm ct} (d-1)}{r_{\rm st}^L} \right| + \frac{1}{d-1} \right]
+ (r_{\rm st}^R)^{d-1} \left[ \log \left| \frac{\alpha' \ell_{\rm ct} (d-1)}{r_{\rm st}^R} \right| + \frac{1}{d-1} \right]
\right\rbrace \, .
\end{aligned}
\label{eq:Ict_case2_app}
\eeq
The relevant observation is that the sum of boundary terms \eqref{eq:IGHY_case2_app} and \eqref{eq:Ict_case2_app} is time-independent, \ie 
\beq
\frac{d I_{\rm bdy}}{dt} (t_0 \leq t \leq -t_0) = 0 \, .
\label{eq:rate_bdy_app_case2}
\eeq

\subsection{Cases 3-5}
\label{app:ssec:case3}

In case 3, the top and bottom joints of the WDW patch always lie behind the cutoff surfaces, see Fig.~\ref{fig:WDW_BH_stretched}.
In the following, we will focus on the left side of the diagram.

\paragraph{GHY terms.}
There are two codimension-one spacelike surfaces, corresponding to the intersection of the WDW patch with the cutoff surfaces located near the singularities.
The induced metric is given by eq.~\eqref{eq:induced_metric_GHY}, and the corresponding GHY terms sum to
\small
\beq
I_{\rm GHY}^{\rm L} =  \frac{\Omega_{d-1}}{16\pi G_N} (r_{\rm max})^{d-2}
\left[ 2(d-1) f(r_{\rm max}) + r f'(r_{\rm max}) \right]  \le  \int_{-\infty}^{-t_L-r^*(r_{\rm max})} du  + \int_{-t_L+r^*(r_{\rm max})}^{\infty} dv  \ri  
\, ,
\eeq
\normalsize
where we expressed the result as an integral over $u(v)$ for convenience.
The important observation is that the time dependence on $t_L$ cancels when summing the two terms.

\paragraph{Joint terms.}
The right joint evaluated at the stretched horizons has the same formal expression $I_{\mathcal{J}}^{\rm st, L}$ computed in eq.~\eqref{eq:joints_stretched_horizons}.
The intersections between the cutoff surfaces and the WDW patch generate two more joints on the left region, given by  
\beq
I_{\mathcal{J}}^{\rm TL} =  \frac{\Omega_{d-1}}{8 \pi G_N} (r_{\rm max})^{d-1} \log \left| \frac{\alpha}{\sqrt{f(r_{\rm max})}} \right| \, ,
\qquad
I_{\mathcal{J}}^{\rm BL} =  \frac{\Omega_{d-1}}{8 \pi G_N} (r_{\rm max})^{d-1} \log \left| \frac{\beta}{\sqrt{f(r_{\rm max})}} \right| \, .
\eeq

\paragraph{Counterterm.}
The parametrization \eqref{eq:list_null_normals_joints} is affine, therefore there are no codimension-one boundary terms for the null surfaces except for the counterterms.
The null counterterms read
\begin{subequations}
\beq
I_{\rm ct}^{\rm TL} = -\frac{\Omega_{d-1}}{8\pi G_N} \left\lbrace (r_{\rm max})^{d-1} \left[ \log \left| \frac{\alpha \ell_{\rm ct} (d-1)}{r_{\rm max}} \right| + \frac{1}{d-1} \right] 
- (r_{\rm st})^{d-1} \left[ \log \left| \frac{\alpha \ell_{\rm ct} (d-1)}{r_{\rm st}} \right| + \frac{1}{d-1} \right]  \right\rbrace \, ,
\eeq
\beq
I_{\rm ct}^{\rm BL} = -\frac{\Omega_{d-1}}{8\pi G_N} \left\lbrace (r_{\rm max})^{d-1} \left[ \log \left| \frac{\beta \ell_{\rm ct} (d-1)}{r_{\rm max}} \right| + \frac{1}{d-1} \right] 
- (r_{\rm st})^{d-1} \left[ \log \left| \frac{\beta \ell_{\rm ct} (d-1)}{r_{\rm st}} \right| + \frac{1}{d-1} \right]  \right\rbrace \, .
\eeq
\end{subequations}
Combining with the joints, the normalization of null normals cancel from the action.

\paragraph{Total sum.}
The sum of all the boundary terms \eqref{eq:bdy_action} on the left side of the diagram is time-independent, without need for any particular assumption on the boundary times $t_L, t_R$.
An analog computation reveals that the right side is also independently time-independent, leading to the result
\beq
\frac{d I_{\rm bdy}}{dt_L} = \frac{dI_{\rm bdy}}{dt_R}=0 \, .
\label{eq:rate_CA_bdy_case3}
\eeq

\vskip 2mm
\noindent
Finally, let us point out that cases 4 and 5 do not present additional new terms, instead they always reduce to contributions similar to other cases.

\bibliographystyle{JHEP}
\bibliography{references.bib}

\providecommand{\href}[2]{#2}\begingroup\raggedright\begin{thebibliography}{100}

\bibitem{tHooft:1993dmi}
G.~'t~Hooft, \emph{{Dimensional reduction in quantum gravity}}, {\emph{Conf.
  Proc. C} {\bfseries 930308} (1993) 284}
  [\href{https://arxiv.org/abs/gr-qc/9310026}{{\ttfamily gr-qc/9310026}}].

\bibitem{Susskind:1994vu}
L.~Susskind, \emph{{The World as a hologram}},
  \href{https://doi.org/10.1063/1.531249}{\emph{J. Math. Phys.} {\bfseries 36}
  (1995) 6377} [\href{https://arxiv.org/abs/hep-th/9409089}{{\ttfamily
  hep-th/9409089}}].

\bibitem{Ryu:2006bv}
S.~Ryu and T.~Takayanagi, \emph{{Holographic derivation of entanglement entropy
  from AdS/CFT}},
  \href{https://doi.org/10.1103/PhysRevLett.96.181602}{\emph{Phys. Rev. Lett.}
  {\bfseries 96} (2006) 181602}
  [\href{https://arxiv.org/abs/hep-th/0603001}{{\ttfamily hep-th/0603001}}].

\bibitem{Maldacena:1997re}
J.M.~Maldacena, \emph{{The Large N limit of superconformal field theories and
  supergravity}}, \href{https://doi.org/10.4310/ATMP.1998.v2.n2.a1}{\emph{Adv.
  Theor. Math. Phys.} {\bfseries 2} (1998) 231}
  [\href{https://arxiv.org/abs/hep-th/9711200}{{\ttfamily hep-th/9711200}}].

\bibitem{Galante:2023uyf}
D.A.~Galante, \emph{{Modave lectures on de Sitter space \& holography}},
  \href{https://doi.org/10.22323/1.435.0003}{\emph{PoS} {\bfseries Modave2022}
  (2023) 003} [\href{https://arxiv.org/abs/2306.10141}{{\ttfamily
  2306.10141}}].

\bibitem{PhysRevD.15.2738}
G.W.~Gibbons and S.W.~Hawking, \emph{Cosmological event horizons,
  thermodynamics, and particle creation},
  \href{https://doi.org/10.1103/PhysRevD.15.2738}{\emph{Phys. Rev. D}
  {\bfseries 15} (1977) 2738}.

\bibitem{Almheiri:2020cfm}
A.~Almheiri, T.~Hartman, J.~Maldacena, E.~Shaghoulian and A.~Tajdini,
  \emph{{The entropy of Hawking radiation}},
  \href{https://doi.org/10.1103/RevModPhys.93.035002}{\emph{Rev. Mod. Phys.}
  {\bfseries 93} (2021) 035002}
  [\href{https://arxiv.org/abs/2006.06872}{{\ttfamily 2006.06872}}].

\bibitem{Bousso:1999dw}
R.~Bousso, \emph{{The Holographic principle for general backgrounds}},
  \href{https://doi.org/10.1088/0264-9381/17/5/309}{\emph{Class. Quant. Grav.}
  {\bfseries 17} (2000) 997}
  [\href{https://arxiv.org/abs/hep-th/9911002}{{\ttfamily hep-th/9911002}}].

\bibitem{Bousso:2000nf}
R.~Bousso, \emph{{Positive vacuum energy and the N bound}},
  \href{https://doi.org/10.1088/1126-6708/2000/11/038}{\emph{JHEP} {\bfseries
  11} (2000) 038} [\href{https://arxiv.org/abs/hep-th/0010252}{{\ttfamily
  hep-th/0010252}}].

\bibitem{Banks:2006rx}
T.~Banks, B.~Fiol and A.~Morisse, \emph{{Towards a quantum theory of de Sitter
  space}}, \href{https://doi.org/10.1088/1126-6708/2006/12/004}{\emph{JHEP}
  {\bfseries 12} (2006) 004}
  [\href{https://arxiv.org/abs/hep-th/0609062}{{\ttfamily hep-th/0609062}}].

\bibitem{Anninos:2011af}
D.~Anninos, S.A.~Hartnoll and D.M.~Hofman, \emph{{Static Patch Solipsism:
  Conformal Symmetry of the de Sitter Worldline}},
  \href{https://doi.org/10.1088/0264-9381/29/7/075002}{\emph{Class. Quant.
  Grav.} {\bfseries 29} (2012) 075002}
  [\href{https://arxiv.org/abs/1109.4942}{{\ttfamily 1109.4942}}].

\bibitem{Banks:2018ypk}
T.~Banks and W.~Fischler, \emph{{The holographic spacetime model of
  cosmology}}, \href{https://doi.org/10.1142/S0218271818460057}{\emph{Int. J.
  Mod. Phys. D} {\bfseries 27} (2018) 1846005}
  [\href{https://arxiv.org/abs/1806.01749}{{\ttfamily 1806.01749}}].

\bibitem{Shaghoulian:2021cef}
E.~Shaghoulian, \emph{{The central dogma and cosmological horizons}},
  \href{https://doi.org/10.1007/JHEP01(2022)132}{\emph{JHEP} {\bfseries 01}
  (2022) 132} [\href{https://arxiv.org/abs/2110.13210}{{\ttfamily
  2110.13210}}].

\bibitem{Susskind:2021esx}
L.~Susskind, \emph{{Entanglement and Chaos in De Sitter Space Holography: An
  SYK Example}}, \href{https://doi.org/10.22128/jhap.2021.455.1005}{\emph{JHAP}
  {\bfseries 1} (2021) 1} [\href{https://arxiv.org/abs/2109.14104}{{\ttfamily
  2109.14104}}].

\bibitem{Susskind:2022dfz}
L.~Susskind, \emph{{Scrambling in Double-Scaled SYK and De Sitter Space}},
  \href{https://arxiv.org/abs/2205.00315}{{\ttfamily 2205.00315}}.

\bibitem{Lin:2022nss}
H.~Lin and L.~Susskind, \emph{{Infinite Temperature's Not So Hot}},
  \href{https://arxiv.org/abs/2206.01083}{{\ttfamily 2206.01083}}.

\bibitem{Susskind:2022bia}
L.~Susskind, \emph{{De Sitter Space, Double-Scaled SYK, and the Separation of
  Scales in the Semiclassical Limit}},
  \href{https://arxiv.org/abs/2209.09999}{{\ttfamily 2209.09999}}.

\bibitem{Susskind:2023hnj}
L.~Susskind, \emph{{De Sitter Space has no Chords. Almost Everything is
  Confined.}}, \href{https://doi.org/10.22128/jhap.2023.661.1043}{\emph{JHAP}
  {\bfseries 3} (2023) 1} [\href{https://arxiv.org/abs/2303.00792}{{\ttfamily
  2303.00792}}].

\bibitem{Susskind:2023rxm}
L.~Susskind, \emph{{A Paradox and its Resolution Illustrate Principles of de
  Sitter Holography}},  \href{https://arxiv.org/abs/2304.00589}{{\ttfamily
  2304.00589}}.

\bibitem{Rahman:2022jsf}
A.A.~Rahman, \emph{{dS JT Gravity and Double-Scaled SYK}},
  \href{https://arxiv.org/abs/2209.09997}{{\ttfamily 2209.09997}}.

\bibitem{Nomura:2017fyh}
Y.~Nomura, P.~Rath and N.~Salzetta, \emph{{Spacetime from Unentanglement}},
  \href{https://doi.org/10.1103/PhysRevD.97.106010}{\emph{Phys. Rev. D}
  {\bfseries 97} (2018) 106010}
  [\href{https://arxiv.org/abs/1711.05263}{{\ttfamily 1711.05263}}].

\bibitem{Nomura:2019qps}
Y.~Nomura, \emph{{Spacetime and Universal Soft Modes --- Black Holes and
  Beyond}}, \href{https://doi.org/10.1103/PhysRevD.101.066024}{\emph{Phys. Rev.
  D} {\bfseries 101} (2020) 066024}
  [\href{https://arxiv.org/abs/1908.05728}{{\ttfamily 1908.05728}}].

\bibitem{Murdia:2022giv}
C.~Murdia, Y.~Nomura and K.~Ritchie, \emph{{Black hole and de Sitter
  microstructures from a semiclassical perspective}},
  \href{https://doi.org/10.1103/PhysRevD.107.026016}{\emph{Phys. Rev. D}
  {\bfseries 107} (2023) 026016}
  [\href{https://arxiv.org/abs/2207.01625}{{\ttfamily 2207.01625}}].

\bibitem{Shaghoulian:2022fop}
E.~Shaghoulian and L.~Susskind, \emph{{Entanglement in De Sitter space}},
  \href{https://doi.org/10.1007/JHEP08(2022)198}{\emph{JHEP} {\bfseries 08}
  (2022) 198} [\href{https://arxiv.org/abs/2201.03603}{{\ttfamily
  2201.03603}}].

\bibitem{Franken:2023pni}
V.~Franken, H.~Partouche, F.~Rondeau and N.~Toumbas, \emph{{Bridging the static
  patches: de Sitter holography and entanglement}},
  \href{https://doi.org/10.1007/JHEP08(2023)074}{\emph{JHEP} {\bfseries 08}
  (2023) 074} [\href{https://arxiv.org/abs/2305.12861}{{\ttfamily
  2305.12861}}].

\bibitem{Franken:2023jas}
V.~Franken, H.~Partouche, F.~Rondeau and N.~Toumbas, \emph{{Closed FRW
  holography: A time-dependent ER=EPR realization}},
  \href{https://arxiv.org/abs/2310.20652}{{\ttfamily 2310.20652}}.

\bibitem{Strominger:2001pn}
A.~Strominger, \emph{{The dS / CFT correspondence}},
  \href{https://doi.org/10.1088/1126-6708/2001/10/034}{\emph{JHEP} {\bfseries
  10} (2001) 034} [\href{https://arxiv.org/abs/hep-th/0106113}{{\ttfamily
  hep-th/0106113}}].

\bibitem{Anninos:2017hhn}
D.~Anninos and D.M.~Hofman, \emph{{Infrared Realization of dS$_2$ in AdS$_2$}},
  \href{https://doi.org/10.1088/1361-6382/aab143}{\emph{Class. Quant. Grav.}
  {\bfseries 35} (2018) 085003}
  [\href{https://arxiv.org/abs/1703.04622}{{\ttfamily 1703.04622}}].

\bibitem{Shyam:2021ciy}
V.~Shyam, \emph{{$ \mathrm{T}\overline{\mathrm{T}} $ +
  \ensuremath{\Lambda}$_{2}$ deformed CFT on the stretched dS$_{3}$ horizon}},
  \href{https://doi.org/10.1007/JHEP04(2022)052}{\emph{JHEP} {\bfseries 04}
  (2022) 052} [\href{https://arxiv.org/abs/2106.10227}{{\ttfamily
  2106.10227}}].

\bibitem{Lewkowycz:2019xse}
A.~Lewkowycz, J.~Liu, E.~Silverstein and G.~Torroba, \emph{{$ T\overline{T} $
  and EE, with implications for (A)dS subregion encodings}},
  \href{https://doi.org/10.1007/JHEP04(2020)152}{\emph{JHEP} {\bfseries 04}
  (2020) 152} [\href{https://arxiv.org/abs/1909.13808}{{\ttfamily
  1909.13808}}].

\bibitem{Coleman:2021nor}
E.~Coleman, E.A.~Mazenc, V.~Shyam, E.~Silverstein, R.M.~Soni, G.~Torroba
  et~al., \emph{{De Sitter microstates from T$ \overline{T} $ +
  \ensuremath{\Lambda}$_{2}$ and the Hawking-Page transition}},
  \href{https://doi.org/10.1007/JHEP07(2022)140}{\emph{JHEP} {\bfseries 07}
  (2022) 140} [\href{https://arxiv.org/abs/2110.14670}{{\ttfamily
  2110.14670}}].

\bibitem{Anninos:2020hfj}
D.~Anninos, F.~Denef, Y.T.A.~Law and Z.~Sun, \emph{{Quantum de Sitter horizon
  entropy from quasicanonical bulk, edge, sphere and topological string
  partition functions}},
  \href{https://doi.org/10.1007/JHEP01(2022)088}{\emph{JHEP} {\bfseries 01}
  (2022) 088} [\href{https://arxiv.org/abs/2009.12464}{{\ttfamily
  2009.12464}}].

\bibitem{Banihashemi:2022htw}
B.~Banihashemi, T.~Jacobson, A.~Svesko and M.~Visser, \emph{{The minus sign in
  the first law of de Sitter horizons}},
  \href{https://doi.org/10.1007/JHEP01(2023)054}{\emph{JHEP} {\bfseries 01}
  (2023) 054} [\href{https://arxiv.org/abs/2208.11706}{{\ttfamily
  2208.11706}}].

\bibitem{Chandrasekaran:2022cip}
V.~Chandrasekaran, R.~Longo, G.~Penington and E.~Witten, \emph{{An algebra of
  observables for de Sitter space}},
  \href{https://doi.org/10.1007/JHEP02(2023)082}{\emph{JHEP} {\bfseries 02}
  (2023) 082} [\href{https://arxiv.org/abs/2206.10780}{{\ttfamily
  2206.10780}}].

\bibitem{Witten:2023qsv}
E.~Witten, \emph{{Algebras, Regions, and Observers}},
  \href{https://arxiv.org/abs/2303.02837}{{\ttfamily 2303.02837}}.

\bibitem{Witten:2023xze}
E.~Witten, \emph{{A Background Independent Algebra in Quantum Gravity}},
  \href{https://arxiv.org/abs/2308.03663}{{\ttfamily 2308.03663}}.

\bibitem{Mirbabayi:2023vgl}
M.~Mirbabayi, \emph{{An Observer's Measure of De Sitter Entropy}},
  \href{https://arxiv.org/abs/2311.07724}{{\ttfamily 2311.07724}}.

\bibitem{Aguilar-Gutierrez:2023odp}
S.E.~Aguilar-Gutierrez, E.~Bahiru and R.~Esp\'\i{}ndola, \emph{{The
  centaur-algebra of observables}},
  \href{https://arxiv.org/abs/2307.04233}{{\ttfamily 2307.04233}}.

\bibitem{Kudler-Flam:2023qfl}
J.~Kudler-Flam, S.~Leutheusser and G.~Satishchandran, \emph{{Generalized Black
  Hole Entropy is von Neumann Entropy}},
  \href{https://arxiv.org/abs/2309.15897}{{\ttfamily 2309.15897}}.

\bibitem{Chapman:2021jbh}
S.~Chapman and G.~Policastro, \emph{{Quantum computational complexity from
  quantum information to black holes and back}},
  \href{https://doi.org/10.1140/epjc/s10052-022-10037-1}{\emph{Eur. Phys. J. C}
  {\bfseries 82} (2022) 128}
  [\href{https://arxiv.org/abs/2110.14672}{{\ttfamily 2110.14672}}].

\bibitem{Maldacena:2013xja}
J.~Maldacena and L.~Susskind, \emph{{Cool horizons for entangled black holes}},
  \href{https://doi.org/10.1002/prop.201300020}{\emph{Fortsch. Phys.}
  {\bfseries 61} (2013) 781} [\href{https://arxiv.org/abs/1306.0533}{{\ttfamily
  1306.0533}}].

\bibitem{Susskind:2014moa}
L.~Susskind, \emph{{Entanglement is not enough}},
  \href{https://doi.org/10.1002/prop.201500095}{\emph{Fortsch. Phys.}
  {\bfseries 64} (2016) 49} [\href{https://arxiv.org/abs/1411.0690}{{\ttfamily
  1411.0690}}].

\bibitem{Susskind:2014rva}
L.~Susskind, \emph{{Computational Complexity and Black Hole Horizons}},
  \href{https://doi.org/10.1002/prop.201500092}{\emph{Fortsch. Phys.}
  {\bfseries 64} (2016) 24} [\href{https://arxiv.org/abs/1403.5695}{{\ttfamily
  1403.5695}}].

\bibitem{Stanford:2014jda}
D.~Stanford and L.~Susskind, \emph{{Complexity and Shock Wave Geometries}},
  \href{https://doi.org/10.1103/PhysRevD.90.126007}{\emph{Phys. Rev. D}
  {\bfseries 90} (2014) 126007}
  [\href{https://arxiv.org/abs/1406.2678}{{\ttfamily 1406.2678}}].

\bibitem{Couch:2016exn}
J.~Couch, W.~Fischler and P.H.~Nguyen, \emph{{Noether charge, black hole
  volume, and complexity}},
  \href{https://doi.org/10.1007/JHEP03(2017)119}{\emph{JHEP} {\bfseries 03}
  (2017) 119} [\href{https://arxiv.org/abs/1610.02038}{{\ttfamily
  1610.02038}}].

\bibitem{Brown:2015bva}
A.R.~Brown, D.A.~Roberts, L.~Susskind, B.~Swingle and Y.~Zhao,
  \emph{{Holographic Complexity Equals Bulk Action?}},
  \href{https://doi.org/10.1103/PhysRevLett.116.191301}{\emph{Phys. Rev. Lett.}
  {\bfseries 116} (2016) 191301}
  [\href{https://arxiv.org/abs/1509.07876}{{\ttfamily 1509.07876}}].

\bibitem{Brown:2015lvg}
A.R.~Brown, D.A.~Roberts, L.~Susskind, B.~Swingle and Y.~Zhao,
  \emph{{Complexity, action, and black holes}},
  \href{https://doi.org/10.1103/PhysRevD.93.086006}{\emph{Phys. Rev. D}
  {\bfseries 93} (2016) 086006}
  [\href{https://arxiv.org/abs/1512.04993}{{\ttfamily 1512.04993}}].

\bibitem{Belin:2021bga}
A.~Belin, R.C.~Myers, S.-M.~Ruan, G.~S\'arosi and A.J.~Speranza, \emph{{Does
  Complexity Equal Anything?}},
  \href{https://doi.org/10.1103/PhysRevLett.128.081602}{\emph{Phys. Rev. Lett.}
  {\bfseries 128} (2022) 081602}
  [\href{https://arxiv.org/abs/2111.02429}{{\ttfamily 2111.02429}}].

\bibitem{Belin:2022xmt}
A.~Belin, R.C.~Myers, S.-M.~Ruan, G.~S\'arosi and A.J.~Speranza,
  \emph{{Complexity equals anything II}},
  \href{https://doi.org/10.1007/JHEP01(2023)154}{\emph{JHEP} {\bfseries 01}
  (2023) 154} [\href{https://arxiv.org/abs/2210.09647}{{\ttfamily
  2210.09647}}].

\bibitem{Omidi:2022whq}
F.~Omidi, \emph{{Generalized volume-complexity for two-sided hyperscaling
  violating black branes}},
  \href{https://doi.org/10.1007/JHEP01(2023)105}{\emph{JHEP} {\bfseries 01}
  (2023) 105} [\href{https://arxiv.org/abs/2207.05287}{{\ttfamily
  2207.05287}}].

\bibitem{Jorstad:2023kmq}
E.~J\o{}rstad, R.C.~Myers and S.-M.~Ruan, \emph{{Complexity=anything:
  singularity probes}},
  \href{https://doi.org/10.1007/JHEP07(2023)223}{\emph{JHEP} {\bfseries 07}
  (2023) 223} [\href{https://arxiv.org/abs/2304.05453}{{\ttfamily
  2304.05453}}].

\bibitem{Bao:2017iye}
N.~Bao, C.~Cao, S.M.~Carroll and L.~McAllister, \emph{{Quantum Circuit
  Cosmology: The Expansion of the Universe Since the First Qubit}},
  \href{https://arxiv.org/abs/1702.06959}{{\ttfamily 1702.06959}}.

\bibitem{Bao:2017qmt}
N.~Bao, C.~Cao, S.M.~Carroll and A.~Chatwin-Davies, \emph{{De Sitter Space as a
  Tensor Network: Cosmic No-Hair, Complementarity, and Complexity}},
  \href{https://doi.org/10.1103/PhysRevD.96.123536}{\emph{Phys. Rev. D}
  {\bfseries 96} (2017) 123536}
  [\href{https://arxiv.org/abs/1709.03513}{{\ttfamily 1709.03513}}].

\bibitem{Niermann:2021wco}
L.~Niermann and T.J.~Osborne, \emph{{Holographic networks for (1+1)-dimensional
  de~Sitter space-time}},
  \href{https://doi.org/10.1103/PhysRevD.105.125009}{\emph{Phys. Rev. D}
  {\bfseries 105} (2022) 125009}
  [\href{https://arxiv.org/abs/2102.09223}{{\ttfamily 2102.09223}}].

\bibitem{Cao:2023gkw}
C.~Cao, W.~Chemissany, A.~Jahn and Z.~Zimbor\'as, \emph{{Overlapping qubits
  from non-isometric maps and de Sitter tensor networks}},
  \href{https://arxiv.org/abs/2304.02673}{{\ttfamily 2304.02673}}.

\bibitem{Jorstad:2022mls}
E.~J\o{}rstad, R.C.~Myers and S.-M.~Ruan, \emph{{Holographic complexity in
  dS$_{d+1}$}}, \href{https://doi.org/10.1007/JHEP05(2022)119}{\emph{JHEP}
  {\bfseries 05} (2022) 119}
  [\href{https://arxiv.org/abs/2202.10684}{{\ttfamily 2202.10684}}].

\bibitem{Auzzi:2023qbm}
R.~Auzzi, G.~Nardelli, G.P.~Ungureanu and N.~Zenoni, \emph{{Volume complexity
  of dS bubbles}},
  \href{https://doi.org/10.1103/PhysRevD.108.026006}{\emph{Phys. Rev. D}
  {\bfseries 108} (2023) 026006}
  [\href{https://arxiv.org/abs/2302.03584}{{\ttfamily 2302.03584}}].

\bibitem{Anegawa:2023wrk}
T.~Anegawa, N.~Iizuka, S.K.~Sake and N.~Zenoni, \emph{{Is action complexity
  better for de Sitter space in Jackiw-Teitelboim gravity?}},
  \href{https://doi.org/10.1007/JHEP06(2023)213}{\emph{JHEP} {\bfseries 06}
  (2023) 213} [\href{https://arxiv.org/abs/2303.05025}{{\ttfamily
  2303.05025}}].

\bibitem{Chapman:2021eyy}
S.~Chapman, D.A.~Galante and E.D.~Kramer, \emph{{Holographic complexity and de
  Sitter space}}, \href{https://doi.org/10.1007/JHEP02(2022)198}{\emph{JHEP}
  {\bfseries 02} (2022) 198}
  [\href{https://arxiv.org/abs/2110.05522}{{\ttfamily 2110.05522}}].

\bibitem{Aguilar-Gutierrez:2023zqm}
S.E.~Aguilar-Gutierrez, M.P.~Heller and S.~Van~der Schueren, \emph{{Complexity
  = Anything Can Grow Forever in de Sitter}},
  \href{https://arxiv.org/abs/2305.11280}{{\ttfamily 2305.11280}}.

\bibitem{Aguilar-Gutierrez:2023tic}
S.E.~Aguilar-Gutierrez, A.K.~Patra and J.F.~Pedraza, \emph{{Entangled universes
  in dS wedge holography}},
  \href{https://doi.org/10.1007/JHEP10(2023)156}{\emph{JHEP} {\bfseries 10}
  (2023) 156} [\href{https://arxiv.org/abs/2308.05666}{{\ttfamily
  2308.05666}}].

\bibitem{Anegawa:2023dad}
T.~Anegawa and N.~Iizuka, \emph{{Shock waves and delay of hyperfast growth in
  de Sitter complexity}},
  \href{https://doi.org/10.1007/JHEP08(2023)115}{\emph{JHEP} {\bfseries 08}
  (2023) 115} [\href{https://arxiv.org/abs/2304.14620}{{\ttfamily
  2304.14620}}].

\bibitem{Baiguera:2023tpt}
S.~Baiguera, R.~Berman, S.~Chapman and R.C.~Myers, \emph{{The cosmological
  switchback effect}},
  \href{https://doi.org/10.1007/JHEP07(2023)162}{\emph{JHEP} {\bfseries 07}
  (2023) 162} [\href{https://arxiv.org/abs/2304.15008}{{\ttfamily
  2304.15008}}].

\bibitem{Aguilar-Gutierrez:2023pnn}
S.E.~Aguilar-Gutierrez, \emph{{C=Anything and the switchback effect in
  Schwarzschild-de Sitter space}},
  \href{https://doi.org/10.1007/JHEP03(2024)062}{\emph{JHEP} {\bfseries 03}
  (2024) 062} [\href{https://arxiv.org/abs/2309.05848}{{\ttfamily
  2309.05848}}].

\bibitem{Gao:2000ga}
S.~Gao and R.M.~Wald, \emph{{Theorems on gravitational time delay and related
  issues}}, \href{https://doi.org/10.1088/0264-9381/17/24/305}{\emph{Class.
  Quant. Grav.} {\bfseries 17} (2000) 4999}
  [\href{https://arxiv.org/abs/gr-qc/0007021}{{\ttfamily gr-qc/0007021}}].

\bibitem{Aalsma:2021kle}
L.~Aalsma, A.~Cole, E.~Morvan, J.P.~van~der Schaar and G.~Shiu, \emph{{Shocks
  and information exchange in de Sitter space}},
  \href{https://doi.org/10.1007/JHEP10(2021)104}{\emph{JHEP} {\bfseries 10}
  (2021) 104} [\href{https://arxiv.org/abs/2105.12737}{{\ttfamily
  2105.12737}}].

\bibitem{Aguilar-Gutierrez:2023ymx}
S.E.~Aguilar-Gutierrez, R.~Esp\'\i{}ndola and E.K.~Morvan-Benhaim, \emph{{A
  teleportation protocol in Schwarzschild-de Sitter space}},
  \href{https://arxiv.org/abs/2308.13516}{{\ttfamily 2308.13516}}.

\bibitem{Bousso:1999cb}
R.~Bousso, \emph{{Holography in general space-times}},
  \href{https://doi.org/10.1088/1126-6708/1999/06/028}{\emph{JHEP} {\bfseries
  06} (1999) 028} [\href{https://arxiv.org/abs/hep-th/9906022}{{\ttfamily
  hep-th/9906022}}].

\bibitem{Caginalp:2019fyt}
R.J.~Caginalp, \emph{{Holographic Complexity in FRW Spacetimes}},
  \href{https://doi.org/10.1103/PhysRevD.101.066027}{\emph{Phys. Rev. D}
  {\bfseries 101} (2020) 066027}
  [\href{https://arxiv.org/abs/1906.02227}{{\ttfamily 1906.02227}}].

\bibitem{Bousso:1998bn}
R.~Bousso, \emph{{Proliferation of de Sitter space}},
  \href{https://doi.org/10.1103/PhysRevD.58.083511}{\emph{Phys. Rev. D}
  {\bfseries 58} (1998) 083511}
  [\href{https://arxiv.org/abs/hep-th/9805081}{{\ttfamily hep-th/9805081}}].

\bibitem{Bousso:1999ms}
R.~Bousso, \emph{{Quantum global structure of de Sitter space}},
  \href{https://doi.org/10.1103/PhysRevD.60.063503}{\emph{Phys. Rev. D}
  {\bfseries 60} (1999) 063503}
  [\href{https://arxiv.org/abs/hep-th/9902183}{{\ttfamily hep-th/9902183}}].

\bibitem{Aguilar-Gutierrez:2021bns}
S.E.~Aguilar-Gutierrez, A.~Chatwin-Davies, T.~Hertog, N.~Pinzani-Fokeeva and
  B.~Robinson, \emph{{Islands in Multiverse Models}},
  \href{https://doi.org/10.1007/JHEP11(2021)212}{\emph{JHEP} {\bfseries 11}
  (2021) 212} [\href{https://arxiv.org/abs/2108.01278}{{\ttfamily
  2108.01278}}].

\bibitem{Langhoff:2021uct}
K.~Langhoff, C.~Murdia and Y.~Nomura, \emph{{Multiverse in an inverted
  island}}, \href{https://doi.org/10.1103/PhysRevD.104.086007}{\emph{Phys. Rev.
  D} {\bfseries 104} (2021) 086007}
  [\href{https://arxiv.org/abs/2106.05271}{{\ttfamily 2106.05271}}].

\bibitem{Yadav:2023qfg}
G.~Yadav, \emph{{Multiverse in Karch-Randall Braneworld}},
  \href{https://doi.org/10.1007/JHEP03(2023)103}{\emph{JHEP} {\bfseries 03}
  (2023) 103} [\href{https://arxiv.org/abs/2301.06151}{{\ttfamily
  2301.06151}}].

\bibitem{Aguilar-Gutierrez:2023zoi}
S.E.~Aguilar-Gutierrez and F.~Landgren, \emph{{A multiverse model in dS wedge
  holography}},  \href{https://arxiv.org/abs/2311.02074}{{\ttfamily
  2311.02074}}.

\bibitem{Hartle:2016tpo}
J.~Hartle and T.~Hertog, \emph{{One Bubble to Rule Them All}},
  \href{https://doi.org/10.1103/PhysRevD.95.123502}{\emph{Phys. Rev. D}
  {\bfseries 95} (2017) 123502}
  [\href{https://arxiv.org/abs/1604.03580}{{\ttfamily 1604.03580}}].

\bibitem{Aguilar-Gutierrez:2024xfi}
S.E.~Aguilar-Gutierrez, \emph{{Holographic complexity of axion-de Sitter
  universes}},  \href{https://arxiv.org/abs/2401.00851}{{\ttfamily
  2401.00851}}.

\bibitem{Carmi:2017jqz}
D.~Carmi, S.~Chapman, H.~Marrochio, R.C.~Myers and S.~Sugishita, \emph{{On the
  Time Dependence of Holographic Complexity}},
  \href{https://doi.org/10.1007/JHEP11(2017)188}{\emph{JHEP} {\bfseries 11}
  (2017) 188} [\href{https://arxiv.org/abs/1709.10184}{{\ttfamily
  1709.10184}}].

\bibitem{Chapman:2018bqj}
S.~Chapman, D.~Ge and G.~Policastro, \emph{{Holographic Complexity for Defects
  Distinguishes Action from Volume}},
  \href{https://doi.org/10.1007/JHEP05(2019)049}{\emph{JHEP} {\bfseries 05}
  (2019) 049} [\href{https://arxiv.org/abs/1811.12549}{{\ttfamily
  1811.12549}}].

\bibitem{Sato:2019kik}
Y.~Sato and K.~Watanabe, \emph{{Does Boundary Distinguish Complexities?}},
  \href{https://doi.org/10.1007/JHEP11(2019)132}{\emph{JHEP} {\bfseries 11}
  (2019) 132} [\href{https://arxiv.org/abs/1908.11094}{{\ttfamily
  1908.11094}}].

\bibitem{Braccia:2019xxi}
P.~Braccia, A.L.~Cotrone and E.~Tonni, \emph{{Complexity in the presence of a
  boundary}}, \href{https://doi.org/10.1007/JHEP02(2020)051}{\emph{JHEP}
  {\bfseries 02} (2020) 051}
  [\href{https://arxiv.org/abs/1910.03489}{{\ttfamily 1910.03489}}].

\bibitem{Auzzi:2021nrj}
R.~Auzzi, S.~Baiguera, S.~Bonansea, G.~Nardelli and K.~Toccacelo, \emph{{Volume
  complexity for Janus AdS$_{3}$ geometries}},
  \href{https://doi.org/10.1007/JHEP08(2021)045}{\emph{JHEP} {\bfseries 08}
  (2021) 045} [\href{https://arxiv.org/abs/2105.08729}{{\ttfamily
  2105.08729}}].

\bibitem{Baiguera:2021cba}
S.~Baiguera, S.~Bonansea and K.~Toccacelo, \emph{{Volume complexity for the
  nonsupersymmetric Janus AdS5 geometry}},
  \href{https://doi.org/10.1103/PhysRevD.104.086030}{\emph{Phys. Rev. D}
  {\bfseries 104} (2021) 086030}
  [\href{https://arxiv.org/abs/2105.12743}{{\ttfamily 2105.12743}}].

\bibitem{Auzzi:2021ozb}
R.~Auzzi, S.~Baiguera, S.~Bonansea and G.~Nardelli, \emph{{Action complexity in
  the presence of defects and boundaries}},
  \href{https://doi.org/10.1007/JHEP02(2022)118}{\emph{JHEP} {\bfseries 02}
  (2022) 118} [\href{https://arxiv.org/abs/2112.03290}{{\ttfamily
  2112.03290}}].

\bibitem{Balasubramanian:2001nb}
V.~Balasubramanian, J.~de~Boer and D.~Minic, \emph{{Mass, entropy and
  holography in asymptotically de Sitter spaces}},
  \href{https://doi.org/10.1103/PhysRevD.65.123508}{\emph{Phys. Rev. D}
  {\bfseries 65} (2002) 123508}
  [\href{https://arxiv.org/abs/hep-th/0110108}{{\ttfamily hep-th/0110108}}].

\bibitem{Ghezelbash:2001vs}
A.M.~Ghezelbash and R.B.~Mann, \emph{{Action, mass and entropy of
  Schwarzschild-de Sitter black holes and the de Sitter / CFT correspondence}},
  \href{https://doi.org/10.1088/1126-6708/2002/01/005}{\emph{JHEP} {\bfseries
  01} (2002) 005} [\href{https://arxiv.org/abs/hep-th/0111217}{{\ttfamily
  hep-th/0111217}}].

\bibitem{Heinicke:2014ipp}
C.~Heinicke and F.W.~Hehl, \emph{{Schwarzschild and Kerr Solutions of
  Einstein's Field Equation -- an introduction}},
  \href{https://doi.org/10.1142/S0218271815300062}{\emph{Int. J. Mod. Phys. D}
  {\bfseries 24} (2014) 1530006}
  [\href{https://arxiv.org/abs/1503.02172}{{\ttfamily 1503.02172}}].

\bibitem{Bousso:1996au}
R.~Bousso and S.W.~Hawking, \emph{{Pair creation of black holes during
  inflation}}, \href{https://doi.org/10.1103/PhysRevD.54.6312}{\emph{Phys. Rev.
  D} {\bfseries 54} (1996) 6312}
  [\href{https://arxiv.org/abs/gr-qc/9606052}{{\ttfamily gr-qc/9606052}}].

\bibitem{Morvan:2022ybp}
E.K.~Morvan, J.P.~van~der Schaar and M.R.~Visser, \emph{{On the Euclidean
  action of de Sitter black holes and constrained instantons}},
  \href{https://doi.org/10.21468/SciPostPhys.14.2.022}{\emph{SciPost Phys.}
  {\bfseries 14} (2023) 022}
  [\href{https://arxiv.org/abs/2203.06155}{{\ttfamily 2203.06155}}].

\bibitem{Anninos:2012qw}
D.~Anninos, \emph{{De Sitter Musings}},
  \href{https://doi.org/10.1142/S0217751X1230013X}{\emph{Int. J. Mod. Phys. A}
  {\bfseries 27} (2012) 1230013}
  [\href{https://arxiv.org/abs/1205.3855}{{\ttfamily 1205.3855}}].

\bibitem{Svesko:2022txo}
A.~Svesko, E.~Verheijden, E.P.~Verlinde and M.R.~Visser, \emph{{Quasi-local
  energy and microcanonical entropy in two-dimensional nearly de Sitter
  gravity}}, \href{https://doi.org/10.1007/JHEP08(2022)075}{\emph{JHEP}
  {\bfseries 08} (2022) 075}
  [\href{https://arxiv.org/abs/2203.00700}{{\ttfamily 2203.00700}}].

\bibitem{Maldacena:2019cbz}
J.~Maldacena, G.J.~Turiaci and Z.~Yang, \emph{{Two dimensional Nearly de Sitter
  gravity}}, \href{https://doi.org/10.1007/JHEP01(2021)139}{\emph{JHEP}
  {\bfseries 01} (2021) 139}
  [\href{https://arxiv.org/abs/1904.01911}{{\ttfamily 1904.01911}}].

\bibitem{Shankaranarayanan:2003ya}
S.~Shankaranarayanan, \emph{{Temperature and entropy of Schwarzschild-de Sitter
  space-time}}, \href{https://doi.org/10.1103/PhysRevD.67.084026}{\emph{Phys.
  Rev. D} {\bfseries 67} (2003) 084026}
  [\href{https://arxiv.org/abs/gr-qc/0301090}{{\ttfamily gr-qc/0301090}}].

\bibitem{Choudhury:2004ph}
T.R.~Choudhury and T.~Padmanabhan, \emph{{Concept of temperature in
  multi-horizon spacetimes: Analysis of Schwarzschild-de Sitter metric}},
  \href{https://doi.org/10.1007/s10714-007-0489-0}{\emph{Gen. Rel. Grav.}
  {\bfseries 39} (2007) 1789}
  [\href{https://arxiv.org/abs/gr-qc/0404091}{{\ttfamily gr-qc/0404091}}].

\bibitem{Visser:2012wu}
M.~Visser, \emph{{Area products for stationary black hole horizons}},
  \href{https://doi.org/10.1103/PhysRevD.88.044014}{\emph{Phys. Rev. D}
  {\bfseries 88} (2013) 044014}
  [\href{https://arxiv.org/abs/1205.6814}{{\ttfamily 1205.6814}}].

\bibitem{Levine:2022wos}
A.~Levine and E.~Shaghoulian, \emph{{Encoding beyond cosmological horizons in
  de Sitter JT gravity}},
  \href{https://doi.org/10.1007/JHEP02(2023)179}{\emph{JHEP} {\bfseries 02}
  (2023) 179} [\href{https://arxiv.org/abs/2204.08503}{{\ttfamily
  2204.08503}}].

\bibitem{Dyson:2002pf}
L.~Dyson, M.~Kleban and L.~Susskind, \emph{{Disturbing implications of a
  cosmological constant}},
  \href{https://doi.org/10.1088/1126-6708/2002/10/011}{\emph{JHEP} {\bfseries
  10} (2002) 011} [\href{https://arxiv.org/abs/hep-th/0208013}{{\ttfamily
  hep-th/0208013}}].

\bibitem{Susskind:2011ap}
L.~Susskind, \emph{{Addendum to Fast Scramblers}},
  \href{https://arxiv.org/abs/1101.6048}{{\ttfamily 1101.6048}}.

\bibitem{Susskind:2021dfc}
L.~Susskind, \emph{{Black Holes Hint towards De Sitter Matrix Theory}},
  \href{https://doi.org/10.3390/universe9080368}{\emph{Universe} {\bfseries 9}
  (2023) 368} [\href{https://arxiv.org/abs/2109.01322}{{\ttfamily
  2109.01322}}].

\bibitem{Susskind:2021omt}
L.~Susskind, \emph{{De Sitter Holography: Fluctuations, Anomalous Symmetry, and
  Wormholes}}, \href{https://doi.org/10.3390/universe7120464}{\emph{Universe}
  {\bfseries 7} (2021) 464} [\href{https://arxiv.org/abs/2106.03964}{{\ttfamily
  2106.03964}}].

\bibitem{Hartman:2013qma}
T.~Hartman and J.~Maldacena, \emph{{Time Evolution of Entanglement Entropy from
  Black Hole Interiors}},
  \href{https://doi.org/10.1007/JHEP05(2013)014}{\emph{JHEP} {\bfseries 05}
  (2013) 014} [\href{https://arxiv.org/abs/1303.1080}{{\ttfamily 1303.1080}}].

\bibitem{Chapman:2016hwi}
S.~Chapman, H.~Marrochio and R.C.~Myers, \emph{{Complexity of Formation in
  Holography}}, \href{https://doi.org/10.1007/JHEP01(2017)062}{\emph{JHEP}
  {\bfseries 01} (2017) 062}
  [\href{https://arxiv.org/abs/1610.08063}{{\ttfamily 1610.08063}}].

\bibitem{Lehner:2016vdi}
L.~Lehner, R.C.~Myers, E.~Poisson and R.D.~Sorkin, \emph{{Gravitational action
  with null boundaries}},
  \href{https://doi.org/10.1103/PhysRevD.94.084046}{\emph{Phys. Rev. D}
  {\bfseries 94} (2016) 084046}
  [\href{https://arxiv.org/abs/1609.00207}{{\ttfamily 1609.00207}}].

\bibitem{Carmi:2016wjl}
D.~Carmi, R.C.~Myers and P.~Rath, \emph{{Comments on Holographic Complexity}},
  \href{https://doi.org/10.1007/JHEP03(2017)118}{\emph{JHEP} {\bfseries 03}
  (2017) 118} [\href{https://arxiv.org/abs/1612.00433}{{\ttfamily
  1612.00433}}].

\bibitem{Parattu:2015gga}
K.~Parattu, S.~Chakraborty, B.R.~Majhi and T.~Padmanabhan, \emph{{A Boundary
  Term for the Gravitational Action with Null Boundaries}},
  \href{https://doi.org/10.1007/s10714-016-2093-7}{\emph{Gen. Rel. Grav.}
  {\bfseries 48} (2016) 94} [\href{https://arxiv.org/abs/1501.01053}{{\ttfamily
  1501.01053}}].

\bibitem{Chakraborty:2016yna}
S.~Chakraborty, \emph{{Boundary Terms of the Einstein\textendash{}Hilbert
  Action}}, \href{https://doi.org/10.1007/978-3-319-51700-1_5}{\emph{Fundam.
  Theor. Phys.} {\bfseries 187} (2017) 43}
  [\href{https://arxiv.org/abs/1607.05986}{{\ttfamily 1607.05986}}].

\bibitem{Chapman:2018dem}
S.~Chapman, H.~Marrochio and R.C.~Myers, \emph{{Holographic complexity in
  Vaidya spacetimes. Part I}},
  \href{https://doi.org/10.1007/JHEP06(2018)046}{\emph{JHEP} {\bfseries 06}
  (2018) 046} [\href{https://arxiv.org/abs/1804.07410}{{\ttfamily
  1804.07410}}].

\bibitem{Chapman:2018lsv}
S.~Chapman, H.~Marrochio and R.C.~Myers, \emph{{Holographic complexity in
  Vaidya spacetimes. Part II}},
  \href{https://doi.org/10.1007/JHEP06(2018)114}{\emph{JHEP} {\bfseries 06}
  (2018) 114} [\href{https://arxiv.org/abs/1805.07262}{{\ttfamily
  1805.07262}}].

\bibitem{MARSDEN1980109}
J.E.~Marsden and F.J.~Tipler, \emph{Maximal hypersurfaces and foliations of
  constant mean curvature in general relativity},
  \href{https://doi.org/https://doi.org/10.1016/0370-1573(80)90154-4}{\emph{Physics
  Reports} {\bfseries 66} (1980) 109}.

\bibitem{Faruk:2023uzs}
M.M.~Faruk, E.~Morvan and J.P.~van~der Schaar, \emph{{Static sphere observers
  and geodesics in Schwarzschild-de Sitter spacetime}},
  \href{https://arxiv.org/abs/2312.06878}{{\ttfamily 2312.06878}}.

\bibitem{Galante:2022nhj}
D.~Galante, \emph{{Geodesics, complexity and holography in (A)dS$_2$}},
  \href{https://doi.org/10.22323/1.406.0359}{\emph{PoS} {\bfseries CORFU2021}
  (2022) 359}.

\bibitem{Chapman:2022mqd}
S.~Chapman, D.A.~Galante, E.~Harris, S.U.~Sheorey and D.~Vegh, \emph{{Complex
  geodesics in de Sitter space}},
  \href{https://doi.org/10.1007/JHEP03(2023)006}{\emph{JHEP} {\bfseries 03}
  (2023) 006} [\href{https://arxiv.org/abs/2212.01398}{{\ttfamily
  2212.01398}}].

\bibitem{Aalsma:2022eru}
L.~Aalsma, M.M.~Faruk, J.P.~van~der Schaar, M.R.~Visser and J.~de~Witte,
  \emph{{Late-time correlators and complex geodesics in de Sitter space}},
  \href{https://doi.org/10.21468/SciPostPhys.15.1.031}{\emph{SciPost Phys.}
  {\bfseries 15} (2023) 031}
  [\href{https://arxiv.org/abs/2212.01394}{{\ttfamily 2212.01394}}].

\bibitem{Blommaert:2023opb}
A.~Blommaert, T.G.~Mertens and S.~Yao, \emph{{Dynamical actions and
  q-representation theory for double-scaled SYK}},
  \href{https://arxiv.org/abs/2306.00941}{{\ttfamily 2306.00941}}.

\bibitem{Narovlansky:2023lfz}
V.~Narovlansky and H.~Verlinde, \emph{{Double-scaled SYK and de Sitter
  Holography}},  \href{https://arxiv.org/abs/2310.16994}{{\ttfamily
  2310.16994}}.

\bibitem{Blommaert:2023wad}
A.~Blommaert, T.G.~Mertens and S.~Yao, \emph{{The q-Schwarzian and Liouville
  gravity}},  \href{https://arxiv.org/abs/2312.00871}{{\ttfamily 2312.00871}}.

\bibitem{Verlinde:2024znh}
H.~Verlinde, \emph{{Double-scaled SYK, Chords and de Sitter Gravity}},
  \href{https://arxiv.org/abs/2402.00635}{{\ttfamily 2402.00635}}.

\bibitem{Verlinde:2024zrh}
H.~Verlinde and M.~Zhang, \emph{{SYK Correlators from 2D Liouville-de Sitter
  Gravity}},  \href{https://arxiv.org/abs/2402.02584}{{\ttfamily 2402.02584}}.

\bibitem{Nielsen1}
M.A.~Nielsen, \emph{A geometric approach to quantum circuit lower bounds},
  \href{https://arxiv.org/abs/quant-ph/0502070}{{\ttfamily quant-ph/0502070}}.

\bibitem{Nielsen2}
M.A.~Nielsen, M.R.~Dowling, M.~Gu and A.C.~Doherty, \emph{Quantum computation
  as geometry}, \href{https://doi.org/10.1126/science.1121541}{\emph{Science}
  {\bfseries 311} (2006) 1133}.

\bibitem{Nielsen3}
M.R.~Dowling and M.A.~Nielsen, \emph{The geometry of quantum computation},
  \href{https://arxiv.org/abs/quant-ph/0701004}{{\ttfamily quant-ph/0701004}}.

\bibitem{Parker:2018yvk}
D.E.~Parker, X.~Cao, A.~Avdoshkin, T.~Scaffidi and E.~Altman, \emph{{A
  Universal Operator Growth Hypothesis}},
  \href{https://doi.org/10.1103/PhysRevX.9.041017}{\emph{Phys. Rev. X}
  {\bfseries 9} (2019) 041017}
  [\href{https://arxiv.org/abs/1812.08657}{{\ttfamily 1812.08657}}].

\bibitem{Carrasco:2023fcj}
R.~Carrasco, J.F.~Pedraza, A.~Svesko and Z.~Weller-Davies, \emph{{Gravitation
  from optimized computation: Einstein and beyond}},
  \href{https://doi.org/10.1007/JHEP09(2023)167}{\emph{JHEP} {\bfseries 09}
  (2023) 167} [\href{https://arxiv.org/abs/2306.08503}{{\ttfamily
  2306.08503}}].

\bibitem{Emparan:2021hyr}
R.~Emparan, A.M.~Frassino, M.~Sasieta and M.~Toma\v{s}evi\'c,
  \emph{{Holographic complexity of quantum black holes}},
  \href{https://doi.org/10.1007/JHEP02(2022)204}{\emph{JHEP} {\bfseries 02}
  (2022) 204} [\href{https://arxiv.org/abs/2112.04860}{{\ttfamily
  2112.04860}}].

\bibitem{Chen:2023tpi}
B.~Chen, Y.~Liu and B.~Yu, \emph{{Holographic complexity of rotating quantum
  black holes}}, \href{https://doi.org/10.1007/JHEP01(2024)055}{\emph{JHEP}
  {\bfseries 01} (2024) 055}
  [\href{https://arxiv.org/abs/2307.15968}{{\ttfamily 2307.15968}}].

\bibitem{Emparan:2022ijy}
R.~Emparan, J.F.~Pedraza, A.~Svesko, M.~Toma\v{s}evi\'c and M.R.~Visser,
  \emph{{Black holes in dS$_{3}$}},
  \href{https://doi.org/10.1007/JHEP11(2022)073}{\emph{JHEP} {\bfseries 11}
  (2022) 073} [\href{https://arxiv.org/abs/2207.03302}{{\ttfamily
  2207.03302}}].

\end{thebibliography}\endgroup
\end{document}